\documentclass[11pt]{article}

\usepackage[utf8]{inputenc}
\usepackage[T1]{fontenc}
\usepackage[a4paper,margin=2.5cm]{geometry}
\usepackage{amsmath,amssymb,amsthm}
\usepackage{booktabs}
\usepackage{graphicx}
\usepackage{array}
\usepackage{multirow}
\usepackage{longtable}
\usepackage{tikz}
\usepackage{algorithm}
\usepackage{algpseudocode}
\usepackage[authoryear,round]{natbib}
\usepackage[hidelinks]{hyperref}

\newcommand{\tablesize}{\footnotesize}

\theoremstyle{definition}

\theoremstyle{remark}

\newcommand{\Zn}{\mathcal{Z}^{n}}
\newcommand{\Znb}{\mathcal{Z}^{n\backslash b}}
\newcommand{\Znnu}{\mathcal{Z}^{n\backslash\boldsymbol{\nu}}}
\newcommand{\cN}{\mathcal{N}}
\newcommand{\ZnN}{\mathcal{Z}^{n\backslash\mathcal{N}}}
\newcommand{\Znlambda}{\mathcal{Z}^{n\backslash\lambda}}
\newcommand{\Tnk}{\mathcal{T}^{n,k}}
\newcommand{\Sn}{\mathcal{S}^{n}}
\newcommand{\C}{\mathbf{C}}

\newcommand{\bnu}{\boldsymbol{\nu}}
\newcommand{\tauX}{\tau_X}
\newcommand{\stirling}[2]{\genfrac{\{}{\}}{0pt}{}{#1}{#2}}
\newcommand{\free}{\ast}
\newcommand{\pre}{\prec}
\DeclareMathOperator*{\argmax}{arg\,max}
\DeclareMathOperator*{\argmin}{arg\,min}

\title{Kemeny median rankings with ties under constraints on the buckets}

\author{Maurizio Romano\thanks{Department of Business and Economics, University of Cagliari, Italy.}
\and Luca Coraggio\thanks{Department of Economics and Statistics, University of Naples Federico II, Italy.}
\and Claudio Conversano\footnotemark[1]
\and Antonio D'Ambrosio\footnotemark[2]\ \thanks{Corresponding author: antdambr@unina.it}}

\date{}

\begin{document}

\maketitle

\begin{abstract}
The median ranking is the ranking that minimizes the sum of the Kemeny distances from the
rankings expressed by a set of judges. In many applications the consensus requires a prescribed structure: a given number of buckets, given sizes of all or some of the
buckets, as in the forced distributions of Q methodology, or the form of a top-$k$ list.
We show that all these problems are a single problem, the search for the median ranking
among the rankings whose pattern of bucket sizes is admissible, and that they depend on
the data only through the combined input matrix, defined by Emond and Mason in 2002. On this matrix we build BOLT, a
deterministic heuristic that combines a set of starting solutions, a local search that
moves one object at a time and can create or remove ties, and the exact best partition of
a fixed order of the objects into buckets with an admissible pattern. We also provide an exact solution for the median top-$k$ list. 
The solutions of BOLT are assessed, on real and simulated data, against
the optimum certified by a binary linear programming formulation, adapted to the Kemeny
distance and to the constraints on the buckets, and against the methods available in the
literature. Applications to Q-sorts, top-$k$ lists and large
and sparse data sets of ratings show how the constrained problems arise in practice, and
how the treatment of the objects not ranked by the judges changes the solution.
\end{abstract}

\noindent\textbf{Keywords:} Median ranking; Kemeny distance; Combined input matrix; Bucket order;
Heuristics; Binary linear programming

\section{Introduction}\label{sec:intro}

Preference rankings are collected in many fields, ranging from the social and behavioral sciences
to marketing, political science, information retrieval and bioinformatics
\citep{marden1996,alvoyu2014}. 
When a set of $m$ judges ranks a set of $n$ objects (or items, or alternatives), a common task is to summarize their preferences into a single ranking, 
called the consensus or central ranking. Among the many ways of defining a consensus ranking, the definition that describes it as the ranking that minimizes the sum of the distances from the ranks assigned by the judges has become the most prominent, likely due to the formal definition of the median ranking provided by \citet{kemeny1959} and \citet{kemenysnell1962}. Over time, the so-called Kemeny problem has become synonymous with the rank aggregation problem, without paying too much attention to the distance used to address it or to the nature of the ranks themselves. The distance defined by \citet{kemenysnell1962}, framed within a formal and rigorous set of axioms, naturally handles ties, both in the judges' rankings and in the median ranking, where they have the meaning of genuine indifference or, as reported by \citet{emondmason2002}, a ``positive statement of agreement''.

Historically, the rank aggregation problem has mostly dealt with finding the central ranking in cases where tied rankings (or weak rankings) were not
considered either by the judges or in the median ranking itself, whether within or outside Kemeny's axiomatic framework \citep[see, for example,][]{cooksaipe1976,younglevenglick1978,Cook1978,barthelemyetal1989,faginetal2004,cook2006,Cook2007,alimeila2012}; in this setting, the problem was shown to be NP-hard \citep{bartholdietal1989,dworketal2001}.

\noindent In this paper, we refer generically to ``rankings,'' using this term to include all possible cases of tied rankings unless explicitly defined otherwise. 
Considerable progress has been made in exact methods for median ranking with ties. \citet{yooescobedo2021} proposed a binary linear programming formulation in which a binary variable indicates whether 
the first object is ranked ahead of or tied with the second object for each ordered pair of objects. Transitivity is imposed by linear 
constraints on triples of objects. The authors also introduced a social choice property for rankings with ties, which allows some instances 
to be decomposed into smaller ones. \citet{akbariescobedo2021} provided lower bounds for the problem involving ties. Additionally, 
\citet{akbariescobedo2023b} extended the exact and heuristic methods to a family of distances with a parameterized penalty for ties. 
This family contains the Kemeny distance as a special case. \citet{kawashimasukegawa2026} studied the linear ordering problem with ties,
which includes the median ranking problem with ties. They provided conditions under which two objects are ranked in a given order or tied in every optimal solution.
When ties are not allowed in the solution but can occur among the rankings expressed by judges, the MorK algorithms of \citet{azzinimunda2020} and \citet{azzinietal2026}, the exact algorithms of \citet{ricoetal2023}, and the space reduction techniques of \citet{phunghamel2025} find the median ranking in the space of permutations. Despite this progress, exact methods remain computationally expensive for instances with a few hundred objects, especially when the judges disagree.

Within the same axiomatic framework, heuristic methods have been proposed.
\citet{amodioetal2016} proposed QUICK and FAST, and \citet{dambrosioetal2017} proposed a differential evolution algorithm called DECoR. \citet{romanoetal2025} proposed a particle swarm optimization algorithm. \citet{badaldas2018} proposed the subiterative convergence algorithms FUR and SIgFUR for complete rankings. \citet{badalsingh2026} extended these algorithms to rankings with ties. \citet{akbariescobedo2023a} proposed Approximate Condorcet Partitioning, which decomposes large instances into smaller subproblems and provides an instance-specific bound on the quality of the solution. \citet{lorenaetal2022} proposed a biased random-key genetic algorithm with a local search whose insertion and exchange moves are evaluated in linear time. Their objective function, taken from \citet{yooetal2020}, is computed from a matrix that coincides with the combined input matrix of \citet{emondmason2002}, in which objects not ranked by a judge do not contribute.
\citet{andrieuetal2023} proposed a framework containing several generalizations of the Kemeny score for incomplete rankings with ties, which include the Kemeny distance for complete rankings. They also proposed an exact algorithm, partitioning methods, and generalized versions of KwikSort \citep{ailoncharikarnewman2008} and of the BioConsert local search \citep{cohenboulakiaetal2011}. Most of these heuristics are randomized, so repeated runs can produce different solutions.

Other methods have a guaranteed approximation ratio. The KwikSort algorithm is a randomized pivoting algorithm, derandomized by \citet{vanzuylenwilliamson2009}; \citet{brancotteetal2015} adapted it to rankings with ties. \citet{faginetal2004} showed that the aggregation of rankings with ties through the median of the positions, followed by an optimal grouping of the ordered objects into buckets, is a constant-factor approximation for a family of distances that contains the Kemeny distance. Some of the methods developed for ``rank aggregation with ties'' optimize a different objective: \citet{cohenboulakiaetal2011} and \citet{brancotteetal2015} assign the same cost to a pair ordered in opposite ways and to a pair tied in only one of the two rankings, which is not the Kemeny distance.

Another related problem is the optimal bucket order problem \citep{gionisetal2006,ukkonenetal2009}, in which the data are summarized by a matrix of precedence fractions. The solution is the bucket order whose matrix is closest to the data in the $L_1$ norm. Its heuristics \citep{aledoetal2017,aledoetal2018,aledoetal2021}, exact formulation \citep{aledoetal2025}, and extension to a set of bucket orders \citep{aledogamezrosete2025} solve a problem with a different goal. For example, considering 100 judges and two objects $a$ and $b$, when 51 of them prefer $a$ to $b$ and 49 prefer $b$ to $a$, the optimal bucket order ties $a$ and $b$, while the median ranking ranks $a$ ahead of $b$ \citep{dambrosioetal2017}.

In many applications, the consensus has a prescribed structure. The median constrained bucket order \citep[MCBO, ][]{dambrosioetal2019} is the median ranking with a prescribed number of buckets. In Q methodology \citep{stephenson1935,brown1980,wattsstenner2012} and performance appraisal, objects must be distributed among ordered categories of prescribed sizes. A top-$k$ list asks for a ranking of the first $k$ objects only \citep{faginetal2003}. Recently, \citet{aledoetal2025} formulated the optimal bucket order problem with a prescribed number of buckets, with prescribed bucket sizes, and with a variant of the top-$k$ structure as an integer linear programming problem.

This paper addresses the median ranking problem, with and without constraints on the buckets, within the Kemeny axiomatic framework and is entirely based on the properties of the combined input matrix. This matrix has been known since \citet{emondmason2002} and is sometimes used under a different name by exact formulations \citep{yooescobedo2021}. Here, we use it as the single object on which all procedures work. The objective function, starting solutions, moves of the local search, best partition of a fixed order into buckets, lower bound, and exact formulation are all written in terms of this matrix, so that their cost does not depend on the number of judges.

The contribution of this study is twofold. First, we describe all the problems through the patterns that the solution is allowed to have, i.e., the sizes of its buckets. The median ranking, the MCBO, the median ranking with all or some prescribed bucket sizes, the median top-$k$ list, and the median ranking in the space of permutations are all the same problem with different sets of admissible patterns.
For the median top-$k$ list, the objective function splits into selecting $k$ objects and a median ranking problem on the selected objects. This leads to an exact procedure. We then propose BOLT, a heuristic that solves all these problems using the same scheme: starting solutions; a local search that moves one object at a time; and the best partition of a fixed order into buckets with an admissible pattern. While none of these components are novel in isolation, our contribution lies in combining them on the combined input matrix, extending them to partial patterns and the median top-$k$ list, and systematically assessing the heuristic against the certified optimum. To this end, we adapt the binary linear programming formulation of \citet{yooescobedo2021} and the constraints of \citet{aledoetal2025} to the Kemeny distance and partial patterns. We use these adaptations, together with the results published for other methods on their own benchmarks, to measure the quality of BOLT's solutions on real and simulated data.

The paper is organized as follows: Section~\ref{sec:median} recalls the median ranking problem and the combined input matrix, the role of which is discussed in Section~\ref{sec:cij}. Section~\ref{sec:constraints} defines the constrained problems. Section~\ref{sec:bolt} describes BOLT, and Section~\ref{sec:exact} describes the exact formulation.
Section~\ref{sec:experiments} reports the experimental evaluation, and Section~\ref{sec:applications} discusses the applications. Section~\ref{sec:conclusions} concludes. Derivations, pseudocode, and full tables are included in the supplementary material.

\section{Median ranking under the Kemeny axiomatic approach}\label{sec:median}

Let $\Zn$ be the universe of the rankings of $n$ objects. Its
cardinality is
\begin{equation}\label{eq:Zn}
|\Zn| = \sum_{b=1}^{n} b!\,\stirling{n}{b},
\end{equation}
where $\stirling{n}{b}$ is the Stirling number of the second kind, i.e., the number of
ways to partition a set of $n$ objects into $b$ nonempty subsets. These subsets are
the buckets, and the factor $b!$ accounts for their order, so that $b!\stirling{n}{b}$ is
the number of rankings with exactly $b$ buckets \citep{dambrosioetal2019}. The number
$|\Zn|$ is known as the $n$-th ordered Bell (or Fubini) number \citep{gross1962,good1975}.
The universe of the complete rankings, i.e., the permutations of the first $n$
integers, is denoted by $\Sn$, with $|\Sn|=n!$.
A ranking can be represented in two ways \citep{marden1996}. The rank vector, or
ranking in the strict sense, assigns to each object its position; the order vector,
or ordering, lists the objects from the most preferred to the least preferred, with
tied objects in curly brackets. For example, with the objects $a,b,c,d$ listed in
alphabetical order, the rank vector $(2,3,1,2)$ and the order vector
$\langle c,\{a,d\},b\rangle$ describe the same ranking, in which $a$ and $d$ are tied
in the second position. In the following we use the word ranking for both, and we
specify the representation only when it matters.

Let $Y$ and $X$ be two rankings of $n$ objects. Following \citet{kemenysnell1962}, the
Kemeny distance between $X$ and $Y$ is
\begin{equation}\label{eq:kemeny}
d(X,Y) = \frac{1}{2}\sum_{i,j=1}^{n} \left| x_{ij} - y_{ij} \right|,
\end{equation}
where $x_{ij}$ and $y_{ij}$ are the elements of the score matrices of $X$ and $Y$:
$x_{ij}=1$ if the $i$th object is ranked ahead of the $j$th object, $x_{ij}=-1$ if it is
ranked behind, and $x_{ij}=0$ if the two objects are tied or if $i=j$. 
\citet{kemenysnell1962} proved that this distance is the unique measure that satisfies
Kemeny's axioms: any distance for rankings must
be a metric, it must be invariant under a relabeling of the objects, it must be
consistent in measurement, and its minimum positive value must be equal to one.

\citet{emondmason2002} defined the rank correlation coefficient $\tauX$, proving that it is unequivocally related to the Kemeny distance in this way:
\begin{equation}\label{eq:taux}
\tauX(X,Y) = 1 - \frac{2\,d(X,Y)}{n(n-1)} = \frac{\sum_{i,j=1}^{n} x_{ij}\,y_{ij}}{n(n-1)}.
\end{equation}
The right-hand side of Eq.~\ref{eq:taux} requires a slight modification of the score
matrix: they assigned the score $1$ to the pair $(i,j)$ if the
$i$th object is either ranked ahead of or tied with the $j$th object, and $-1$
otherwise, with $x_{ii}=0$. Hence, a tie between the objects $i$ and $j$ gives $x_{ij}=x_{ji}=1$. From now
on, the elements of the score matrices are coded in this way.

Let $X^{(1)},\dots,X^{(r)}$ be a set of distinct rankings of $n$ objects, each with
an associated weight $w_h$, with $\sum_{h=1}^{r} w_h = m$, where $m$ is the number of
judges. The weights can also be of a different nature, for example relative frequencies or
importance weights: since the median ranking does not change when all the weights are
multiplied by the same positive constant, we can always rescale them so that they sum
to $m$. The combined input matrix $\C$ \citep{emondmason2002} has elements
\begin{equation}\label{eq:cij}
c_{ij} = \sum_{h=1}^{r} w_h\, x^{(h)}_{ij},
\end{equation}
i.e., it is the weighted sum of the score matrices of the rankings of the judges.
The average $\tauX$ of a ranking $Y$ over the judges is
\begin{equation}\label{eq:avgtaux}
\bar{\tau}_X(Y) = \frac{\sum_{i,j=1}^{n} c_{ij}\, y_{ij}}{m\,n(n-1)},
\end{equation}
and the median ranking is that ranking (or those rankings) $\hat{Y}$ for which
\begin{equation}\label{eq:median}
\hat{Y} = \argmin_{Y\in\Zn} \sum_{h=1}^{r} w_h\, d(X^{(h)},Y)
= \argmax_{Y\in\Zn} \bar{\tau}_X(Y).
\end{equation}
It is worth noting that the data enter the problem only through $\C$: whatever
the number of judges, the dimensionality of the problem is governed by the number
of objects. From now on we use the average $\tauX$ as the objective function of all the
problems considered in this paper. Objects not ranked by some judges are discussed in
Section~\ref{sec:missing}.

The sum of distances corresponding to $\bar{\tau}_X(Y)$ is
$\sum_h w_h d(X^{(h)},Y)=m\,n(n-1)\left(1-\bar{\tau}_X(Y)\right)/2$, so that a difference
of $\delta$ in the average $\tauX$ corresponds to a difference of
$\delta\, m\,n(n-1)/2$ in the sum of the distances.

\section{The central role of the combined input matrix}\label{sec:cij}

All problems in this paper share the same objective function (the average $\tauX$ in Eq.~\ref{eq:avgtaux}) and differ only in the subset of $\Zn$ searched. In this section, we rewrite the objective function in a form that makes the role of the buckets explicit. We then draw the consequences used by all the paper's procedures: the heuristic in Section~\ref{sec:bolt}; the lower bound and the exact formulation in Section~\ref{sec:exact}.

\subsection{Only the ordered pairs count}\label{sec:ordered}

Let $Y\in\Zn$ and let $i\pre_Y j$ indicate that the object $i$ belongs to a bucket of $Y$
that precedes the bucket of the object $j$. Recall that, with the coding of
\citet{emondmason2002}, $y_{ij}=y_{ji}=1$ if $i$ and $j$ are tied in $Y$, $y_{ij}=1$ and
$y_{ji}=-1$ if $i\pre_Y j$, and $y_{ii}=0$. Since the diagonal elements of the score
matrix of $Y$ are zero, $\sum_{i,j}c_{ij}y_{ij}$ is the sum, over the unordered pairs
$\{i,j\}$ of distinct objects, of $c_{ij}y_{ij}+c_{ji}y_{ji}$, which is equal to
$c_{ij}+c_{ji}$ if $i$ and $j$ are tied in $Y$, and to $c_{ij}-c_{ji}=(c_{ij}+c_{ji})-2c_{ji}$
if $i\pre_Y j$ (the case $j\pre_Y i$ is symmetric). Hence, for every ranking $Y\in\Zn$ and
every combined input matrix $\C$,
\begin{equation}\label{eq:ordered}
\sum_{i,j=1}^{n} c_{ij}\,y_{ij} = \sum_{i<j}\left(c_{ij}+c_{ji}\right)
- 2\sum_{(i,j):\ i\pre_Y j} c_{ji}.
\end{equation}
The first term on the right-hand side of Eq.~\ref{eq:ordered} does not depend on $Y$.
Hence, maximizing the average $\tauX$ over any subset of $\Zn$ is equivalent to
minimizing the penalty
\begin{equation}\label{eq:penalty}
P(Y) = \sum_{(i,j):\ i\pre_Y j} c_{ji}
\end{equation}
over the same subset. Only the pairs of objects placed in different buckets
contribute to the penalty, and each of them contributes exactly the consensus in favor of
the opposite decision: $c_{ji}$ is the number of judges that rank $j$ ahead of or
tied with $i$, minus the number of judges that rank $i$ ahead of $j$. The pairs of
tied objects do not contribute. The penalty in Eq.~\ref{eq:penalty} is
closely related to the penalty used in the branch-and-bound algorithm of
\citet{emondmason2002}, which is computed incrementally by considering the elements
$c_{ij}$ and $c_{ji}$ of the pairs of objects processed so far.

\subsection{What the combined input matrix tells us}\label{sec:consequences}

Eq.~\ref{eq:ordered} has three consequences that are used throughout the paper.

\noindent First, once $\C$ has been computed, the cost of evaluating a candidate solution, of
comparing two solutions or of changing a solution locally does not depend on the number
of judges $m$, nor on the number of distinct rankings $r$. The rankings of the judges are
used only, optionally, as starting solutions of the heuristic.

\noindent Second, each unordered pair $\{i,j\}$ contributes to the penalty of Eq.~\ref{eq:penalty}
one of three values: $0$ if $i$ and $j$ are tied, $c_{ji}$ if $i\pre_Y j$, and $c_{ij}$ if
$j\pre_Y i$. Hence, for every $Y\in\Zn$,
\begin{equation}\label{eq:lb}
P(Y) \geq \sum_{i<j} \min\left\{0,\ c_{ij},\ c_{ji}\right\} = P_{LB},
\end{equation}
and $P_{LB}$ is a lower bound for the penalty of the median ranking, and therefore an
upper bound for its average $\tauX$. The bound is attained if and only if the best
choices can be made, pair by pair, so that they define a ranking. The
same bound, written in terms of the Kemeny distance, is used by
\citet{akbariescobedo2021,akbariescobedo2023a} to measure the quality of their
solutions; we use it in the same way in Section~\ref{sec:experiments}, and as the
starting bound of the exact procedure of Section~\ref{sec:exact}.

\noindent Third, Eq.~\ref{eq:ordered} shows that, for a pair taken separately, a tie is the best
choice only when both $c_{ij}$ and $c_{ji}$ are positive, i.e., when the number of judges
that rank $i$ ahead of or tied with $j$ exceeds the number of those that rank $j$ ahead
of $i$, and vice versa. When the judges express rankings without ties, $c_{ij}=-c_{ji}$ and a tie is
never better than the best order of the pair. In this case there is always a
median ranking without ties: within each bucket of a median ranking, the objects can be
ordered in some way and in the reverse way, and the contributions of the pairs of the
bucket to Eq.~\ref{eq:ordered} change sign between the two orders, so that one of them
is at least as good as the tie, while the other pairs do not change. A median ranking
with ties can still exist, but only together with an equally good permutation.

\subsection{Objects not ranked by the judges}\label{sec:missing}

In many experiments each judge could rank only some of the objects. Since the data
enter the problem only through $\C$, the treatment of the objects not ranked is a choice on $\C$ alone: the objective function, the sub-universes of
Eq.~\ref{eq:median-pattern} and the algorithm of Section~\ref{sec:algorithm} are
the same with any convention.

Let $X^{(h)}$ be the $h$-th ranking. The objects that are not ranked in $X^{(h)}$ are
placed, tied with each other, in a bucket that follows all the ranked objects, and
$x^{(h)}_{ij}$ denotes the elements of the score matrix of $X^{(h)}$, defined as in
Section~\ref{sec:median}. Placing the objects not ranked in the last bucket is just a way
to write $X^{(h)}$: which pairs of objects are actually taken into account depends on the
convention adopted, as explained below. The combined input matrix can then be written as
\begin{equation}\label{eq:cijdelta}
c_{ij} = \sum_{h=1}^{r} w_h\,I^{(h)}_{ij}\,x^{(h)}_{ij},
\end{equation}
where $I^{(h)}_{ij}=I^{(h)}_{ji}$ is equal to $1$ if the pair of objects $i$ and $j$
is counted in $X^{(h)}$, and to $0$ otherwise. With complete data all the pairs are
counted, and Eq.~\ref{eq:cijdelta} coincides with Eq.~\ref{eq:cij}: what has to be
chosen is only which pairs are counted. Three conventions can be used; the choice depends
on what a missing judgement means in the data at hand, and it has to be stated together
with the results.

The first convention \citep{emondmason2000,emondmason2002,yooetal2020} treats the objects
not ranked by a judge as \emph{missing}: $I^{(h)}_{ij}=1$ only if both $i$ and $j$
are ranked in $X^{(h)}$. It is appropriate when not ranking an object means that the judge did not
evaluate it, for example when a user ranks only the movies that he or she has watched.

The second convention counts all the pairs: $I^{(h)}_{ij}=1$ for every $i$ and $j$.
The judge is assumed to prefer every ranked object to every object that he or she did not
rank, and the objects not ranked are tied with each other. This is the usual way of treating
a partial ranking or a top-$k$ list as a bucket order \citep{faginetal2006,ailon2010},
used when not ranking an object is itself a judgement. With this convention an object ranked
by many judges is ranked ahead of all the objects that those judges did not rank, so that
the solutions also reflect how often each object is ranked.

Within the top-$k$ lists framework, however, the second convention says more than the
judges said. A respondent who chooses the three most trusted sources among seven tells us
that each of them is preferred to each of the other four, but nothing about the order of
these four: their tie is a consequence of the coding, not a judgement. The third
convention uses exactly the information given by the judge: $I^{(h)}_{ij}=0$ only if
neither $i$ nor $j$ is ranked in $X^{(h)}$. The idea goes back to
\citet[Section~8.1]{emondmason2002}, who noted that two subsets of objects can be ranked one
ahead of the other while their objects are left ``unranked'', as opposed to tied, with
respect to each other. It is also the information carried by a top-$k$ list in the
approach of \citet{critchlow1985}, in which a partial ranking is identified with the set
of the complete rankings compatible with it: in the average of their score matrices, a
pair of objects not ranked by the judge has value zero, and every other pair has value
$x^{(h)}_{ij}$, which is exactly what the judge adds to Eq.~\ref{eq:cijdelta} with the
third convention. To the best of our knowledge, this convention has not been developed so
far within the Kemeny axiomatic framework, nor compared with the other two.

The second and the third convention differ only in the pairs of objects that a judge
leaves unranked. Let $c_{ij}$ be computed with the third convention, and let $u_{ij}$ be
the number of judges who rank neither $i$ nor $j$. With the second convention, the
combined input matrix has elements $c_{ij}+u_{ij}$ for $i\neq j$.
Since $y_{ij}+y_{ji}=0$ when $Y$ orders $i$ and $j$,
and $y_{ij}+y_{ji}=2$ when $Y$ ties them, for any ranking $Y$ we have
\begin{equation}\label{eq:conv23}
\sum_{i\neq j}(c_{ij}+u_{ij})\,y_{ij}=\sum_{i\neq j}c_{ij}\,y_{ij}+2\sum_{i<j,\;y_{ij}=y_{ji}=1}u_{ij}.
\end{equation}
Hence, when ties are not allowed in the solution, the two conventions have the same median ranking. Otherwise,
the second convention rewards the solutions that tie the objects rarely
ranked by the judges, independently of their preferences. With a median top-$k$ list, whose last bucket always contains $n-k$ objects, the two conventions differ only in which objects are placed in that bucket.

\section{Constraints on the buckets}\label{sec:constraints}

In this section we describe the universe of rankings $\Zn$ through the sizes of the buckets; this description allows us to define a family of constrained problems as special cases of a single one.

\subsection{Patterns and shapes}\label{sec:patterns}

A ranking $Y\in\Zn$ with $b$ buckets can be written as an ordered sequence of
nonempty and disjoint subsets of the objects, $B_1,\dots,B_b$, whose union is the set
of all the objects. The objects of $B_1$ are tied in the first position, the objects of
$B_2$ are tied in the second position, and so on. The rank vector of $Y$ assigns
the integer $t$ to all the objects of $B_t$.

We define the \emph{pattern} of a ranking $Y\in\Zn$ with buckets
$B_1,\dots,B_b$ as the vector of the sizes of its buckets, in the order of the buckets,
\begin{equation}\label{eq:pattern}
\bnu(Y) = \left(|B_1|,|B_2|,\dots,|B_b|\right),
\end{equation}
and its \emph{shape}, $\lambda(Y)$, as the vector obtained by sorting the
elements of $\bnu(Y)$ in non-increasing order.

The pattern of a ranking is a sequence of $b$ positive integers whose sum is $n$, i.e.,
a \emph{composition} of $n$ into $b$ parts; its shape is an \emph{integer partition} of
$n$ into $b$ parts, i.e., a way of writing $n$ as a sum of $b$ positive integers
regardless of their order \citep{andrews1976,stanley2012}. A composition of
$n$ is obtained by writing $n$ units in a row and deciding, for each of the $n-1$ gaps
between two consecutive units, whether to cut or not: hence there are $2^{n-1}$
compositions of $n$, $\binom{n-1}{b-1}$ of them having $b$ parts.

For example, with $n=10$ objects, the ranking $\langle c,\{a,e\},\{b,d,f,g\},\{h,i,j\}\rangle$ has $b=4$ buckets, rank vector $(2,3,1,3,2,3,3,4,4,4)$, pattern $(1,2,4,3)$ and shape $(4,3,2,1)$. A complete ranking has pattern $(1,1,\dots,1)$; the ranking in which all the objects are
tied has pattern $(n)$.
We write $\lambda\vdash n$ to indicate that $\lambda$ is an
integer partition of $n$, $\ell(\lambda)$ for the number of its parts and
$m_j(\lambda)$ for the number of its parts equal to $j$.

For a given composition $\bnu=(\nu_1,\dots,\nu_b)$ of $n$ and a given partition
$\lambda$ of $n$, we define the sub-universes
\begin{equation}\label{eq:subuniverses}
\Znnu = \left\{ Y\in\Zn : \bnu(Y)=\bnu \right\}, \qquad
\Znlambda = \left\{ Y\in\Zn : \lambda(Y)=\lambda \right\}.
\end{equation}
The universe of rankings is then organized as a hierarchy:
\begin{equation}\label{eq:hierarchy}
\Zn = \bigcup_{b=1}^{n} \Znb, \qquad
\Znb = \bigcup_{\lambda\vdash n,\ \ell(\lambda)=b} \Znlambda, \qquad
\Znlambda =  \bigcup_{\bnu:\ \lambda(\bnu)=\lambda} \Znnu,
\end{equation}
where we also write $\lambda(\bnu)$ for the shape of a pattern $\bnu$, and all the unions are disjoint.

Let $\bnu=(\nu_1,\dots,\nu_b)$ be a composition of $n$ into $b$ parts. A ranking with
pattern $\bnu$ is determined by the choice of the $\nu_1$ objects of the first bucket,
then of the $\nu_2$ objects of the second bucket among the remaining ones, and so on, so
that
\begin{equation}\label{eq:card-pattern}
|\Znnu| = \binom{n}{\nu_1,\nu_2,\dots,\nu_b}
= \frac{n!}{\nu_1!\,\nu_2!\cdots\nu_b!}.
\end{equation}
The multinomial coefficient does not depend on the order of the parts: all the
patterns with shape $\lambda$, whose number is $b!/\prod_{j\geq 1} m_j(\lambda)!$, contain
the same number of rankings, and
\begin{equation}\label{eq:card-shape}
|\Znlambda| = |\Znnu| \cdot \frac{b!}{\prod_{j\geq 1} m_j(\lambda)!},
\end{equation}
where $\bnu$ is any pattern with shape $\lambda$. Summing over the partitions of $n$ into
$b$ parts gives
\begin{equation}\label{eq:card-buckets}
|\Znb| = \sum_{\lambda\vdash n,\ \ell(\lambda)=b} |\Znlambda| = b!\,\stirling{n}{b},
\end{equation}
where the second equality is the count recalled in Section~\ref{sec:median}
\citep{dambrosioetal2019}, which Eq.~\ref{eq:card-buckets} refines.
Table~A.1 of the supplementary material shows the decomposition for $n=5$ and $n=6$. The
decomposition shows how much the search space shrinks when the structure of the solution
is prescribed: with $n=10$ objects the universe contains $102{,}247{,}563$ rankings, the
rankings with $b=5$ buckets are $5{,}103{,}000$, distributed among only $7$ shapes and
$126$ patterns, and a single pattern such as $(1,1,2,3,3)$ contains $50{,}400$ rankings.

\subsection{A single definition for all the problems}\label{sec:constrained}

The hierarchy in Eq.~\ref{eq:hierarchy} suggests that all the problems considered in this
paper can be described by the patterns that the solution is allowed to have. Let $\cN$ be
a nonempty set of compositions of $n$, the \emph{admissible patterns}, and let
\begin{equation}\label{eq:ZnN}
\ZnN = \left\{ Y\in\Zn : \bnu(Y)\in\cN \right\}
\end{equation}
be the set of the rankings whose pattern is admissible.
We define the \emph{median ranking with admissible patterns $\cN$}
as that ranking (or those rankings) $\hat{Y}$ for which
\begin{equation}\label{eq:median-pattern}
\hat{Y} = \argmin_{Y\in\ZnN} \sum_{h=1}^{r} w_h\, d(X^{(h)},Y)
= \argmax_{Y\in\ZnN} \bar{\tau}_X(Y).
\end{equation}

All the problems of this paper are special cases of Eq.~\ref{eq:median-pattern},
obtained with different sets $\cN$:
\begin{itemize}
\item if $\cN$ contains all the $2^{n-1}$ compositions of $n$, then $\ZnN=\Zn$ and the
solution is the median ranking of Eq.~\ref{eq:median}: nothing is forbidden;
\item if $\cN$ contains the compositions with $b$ parts, then $\ZnN=\Znb$ and the solution
is the MCBO with $b$ buckets as defined by \citet{dambrosioetal2019};
\item if $\cN=\{\bnu\}$ contains a single pattern, then $\ZnN=\Znnu$: the sizes of all the
buckets are prescribed, as in the forced distributions of Q methodology and of
performance appraisal, or as in the pattern $(k,n-k)$ that describes the election of a
committee of $k$ members;
\item if $\cN=\{(1,\dots,1)\}$, then $\ZnN=\Sn$ and the solution is the median ranking in
the space of the permutations;
\item if $\cN=\{(1,\dots,1,n-k)\}$, with $k$ elements equal to one, the solution is a
\emph{median top-$k$ list} (Section~\ref{sec:topkdecomp}).
\end{itemize}
Between these cases lie the \emph{partial patterns}, in which only some bucket sizes are
prescribed. We write them as patterns whose elements are either positive integers or a
symbol, $\free$, which indicates a bucket of free size: a partial pattern with $b$
elements, $\bnu=(\nu_1,\dots,\nu_b)$ with $\nu_t\in\{1,2,\dots,n\}\cup\{\free\}$, is a short
notation for the set $\cN$ of the compositions of $n$ into $b$ parts that coincide with
$\bnu$ in all the prescribed elements, and we still write $\Znnu$ for $\ZnN$. For
example, the pattern $(1,1,\free,\free,\free)$ asks for five buckets, the first two of
which contain one object each, and the pattern $(2,\free,2)$ asks for three buckets, the
first and the last of which contain two objects each. If $\bnu$ has $f$ free elements and
the sum of its prescribed elements is $s$, the free elements share the remaining $n-s$
objects, so that $\cN$ contains $\binom{n-s-1}{f-1}$ compositions. A pattern without the
symbol $\free$ is called \emph{complete}.

The procedures of Sections~\ref{sec:bolt} and \ref{sec:exact} cannot handle an arbitrary set
$\cN$, but only the sets that can be described position by position: a set $A_t$ of
admissible sizes for the $t$th bucket, $t=1,2,\dots$, and a set $\mathcal{B}$ of admissible
numbers of buckets. All the problems listed above belong to this class: for example,
the median ranking has $A_t=\{1,\dots,n\}$ for all $t$ and $\mathcal{B}=\{1,\dots,n\}$, and
the MCBO with $b$ buckets has $A_t=\{1,\dots,n\}$ and $\mathcal{B}=\{b\}$.

\subsection{The median top-$k$ list}\label{sec:topkdecomp}

A top-$k$ list is a ranking in which the first $k$ objects are ranked in a strict order
and the remaining $n-k$ objects are not distinguished. Since the solution must belong to
$\Zn$, and the Kemeny distance can leave two objects undistinguished only by tying them,
a top-$k$ list is a bucket order with pattern $(1,\dots,1,n-k)$, with $k$ elements equal
to one. In this respect, the solution differs from the input rankings
(Section~\ref{sec:missing}): the pairs not compared by a judge can be removed from $\C$,
while the solution has to say something about every pair. We denote by $\Tnk$ the
sub-universe of the top-$k$ lists, with $1\leq k\leq n-2$; by Eq.~\ref{eq:card-pattern},
$|\Tnk|=n!/(n-k)!$.

Unlike the aggregation of top-$k$ lists
\citep{faginetal2003,faginetal2006,ailon2010,mathieumauras2020}, in which the top-$k$
lists are the \emph{input} of the problem, here they are the \emph{output}, and the input
rankings can be of any nature.

For the top-$k$ lists, Eq.~\ref{eq:ordered} takes a particularly simple form.
Let $\gamma_a=\sum_{j\neq a} c_{ja}$ be the total of the $a$th column of $\C$, and
let $Y\in\Tnk$ be a top-$k$ list in which the set $K$ of the first $k$ objects
is ranked as $a_1\pre_Y a_2\pre_Y\cdots\pre_Y a_k$. Then
\begin{equation}\label{eq:topk}
\sum_{i,j=1}^{n} c_{ij}\,y_{ij} = \sum_{i<j}\left(c_{ij}+c_{ji}\right)
- 2\sum_{a\in K}\gamma_a + 2\sum_{1\leq s<t\leq k} c_{a_s a_t}.
\end{equation}
The derivation is given in the supplementary material.

Eq.~\ref{eq:topk} shows that the median top-$k$ list solves two nested
problems. The first one is the selection of the set $K$, which is driven by a term that
depends on each selected object separately, $-2\gamma_a$: the larger the total of the
$a$th column of $\C$, i.e., the more the judges rank the other objects ahead of or tied
with $a$, the less convenient is the selection of $a$. The objects in the last bucket
contribute only through these column totals. The second problem is the order of the
selected objects, which depends only on the restriction of $\C$ to the objects of $K$.

Given $K$, the best order of its objects is the median ranking, in the space of the
permutations of $K$, of the combined input matrix restricted to the rows and columns of
$K$: by Eq.~\ref{eq:ordered} applied to the restricted matrix, the last term of
Eq.~\ref{eq:topk} differs from the numerator of its average $\tauX$ only by a constant
that does not depend on the order.

Two consequences are worth noting. First, when all the judges express rankings
without ties, $c_{ja}=-c_{aj}$ and $\gamma_a=-\sum_{j\neq a}c_{aj}$, so that the selection term
is a Borda-type score: the total number of pairwise comparisons won by $a$ minus the
number of those lost. When the judges express tied rankings, instead, a tie between $a$
and $j$ counts against the selection of $a$, because it increases $c_{ja}$. Second, since
$|c_{ij}|\leq m$, the contribution of the interaction term in Eq.~\ref{eq:topk} is bounded
by $k(k-1)m$, while the selection term grows with $n$. For large $n$ and small $k$, the
set $K$ is therefore largely determined by the column totals of $\C$, and only the objects
whose column totals are close to the threshold need to be examined jointly.

\section{BOLT: a heuristic working on the combined input matrix}\label{sec:bolt}\label{sec:algorithm}

BOLT (Bucket-Order Local search with optimal Tie partitioning) searches the median
ranking, and its constrained versions, by combining three ingredients, all of which work
only on the combined input matrix $\C$: a set of starting solutions computed from $\C$
(and, optionally, the rankings of the judges), a local search that moves one object at a
time and can create or remove ties, and an exact procedure that, for a fixed order of the
objects, finds the best partition of the order into buckets. The ingredients are not new
when taken one at a time: the insertion local search is classical for the linear ordering
problem \citep{schiavinottostutzle2004,ceberioetal2015}, and its version with ties has
been used by \citet{cohenboulakiaetal2011} and \citet{lorenaetal2022}; the optimal
segmentation of an ordered sequence goes back to \citet{fisher1958}, and it was used by
\citet{faginetal2004} for the aggregation of rankings with ties under the footrule
distance. What BOLT does is to combine them on $\C$ for the Kemeny distance and to extend
the partition step to all the constraints of Section~\ref{sec:constraints}.

Throughout this section we write $d_{ij}=c_{ij}-c_{ji}$ and $\sigma_{ij}=c_{ij}+c_{ji}$.
By Eq.~\ref{eq:ordered}, a pair placed with $i$ ahead of $j$ contributes $d_{ij}$ to
$\sum_{i,j}c_{ij}y_{ij}$, and a tied pair contributes $\sigma_{ij}$.

\subsection{Starting solutions}\label{sec:seeds}

BOLT starts from the following rankings, the first five of which are computed from $\C$
in at most $O(n^2)$ operations per iteration:
\begin{enumerate}
\item the net flow (Borda-type) ranking, which orders the objects by the row sums of the
matrix $\mathbf{D}=(d_{ij})$, objects with equal sums being tied;
\item the Copeland ranking, which orders the objects by the number of pairwise
majorities won minus the number of those lost, i.e., by the row sums of the signs of
$\mathbf{D}$;
\item a Markov chain ranking, which orders the objects by the stationary distribution of
the chain that moves from an object to the objects preferred to it by a majority, with
transition probabilities proportional to the positive elements of $\mathbf{D}$ and a
damping factor of $0.85$ (computed by power iterations);
\item a bucket pivot ranking, obtained by the recursive procedure of
\citet{gionisetal2006} with a deterministic choice of the pivot (the object with the
largest $\sum_j|d_{ij}|$ among the objects to be split) and the rule of
\citet{aledoetal2017} for the objects tied with the pivot, adapted to the Kemeny
distance: an object $j$ is tied with the pivot $p$ if $\sigma_{pj}>|d_{pj}|$, i.e., if the
tie is better than both orders of the pair;
\item the starting solution of the branch-and-bound algorithm implemented in
\citet{consrank}, which orders the objects by the number of objects $j$ such that
$c_{ij}>0$ and $c_{ji}\neq 0$, objects with equal numbers being tied: unlike the
Copeland ranking, a pair for which the tie is the best choice counts for both objects;
\item the distinct rankings of the judges, when they are available. Since each starting
solution goes through the local search, their number can be limited by keeping only the
most frequent ones, while $\C$ is always computed from all the rankings.
\end{enumerate}
Moreover, from each of these rankings, with its ties broken by the index of the objects,
the local search without ties described below returns a complete ranking, which is added
to the starting solutions. These complete rankings are local optima in the space of the
permutations, and they are useful when the median ranking has few ties, as it happens
when the judges are heterogeneous (Section~\ref{sec:simexp}).

\subsection{The tie-aware insertion local search}\label{sec:insertion}

Let $Y$ be a ranking with $b$ buckets. The insertion move takes an object $z$ out of its
bucket and puts it back either alone, in one of the gaps between two consecutive buckets
(or before the first bucket, or after the last one), or tied with the objects of an
existing bucket. Let $B'_1,\dots,B'_{b'}$ be the buckets of the ranking obtained by
removing $z$ from $Y$, with $b'=b-1$ if $z$ was alone in its bucket and $b'=b$ otherwise,
and let
\[
U_{zt}=\sum_{j\in B'_t} c_{zj}, \qquad V_{zt}=\sum_{j\in B'_t} c_{jz}, \qquad t=1,\dots,b',
\]
be the sums of the elements of the $z$th row and of the $z$th column of $\C$ over the
objects of $B'_t$. By Eq.~\ref{eq:penalty}, $z$ contributes to the penalty the elements of
its row for the objects placed ahead of it and those of its column for the objects placed
behind it, while the objects tied with $z$ do not contribute. Placed alone in the gap after
the first $t$ buckets, $z$ contributes $\sum_{s\leq t}U_{zs}+\sum_{s>t}V_{zs}$, with
$t=0,\dots,b'$; tied with the objects of $B'_t$, it contributes
$\sum_{s<t}U_{zs}+\sum_{s>t}V_{zs}$, with $t=1,\dots,b'$. With the cumulative sums, all the
$2b'+1$ positions of $z$ are evaluated in $O(n)$ operations, as in the local search of
\citet[Section IV]{lorenaetal2022}, and a pass over all the objects costs $O(n^2)$.
In each pass the objects are taken in the order they have at the beginning of the pass, and
each of them is moved to its best position if this improves the current solution. The
passes are repeated until no object can be moved with an improvement. If the objects are
placed only in the gaps, so that no tie is ever created, the move is the classical
insertion of the linear ordering problem: we call this version the local search without
ties.

\subsection{The best partition of a fixed order}\label{sec:dp}

Let $\mathbf{o}=\langle o_1,\dots,o_n\rangle$ be an order vector without ties, in which
$o_p$ is the object in the $p$th position. A ranking is compatible with $\mathbf{o}$ if its
buckets are blocks of consecutive objects of $\mathbf{o}$, so that a ranking with $b$
buckets compatible with $\mathbf{o}$ is defined by the positions $0=e_0<e_1<\dots<e_b=n$
of the last object of each bucket. As for the compositions of Section~\ref{sec:patterns},
there are $2^{n-1}$ rankings compatible with $\mathbf{o}$, one for each pattern. Since every
ranking is compatible with at least one order, the search for the solution can be split
into the search for a good order and, for a given order, the search for its best partition
into buckets with an admissible pattern, which does not require the enumeration of the
$2^{n-1}$ partitions. Let
\begin{equation}\label{eq:blockgain}
g(s,e) = \sum_{s\leq p<q\leq e} c_{o_qo_p}, \qquad 1\leq s\leq e\leq n,
\end{equation}
be what the pairs of objects from the $s$th to the $e$th position would contribute to the
penalty if they were not tied. Since the tied pairs do not contribute, by
Eq.~\ref{eq:penalty} the penalty of the ranking compatible with $\mathbf{o}$ with cut
points $e_0,\dots,e_b$ is
\begin{equation}\label{eq:penaltyorder}
P(Y) = \sum_{1\leq p<q\leq n} c_{o_qo_p} - \sum_{t=1}^{b} g(e_{t-1}+1,e_t),
\end{equation}
and the best partition of $\mathbf{o}$ is the one that maximizes the sum of the quantities
$g$ of its buckets.

Let the admissible patterns be described, as in Section~\ref{sec:constrained}, by the
sets $A_t$ and $\mathcal{B}$: for a complete or partial pattern $\bnu$ with $b$ elements,
$A_t=\{\nu_t\}$ if the size of the $t$th bucket is prescribed, $A_t=\{1,\dots,n\}$ if it is
free, and $\mathcal{B}=\{b\}$. Let $D_t(e)$ be the best sum of the quantities $g$ over the
admissible partitions of the first $e$ objects of $\mathbf{o}$ into $t$ buckets. Since the
sum is additive over the buckets, the best partition whose last bucket goes from position
$s$ to position $e$ is obtained by adding this bucket to the best partition of the first
$s-1$ objects into $t-1$ buckets. Hence, with $D_0(0)=0$ and $D_0(e)=-\infty$ for $e>0$,
for $t=1,\dots,\max\mathcal{B}$ and $e=1,\dots,n$,
\begin{equation}\label{eq:dp}
D_t(e) = \max_{s:\ e-s+1\in A_t} \left\{ D_{t-1}(s-1) + g(s,e) \right\},
\end{equation}
with $D_t(e)=-\infty$ when no admissible $s$ exists. The best partition of $\mathbf{o}$
has value $\max_{t\in\mathcal{B}}D_t(n)$, and it is recovered by storing the maximizing $s$
at each step; if this value is $-\infty$, no ranking compatible with $\mathbf{o}$ belongs to
$\ZnN$. For a fixed $e$, all the values $g(s,e)$ are obtained in $O(n)$ operations from the
row of $\C$ corresponding to $o_e$, since $g(s,e)=g(s,e-1)+\sum_{p=s}^{e-1}c_{o_eo_p}$.
Once all the values $g$ are computed, in $O(n^2)$ operations, a free bucket costs $O(n^2)$
operations and a prescribed bucket, which admits a single $s$ for each $e$, costs $O(n)$,
so that the recursion costs at most $O(\max\mathcal{B}\cdot n^2)$.

The recursion in Eq.~\ref{eq:dp} gives the partition step of every problem of this paper.
For the median ranking the number of buckets does not matter, the index $t$ can be
dropped, and the recursion becomes $D(e)=\max_{s}\{D(s-1)+g(s,e)\}$, which costs $O(n^2)$:
this is the optimal tie partitioning (OTP) of BOLT. With the pattern
$(\free,\dots,\free)$ it gives the best ranking with exactly $b$ buckets compatible with
$\mathbf{o}$, which is the step needed for the MCBO. With a complete pattern there is a
single admissible partition, obtained by cutting $\mathbf{o}$ at the cumulative sums of the
pattern, so that the solution depends only on the order of the objects; with the pattern
$(1,\dots,1)$ the partition is the order itself. In all the cases the partition is optimal
only for the given order: the best ranking compatible with $\mathbf{o}$ is not, in general,
the solution of the problem, and the search for a good order is left to the local search.

\subsection{Local searches that preserve the admissible patterns}\label{sec:ls}

For the median ranking every pattern is admissible, and the insertion local search of
Section~\ref{sec:insertion} can be used as it is. When $\cN$ does not contain all the
patterns, its moves can lead to a pattern that is not admissible, and they must be
restricted or replaced.

\paragraph{A fixed number of buckets.} If the object is alone in its bucket, removing it
deletes the bucket, so that it can only be put back alone in a gap. Otherwise, removing it
does not change the number of buckets, and it can only be tied with the objects of another
bucket. The change of the penalty is computed as in Section~\ref{sec:insertion}.

\paragraph{A prescribed pattern.} When the sizes of the buckets are prescribed, moving a
single object changes the size of two buckets, so that no insertion move is admissible. We
use instead the \emph{cascade move}. Let $Y$ have buckets $B_1,\dots,B_b$, with the objects
of each bucket listed by their index, let $z\in B_u$ and let $t\neq u$ be the target
bucket. The object $z$ is moved into $B_t$ and, if $t>u$, the first object of each $B_r$,
$r=u+1,\dots,t$, is moved into $B_{r-1}$; if $t<u$, the last object of each $B_r$,
$r=t,\dots,u-1$, is moved into $B_{r+1}$. Each of the buckets from $B_u$ to $B_t$ loses one
object and gains one object, while the other buckets do not change, so that the pattern of
$Y$ is preserved. For each object there are $b-1$ target buckets, and the neighborhood of a
solution contains $n(b-1)$ cascade moves. Let $U_{zu}$ and $V_{zu}$ be the sums of
Section~\ref{sec:insertion}, computed over the objects of $B_u$ other than $z$, so that the
contribution of $z\in B_w$ to the penalty is $\sum_{u<w}U_{zu}+\sum_{u>w}V_{zu}$. With the
matrices $U$ and $V$, of size $n\times b$, the change of the penalty of a cascade move is
computed in $O(b^2)$ operations, and after an accepted move the two matrices are updated in
$O(n)$ operations for each moved object.

\paragraph{The exchange move.} A second move that preserves the pattern exchanges two
objects $x\in B_u$ and $y\in B_t$, with $u\neq t$, while the other objects do not move. By
Eq.~\ref{eq:penalty}, the change of the penalty is the sum of the changes due to $x$ and to
$y$, each of which depends only on the sums of its row and of its column of $\C$ over the
buckets between $B_u$ and $B_t$, and of a correction that depends only on $c_{xy}$ and
$c_{yx}$. With the cumulative sums of $U$ and $V$ over the buckets, each exchange is
evaluated in $O(1)$ operations, and a pass over all the pairs of objects costs $O(n^2)$.
The exchange moves are applied when the insertion moves (fixed number of buckets) or the
cascade moves (prescribed pattern) no longer improve the solution, and the two kinds of
moves are alternated until no move improves the penalty.

\paragraph{A partial pattern.} The cascade and exchange moves keep the current sizes of all
the buckets, and the sizes of the free buckets are changed by the partition step of
Eq.~\ref{eq:dp}. When all the elements of the pattern are free, i.e., for the MCBO, keeping
the current sizes is more restrictive than necessary, and the moves for a fixed number of
buckets are used instead.

\subsection{The complete algorithm}\label{sec:complete}

Algorithm~\ref{alg:bolt} summarizes BOLT for any set $\cN$ of admissible patterns. From each
starting solution, the local search without ties improves the order of the objects; then the
best partition of the order with an admissible pattern (Eq.~\ref{eq:dp}) and the local
search of Section~\ref{sec:ls} are alternated until the penalty does not decrease. The
starting solutions are processed independently, and BOLT returns all the solutions with the
largest average $\tauX$. For the median ranking the two steps are the OTP and the tie-aware
insertion local search, which is also applied once before the first OTP. For the median
ranking in the space of the permutations the partition step is the identity, and only the
local search without ties is used, from the starting solutions with their ties broken.

\begin{algorithm}[h!]
\caption{BOLT}\label{alg:bolt}\label{alg:bolt-pattern}
\begin{algorithmic}[1]
\Require combined input matrix $\C$, number of judges $m$, admissible patterns $\cN$
(by default all the patterns); optionally, the rankings of the judges
\State compute the starting solutions (Section~\ref{sec:seeds})
\For{each starting solution}
  \State improve the order of the objects by the local search without ties; let $\mathbf{o}$ be the order vector of the result (ties broken by the index of the objects)
  \Repeat
    \State $Y\gets$ best partition of $\mathbf{o}$ with a pattern in $\cN$ (Eq.~\ref{eq:dp})
    \State improve $Y$ by the local search that preserves the patterns in $\cN$ (Section~\ref{sec:ls})
    \State $\mathbf{o}\gets$ order vector of $Y$ (buckets from the first to the last, objects by index)
  \Until{the penalty does not decrease}
\EndFor
\Ensure the solution(s) $\hat{Y}\in\ZnN$ with the largest average $\tauX$
\end{algorithmic}
\end{algorithm}

Each pass of the insertion or exchange moves costs $O(n^2)$ operations and each pass of the
cascade moves $O(nb^3)$, while the OTP costs $O(n^2)$ and the partition step with
constraints at most $O(\max\mathcal{B}\cdot n^2)$. The number of passes and of alternations
is not bounded a priori, and the cost of BOLT is reported in the experiments through the
computing times. BOLT is deterministic: there are no random starting solutions and no random
choices, so that repeated runs return the same solutions. The special cases were checked on
small problems against the optimum obtained by enumerating all the admissible rankings. On
$60$ data sets with $n=4$ to $7$ objects and $m=3$ to $12$ judges, with tied input rankings,
in part with missing objects and with non-integer weights, BOLT always returned an
admissible ranking, and it returned the optimum in all the $60$ problems without
constraints, in all the $60$ problems in the space of the permutations and in all the $220$
problems with $b$ buckets ($b=2,\dots,n-1$), in $59$ of the $60$ problems with a complete
pattern, in $53$ of the $54$ with a partial pattern and in $219$ of the $220$ top-$k$ lists
($k=1,\dots,n-2$) (Table~A.2 of the supplementary material).

\subsection{An exact procedure for the median top-$k$ list}\label{sec:exacttopk}

For the median top-$k$ list, Eq.~\ref{eq:topk} allows the global optimum to be computed
exactly in many practical cases. The median top-$k$ list maximizes
\begin{equation}\label{eq:topkobj}
F(a_1,\dots,a_k) = -2\sum_{s=1}^{k}\gamma_{a_s} + 2\sum_{1\leq s<t\leq k} c_{a_sa_t}.
\end{equation}
In a top-$k$ list containing the object $a$, each pair formed by $a$ and one of the other
$k-1$ selected objects $j$ contributes to the second term of Eq.~\ref{eq:topkobj} either
$2c_{aj}$ or $2c_{ja}$. Let $\eta^{+}_a$ be the sum of the $k-1$ largest values of
$\max(c_{aj},c_{ja})$, and $\eta^{-}_a$ the sum of the $k-1$ smallest values of
$\min(c_{aj},c_{ja})$, over $j\neq a$: the terms of $F$ that involve $a$ are at most
$-2\gamma_a+2\eta^{+}_a$ and at least $-2\gamma_a+2\eta^{-}_a$. If $-2\gamma_a+2\eta^{+}_a$
is smaller than the $k$th largest value of $-2\gamma_j+2\eta^{-}_j$, the object $a$ cannot
belong to an optimal top-$k$ list: among the $k$ objects with the largest values of
$-2\gamma_j+2\eta^{-}_j$, at least one, say $j$, is not in the list, and replacing $a$ with
$j$ in the same position strictly increases $F$. The objects that pass this test are the
\emph{candidates}, and the test is repeated with $\eta^{+}_a$ and $\eta^{-}_a$ computed over
the candidates only, until the set of candidates does not change. When the candidates are
still many, a second test uses the value $F_0$ of the solution of Algorithm~\ref{alg:bolt}:
since $F$ is at most the sum of $-2\gamma_x+\eta^{+}_x$ over the objects $x$ of the list, the
object $a$ is kept only if $-2\gamma_a+\eta^{+}_a$, plus the $k-1$ largest values of
$-2\gamma_x+\eta^{+}_x$ over the other candidates, is at least $F_0$. Since the interaction
term is bounded (Section~\ref{sec:topkdecomp}), for large $n$ and small $k$ the candidates
are usually few more than $k$.

Given the set $S$ of candidates, let $\Phi(\emptyset)=0$ and, for every subset
$R\subseteq S$ with at most $k$ objects,
\begin{equation}\label{eq:topkdp}
\Phi(R) = \max_{a\in R}\left\{\Phi(R\setminus\{a\}) - 2\gamma_a + 2\sum_{j\in R\setminus\{a\}} c_{ja}\right\},
\end{equation}
where $a$ is the last object of the list formed by the objects of $R$. For $|R|=k$,
$\Phi(R)$ is the maximum of Eq.~\ref{eq:topkobj} over the orders of the objects of $R$, and
the median top-$k$ list is given by the subset with $k$ objects that maximizes $\Phi$; the
co-optimal solutions are obtained by backtracking (at most $1000$ in our implementation).
Since the number of subsets grows as $2^{|S|}$, the recursion is applied when the candidates
are at most $22$, and the solution is then \emph{certified}, i.e., it is the global optimum
of Eq.~\ref{eq:median-pattern} over $\Tnk$; otherwise, the solution of
Algorithm~\ref{alg:bolt} is returned.

\section{Exact solutions and certification}\label{sec:exact}

A heuristic can be assessed only against a reference. In this section we describe the
exact procedure that we use to certify the solutions of BOLT or, when the optimum is out
of reach, to bound their distance from it. The procedure is much slower than BOLT, and its
computing time grows quickly with the number of objects and with the disagreement among
the judges: it is a tool for the assessment of the heuristic, which can also be used in
practice when the size of the problem allows it.

\subsection{The unconstrained problem}\label{sec:blp}

We use the binary linear programming formulation of \citet{yooescobedo2021}. For each
ordered pair of distinct objects $(i,j)$, the binary variable $y_{ij}$ is equal to $1$ if
$i$ is ranked ahead of or tied with $j$, and to $0$ otherwise, so that a tie corresponds
to $y_{ij}=y_{ji}=1$. With this coding, the element of the score matrix of the solution is
$2y_{ij}-1$, and by Eq.~\ref{eq:avgtaux} the median ranking solves
\begin{align}
\max\ & \sum_{i\neq j} c_{ij}\,(2y_{ij}-1) \label{eq:blp}\\
\text{s.t. } & y_{ij}+y_{ji}\geq 1 && i<j, \label{eq:blp-total}\\
& y_{ij}+y_{jk}-y_{ik}\leq 1 && i,j,k \text{ distinct}, \label{eq:blp-trans}\\
& y_{ij}\in\{0,1\} && i\neq j. \nonumber
\end{align}
Constraints~\ref{eq:blp-total} make every pair comparable, and
constraints~\ref{eq:blp-trans} impose transitivity. When the solution is required to be a
complete ranking, constraints~\ref{eq:blp-total} become equalities, $y_{ij}+y_{ji}=1$, and
the formulation is the classical one for the linear ordering problem.

The formulation has $n(n-1)$ variables and $O(n^3)$ transitivity constraints, which become
too many already with a few hundred objects. We therefore add them lazily: the problem is
first solved without transitivity constraints, the violated triples are added, and the
problem is solved again, until the solution is transitive. Each round solves the linear
relaxation and, once the relaxation is transitive, the integer problem. Since the problem
of each round contains only some of the transitivity constraints, its solution gives an
upper bound for the average $\tauX$ of the median ranking, and it is the median ranking
when it is transitive. The first round gives exactly the bound $P_{LB}$ of
Eq.~\ref{eq:lb}, and the solution of BOLT is given to the solver as the starting solution.
\citet{aledoetal2025} use a similar lazy scheme, while \citet{andrieuetal2023} use all the
transitivity constraints and reduce the size of the problems by partitioning. All the
problems are solved by the open-source solver HiGHS \citep{huangfuhall2018} with a time
limit: when the limit is reached, we report the best solution found and the best bound,
i.e., a certified gap.

\subsection{Constraints on the buckets}\label{sec:blpconstraints}

The constraints of Section~\ref{sec:constraints} are imposed through additional binary
variables, following the assignment formulation of \citet{aledoetal2025}, which was written
for the optimal bucket order problem and whose constraints do not depend on the objective
function. Let $\delta_{it}=1$ if the object $i$ is in the $t$th bucket, for $t=1,\dots,b$,
and let $\Delta_{it}=\sum_{s\leq t}\delta_{is}$. Then
\begin{align}
& \sum_{t=1}^{b} \delta_{it}=1 && i=1,\dots,n, \label{eq:assign}\\
& \sum_{i=1}^{n} \delta_{it}=\nu_t \ \ (\text{prescribed}),\qquad
  \sum_{i=1}^{n} \delta_{it}\geq 1 \ \ (\text{free}) && t=1,\dots,b, \label{eq:sizes}\\
& y_{ij}+y_{ji}\geq \delta_{it}+\delta_{jt} && i<j,\ t=1,\dots,b, \label{eq:sametie}\\
& y_{ji}+\Delta_{it}-\Delta_{jt}\leq 1 && i\neq j,\ t=1,\dots,b-1. \label{eq:order}
\end{align}
Constraints~\ref{eq:sametie} tie the objects of the same bucket, and
constraints~\ref{eq:order} forbid $j$ to be ranked ahead of or tied with $i$ when $i$ is in
one of the first $t$ buckets and $j$ is not. With these constraints the transitivity
constraints are redundant, and the variables $y_{ij}$ are integral as soon as the variables
$\delta_{it}$ are \citep[Propositions 2 and 3]{aledoetal2025}; we still add them lazily,
because they strengthen the relaxation. A partial pattern uses the equality in
Eq.~\ref{eq:sizes} only for the prescribed sizes, the MCBO uses the inequality for all of
them, and the median top-$k$ list is the complete pattern $(1,\dots,1,n-k)$.

The numbers of objects strictly ahead of, tied with and strictly behind the object $i$ are
linear functions of the variables $y$:
\begin{equation}\label{eq:ahead}
\alpha_i=\sum_{j\neq i}(1-y_{ij}),\qquad \theta_i=\sum_{j\neq i}(y_{ij}+y_{ji}-1),\qquad
\beta_i=\sum_{j\neq i}(1-y_{ji}),
\end{equation}
since $y_{ij}=0$ means that $j$ is strictly ahead of $i$. We use them to strengthen the
formulation. With a complete pattern $\bnu$, let $N_t=\nu_1+\dots+\nu_{t-1}$ be the number
of objects ahead of the $t$th bucket. An object is in the $t$th bucket if and only if it has
$N_t$ objects ahead of it and $\nu_t-1$ objects tied with it, so that
constraints~\ref{eq:sametie}--\ref{eq:order} can be replaced by
\begin{equation}\label{eq:positional}
\alpha_i=\sum_{t=1}^{b}N_t\,\delta_{it},\qquad \theta_i=\sum_{t=1}^{b}(\nu_t-1)\,\delta_{it},
\qquad i=1,\dots,n,
\end{equation}
together with Eqs.~\ref{eq:assign}--\ref{eq:sizes}. We call this the positional
formulation; in our experiments it was the fastest one for complete patterns. With a
partial pattern, the same equalities hold only for the objects of the prescribed buckets,
and they become pairs of inequalities that are not binding when the object is in a free
bucket: on $\theta_i$ for all the prescribed buckets, on $\alpha_i$ for those that precede
the first free bucket, and on $\beta_i$ for those that follow the last one. Finally, two
cuts bound the ties: no bucket can contain more than $\nu_{\max}$ objects, the largest
prescribed size or the largest size left to a free bucket, so that
$\theta_i\leq\nu_{\max}-1$ for every object, and the total number of tied pairs lies
between the smallest and the largest value allowed by the admissible patterns.

The formulations were checked against the enumeration of the admissible rankings on $2312$
problems with tied rankings, in part with missing values and weights: $1892$ with $n=4$ to $6$ objects and all
the complete and partial patterns, and $420$ with $n=8$ to $12$ ($240$ with the complete
and partial patterns of the simulation study of the MCBO, $180$ with a free number of
buckets). The problems with prescribed sizes were solved with and without the
strengthening inequalities, and all the $4444$ solutions were optimal.

\section{Experimental evaluation}\label{sec:experiments}

\subsection{Setting}\label{sec:setting}

The experiments address three questions: how close BOLT is to the optimum, how it compares
with the methods published for the same problem, and which of its components matter. The
reference is the certified optimum of Section~\ref{sec:exact} whenever it can be computed,
and the certified bound otherwise. All the computations were performed on a desktop
computer with an Intel Core i5-6500 processor (four cores, 3.2 GHz) and 16 GB of RAM, with
R 4.6.1, our implementation for BOLT and HiGHS 1.14.0, through the R package highs 1.14.0.2,
for the exact procedure. The computing times of the heuristics are elapsed times of a
single process on an otherwise idle machine. The exact procedure added at most
$200{,}000$ violated triples per round and used up to four threads; its time limit is
given for each experiment. BOLT is deterministic and was run once, while the randomized
methods were run $100$ times and the best solution is reported. The quality of a solution
$Y$ is measured by the difference $\bar\tau_X(Y^\ast)-\bar\tau_X(Y)\geq 0$ between the
average $\tauX$ of the optimum $Y^\ast$ and that of $Y$, which we call the gap from the
optimum. When the optimum is not known, the gap is measured from the certified bound, and
it is therefore an upper bound of the true gap.

The methods implemented by us, all in R and working on $\C$, are: the BioConsert local
search \citep{cohenboulakiaetal2011}, started from the same starting solutions of BOLT,
which isolates the contribution of the moves and of the OTP from that of the starting
solutions; the NonStrictKwikSort of \citet{brancotteetal2015} and, for complete rankings,
the KwikSort of \citet{ailoncharikarnewman2008}; the aggregation of \citet{faginetal2004}
through the median of the positions and the optimal grouping under the footrule distance.
For SIgFUR and FUR \citep{badaldas2018,badalsingh2026} we ran on our machine the MATLAB
code made available by the authors\footnote{\url{https://github.com/prakashvs613/RankAggSIgFUR-tied-matlab}},
which returned all the published values that we could check. For ACP and MorK, whose code
is not available, we compare BOLT with the published results.

\subsection{Real and simulated data with certified optimum}\label{sec:realexp}

The first benchmark contains $83$ problems for which the exact procedure certified the
optimum, with a time limit of $600$ seconds: the EMD data of \citet{emondmason2002}, the
five data sets of \citet{badalsingh2026} (machine learning algorithms, universities with
$229$ objects and their research and teaching rankings, cities with $240$ objects), $13$
PrefLib data sets with complete rankings with ties \citep{matteiwalsh2013} and $64$
simulated problems with $n$ from $20$ to $90$ objects. Table~\ref{tab:real} reports the
number of optima and the gaps.

\begin{table}[h!]
\centering\tablesize
\caption{Problems with certified optimum ($83$): number of problems in which each method
returns the optimum, gap from the optimum in terms of $\bar\tau_X$ (mean, 90\% quantile,
maximum) and largest elapsed time. Randomized methods: best of $100$ runs.}
\label{tab:real}
\begin{tabular}{lrrrrr}
\toprule
Method & optima & mean gap & 90\% quantile & max gap & max time (s)\\
\midrule
BOLT & 72 & 7.0e-06 & 3.0e-05 & 8.3e-05 & 12.7\\
BioConsert (same starts) & 66 & 2.2e-05 & 1.2e-04 & 2.1e-04 & 193.0\\
NonStrictKwikSort & 17 & 2.5e-03 & 6.6e-03 & 8.2e-03 & 3.6\\
KwikSort & 4 & 6.3e-02 & 5.4e-02 & 9.0e-01 & 3.0\\
Fagin & 7 & 3.6e-02 & 1.0e-01 & 1.7e-01 & 0.6\\
\bottomrule
\end{tabular}
\end{table}

BOLT returns the optimum in $72$ of the $83$ problems, and in the others its gap is at most
$8.3\times10^{-5}$. Compared with BioConsert started from the same solutions, BOLT is
better in $14$ problems and worse in $3$: most of the gain comes from the starting
solutions, but the OTP and the larger set of moves add to it. The longest time of BOLT,
$12.7$ seconds, was spent on a PrefLib data set with only $12$ objects but thousands of
distinct rankings, all used as starting solutions (Section~\ref{sec:seeds}).

\subsection{Published benchmarks}\label{sec:pubexp}

\paragraph{Approximate Condorcet Partitioning} \citet{akbariescobedo2023a} reported, for
$85$ PrefLib data sets with complete rankings with ties and more than $100$ objects (up to
$2{,}819$), the gap of ACP from the lower bound $P_{LB}$ of Eq.~\ref{eq:lb}. We checked
that our lower bound coincides with theirs on all the $85$ data sets, so that the gaps are
comparable; since they reported the gaps with two decimals, the objective of ACP is known
up to $\pm0.005\%$ of the bound. BOLT is better than the best published ACP solution in
$75$ data sets and equal within the rounding in $10$; the median gap from the lower bound
is $0.68\%$ for BOLT and $0.96\%$ for ACP. Among the $11$ data sets with at most $477$
objects, BOLT attains the lower bound, and is therefore optimal, in $9$. On the other two,
ED-10-50 ($170$ objects) and ED-10-49 ($351$ objects), the exact procedure, with a time
limit of four hours, found the optimum in $71$ seconds and in $37$ minutes: BOLT, in less
than one second, returned the optimum in the first and a solution $22$ points of distance
above it in the second. The computing times of ACP are not comparable, since they come from
a different machine and implementation; BOLT took at most $142$ seconds, on a data set with
$2{,}446$ objects.

\paragraph{SIgFUR and FUR} Table~\ref{tab:bs} compares BOLT with the methods of
\citet{badaldas2018} and \citet{badalsingh2026}, run on our machine.

\begin{table}[h!]
\centering\tablesize
\caption{Data sets of \citet{badalsingh2026}: average $\tauX$ of the solutions and
elapsed time on the same machine. The optimum is certified by the exact procedure, except
for the data set with $1{,}173$ objects, for which the certified bound is reported.}
\label{tab:bs}
\begin{tabular}{lrlrr}
\toprule
Data set ($n\times m$) & Optimum & Method & $\bar\tau_X$ & Time (s)\\
\midrule
ML ($12\times 8$) & 0.72159 & FUR, SIgFUR & 0.72159 & 2.6, 37.1\\
 & & BOLT & 0.72159 & 0.02\\
Universities ($229\times 10$) & 0.72026 & SIgFUR (no time limit) & 0.72026 & 8698\\
 & & BOLT & 0.72025 & 0.56\\
Cities ($240\times 4$) & 0.74793 & SIgFUR & 0.74789 & 5151\\
 & & BOLT & 0.74786 & 0.26\\
EMD ($15\times 21$) & 0.16556 & FUR, SIgFUR, BS & 0.16556 & 4.8--466\\
 & & BOLT & 0.16556 & 0.02\\
University SOI ($1173\times 19$) & $\leq0.57415$ & best BS & 0.55303 & 1727\\
 & & BOLT & 0.56299 & 40\\
\bottomrule
\end{tabular}
\end{table}

On the universities and cities data sets the best SIgFUR solutions are slightly better than
BOLT, by $2$ and $4$ points of distance, while on the research and teaching rankings BOLT is
better than FUR. On the largest data set, for which the exact procedure was stopped with a
certified bound, BOLT is better than the best published solution by about $2\%$ of the sum
of distances. The times differ by three to four orders of magnitude, also because of the
different languages (MATLAB against R with C++).

\paragraph{MorK} \citet{azzinietal2026} studied the median ranking in the space of the
permutations on $3600$ simulated problems with $n$ from $10$ to $50$ objects. Since their
problems are not available, we generated $3600$ problems from the same distribution and
solved them exactly with the formulation of Section~\ref{sec:blp} with $y_{ij}+y_{ji}=1$:
all of them were solved, in at most $26$ seconds each. BOLT for complete rankings returned
the optimum in $3017$ problems. Following \citet{azzinietal2026}, we measure the gap by the
relative difference between the score of the optimum and that of the solution: the mean
gap of BOLT over the values of $n$ is $2.66\times10^{-5}$, against $9.78\times10^{-5}$ published for
MorK-Heuristic, and it is smaller in $39$ of the $41$ values of $n$. The generation of the
problems, the measure and the results for each $n$ are given in the supplementary material.

\subsection{Simulated data with ties}\label{sec:simexp}

\paragraph{Design} We generated $1080$ problems with ties from a Mallows model with the
Kendall distance \citep{mallows1957}, sampled through the representation of
\citet{flignerverducci1986}, and from a mixture of two such models with the same
dispersion, equal weights and a random second center. The ties were introduced by closing
each gap between two consecutive objects of a sampled permutation with probability
$p_{tie}$. The design crosses the number of objects $n\in\{50,100,200\}$, the number of
judges $m\in\{10,50\}$, the dispersion $\theta\in\{0.02,0.05,0.2\}$ (the smaller $\theta$,
the more the judges disagree), $p_{tie}\in\{0.1,0.3,0.6\}$ and the two models, with $10$
replications per cell. The exact procedure, with a time limit of $900$ seconds and four
threads, was run on all the problems with $n\leq100$ and on one replication per cell with
$n=200$ (plus ten further problems with $n=200$); it certified the optimum with ties in
$733$ problems and the optimum in the space of the permutations in $760$.

\paragraph{Accuracy} Table~\ref{tab:sim} reports the results on the problems with
certified optimum.

\begin{table}[h!]
\centering\tablesize
\caption{Simulated problems with ties and certified optimum: number of optima and gap
from the optimum in terms of $\bar\tau_X$. Top: solutions with ties ($733$ problems);
bottom: complete rankings ($760$ problems). Mean elapsed time per problem for
$n=50,100,200$ (single process, idle machine).}
\label{tab:sim}
\begin{tabular}{lrrrr}
\toprule
Method & optima & mean gap & max gap & time (s), $n=50/100/200$\\
\midrule
BOLT & 433 & 6.1e-05 & 9.8e-04 & 0.08 / 0.23 / 0.70\\
BioConsert (same starts) & 275 & 1.3e-04 & 1.1e-03 & 2.7 / 5.7 / 14.7\\
NonStrictKwikSort (100 runs) & 9 & 6.8e-03 & 2.4e-02 & 0.20 / 0.52 / 1.65\\
Fagin & 0 & 5.8e-02 & 2.2e-01 & 0.01 / 0.01 / 0.02\\
\midrule
BOLT (complete rankings) & 409 & 8.6e-05 & 9.8e-04 & 0.03 / 0.06 / 0.19\\
KwikSort (100 runs) & 41 & 5.6e-03 & 2.2e-02 & 0.23 / 0.61 / 2.02\\
\bottomrule
\end{tabular}
\end{table}

BOLT is the most accurate method, and its gaps are one to two orders of magnitude smaller
than those of the pivoting methods. It does not always find the optimum: the share of
optima decreases with the disagreement among the judges and with $n$, from $99\%$ with
$n=50$ and $\theta=0.2$ to $7\%$ with $n=100$ and $\theta=0.02$, and to none of the
certified problems with $n=200$ and $\theta\leq0.05$. With nearly random judges the
objective function is flat, many solutions have an average $\tauX$ within $10^{-4}$ of the
optimum, and the optimum has few ties. Without the complete rankings among the starting
solutions, BOLT missed the optimum in $336$ problems, and in $331$ of them its solution had
fewer buckets than the optimum (median $68$ against $93$); the two solutions differed in
about $2\%$ of the pairs, spread over most of the objects. The other local moves that we
tried (moves of whole buckets, splits of buckets) did not close the gap, while the complete
rankings among the starting solutions (Section~\ref{sec:seeds}) raised the number of optima
from $397$ to $433$, at the cost of about $2.4$ times the computing time.

\paragraph{Ablation} Table~\ref{tab:ablation} reports the results of BOLT when one of its
components is removed, or when it is run from a single starting solution; since the
starting solutions are processed independently, each variant takes the best solution over
a subset of them. The rankings of the judges are the most useful starting solutions in
these data, and the best judge alone is as good as all of them together, while the
starting solutions computed from $\C$ are needed when the rankings of the judges are not
available. Removing the local search or the OTP reduces the number of optima by about a
third, and no single starting solution is enough. The same ablation on the $83$ problems of
Section~\ref{sec:realexp} (Table~B.1 of the supplementary material) leads to the same
conclusions; there, the complete rankings do not add optima, but they reduce the largest
gap from $1.6\times10^{-4}$ to $8.3\times10^{-5}$.

\begin{table}[h!]
\centering\tablesize
\caption{Ablation of BOLT on the simulated problems with ties: number of optima and gap
from the optimum (in terms of $\bar\tau_X$) on the $733$ problems with certified optimum,
and number of the $1080$ problems in which the variant is worse than BOLT. Top: removal
of a component; bottom: a single starting solution (with its complete ranking from the
local search without ties).}
\label{tab:ablation}
\centering
\begin{tabular}{@{}lrrrr@{}}
\toprule
\multicolumn{1}{c}{\multirow{3}{*}{Variant}} & \multicolumn{1}{c}{\multirow{3}{*}{\begin{tabular}[c]{@{}c@{}}optima \\ (of 733)\end{tabular}}} & \multicolumn{1}{c}{\multirow{3}{*}{\begin{tabular}[c]{@{}c@{}}mean \\ gap\end{tabular}}} & \multicolumn{1}{c}{\multirow{3}{*}{\begin{tabular}[c]{@{}c@{}}max \\ gap\end{tabular}}} & \multicolumn{1}{c}{\multirow{3}{*}{\begin{tabular}[c]{@{}c@{}}worse than BOLT \\ (of 1080)\end{tabular}}} \\
\multicolumn{1}{c}{} & \multicolumn{1}{c}{} & \multicolumn{1}{c}{} & \multicolumn{1}{c}{} & \multicolumn{1}{c}{} \\
\multicolumn{1}{c}{} & \multicolumn{1}{c}{} & \multicolumn{1}{c}{} & \multicolumn{1}{c}{} & \multicolumn{1}{c}{} \\ \midrule
All the components & 433 & 6.1e-05 & 9.8e-04 & 0 \\
\multirow{2}{*}{\begin{tabular}[c]{@{}l@{}}Without the complete \\ rankings\end{tabular}} & \multirow{2}{*}{397} & \multirow{2}{*}{7.5e-05} & \multirow{2}{*}{1.1e-03} & \multirow{2}{*}{218} \\
 &  &  &  &  \\
Only the complete rankings & 393 & 7.6e-05 & 9.8e-04 & 237 \\
\multirow{2}{*}{\begin{tabular}[c]{@{}l@{}}Only the starting solutions \\ from $\C$\end{tabular}} & \multirow{2}{*}{303} & \multirow{2}{*}{1.1e-04} & \multirow{2}{*}{1.1e-03} & \multirow{2}{*}{512} \\
 &  &  &  &  \\
\multirow{2}{*}{\begin{tabular}[c]{@{}l@{}}Only the rankings of the \\ judges\end{tabular}} & \multirow{2}{*}{405} & \multirow{2}{*}{7.8e-05} & \multirow{2}{*}{9.8e-04} & \multirow{2}{*}{177} \\
 &  &  &  &  \\
Without OTP & 281 & 1.1e-04 & 9.8e-04 & 362 \\
Only OTP (no local search) & 284 & 1.2e-04 & 1.1e-03 & 538 \\ \midrule
Net flow alone & 153 & 2.9e-04 & 2.3e-03 & 856 \\
Copeland alone & 199 & 2.1e-04 & 2.0e-03 & 766 \\
Markov chain alone & 162 & 2.7e-04 & 2.1e-03 & 848 \\
Bucket pivot alone & 158 & 3.4e-04 & 4.9e-03 & 861 \\
\multirow{2}{*}{\begin{tabular}[c]{@{}l@{}}Branch-and-bound start \\ alone\end{tabular}} & \multirow{2}{*}{199} & \multirow{2}{*}{2.2e-04} & \multirow{2}{*}{2.4e-03} & \multirow{2}{*}{769} \\
 &  &  &  &  \\
Best judge alone & 405 & 7.8e-05 & 9.8e-04 & 177 \\
Median judge alone & 135 & 2.8e-04 & 2.0e-03 & 914 \\
Worst judge alone & 63 & 6.6e-04 & 4.6e-03 & 1001 \\ \bottomrule
\end{tabular}
\end{table}

\subsection{Constrained problems}\label{sec:conexp}

For the constrained problems, Table~\ref{tab:cert} reports the certification of the BOLT
solutions with the formulation of Section~\ref{sec:blpconstraints}, started from the
BOLT solution, on ten problems for each size and constraint: rankings of $m=100$ judges
drawn from a mixture of two Mallows models with $\theta=0.2$, $n=20$ and $50$ objects, a
time limit of $600$ seconds and four threads. Four constraints were considered: a
complete pattern with seven buckets, whose sizes follow the profile $(1,2,3,4,3,2,1)$
scaled to $n$; the partial pattern with the same first and last sizes and the five
central sizes free; $b=5$ buckets; the top-$5$ list.

With $n=20$ all the $40$ problems were solved to optimality, and BOLT returned the
optimum in $38$ of them; the two exceptions have a partial pattern, with a gap of at most
$4.2\times10^{-4}$. With $n=50$ the top-$5$ lists were all solved in about one second,
and nine of the ten complete patterns within the time limit; the
problems with $b=5$ buckets and, above all, with a partial pattern are much harder, and
only $7$ and $3$ of them were solved within the time limit. Among the $29$ certified
problems with $n=50$, BOLT returned the optimum in $27$. In the problems not solved, the
exact formulation improved the BOLT solution in four problems with a partial pattern,
always by less than $2.3\times10^{-4}$, and it never improved it with $b=5$ buckets; the
certified gap, between the bound and BOLT, reached $4.8\times10^{-3}$ with a partial
pattern and $8.3\times10^{-3}$ with $b=5$ buckets. As without constraints, the exact
formulation is weak when the number of buckets, or some of the sizes, are free, and the
remaining gaps are mostly due to the bound. BOLT took about $0.01$ seconds in every
problem. A larger simulation study of the three kinds of constraints, with up to $5000$
objects and a comparison with the QUICK and DECoR algorithms of \citet{dambrosioetal2019},
is reported in the supplementary material.

\begin{table}[h!]
\centering\tablesize
\caption{Certification of BOLT for constrained problems (ten problems for each $n$ and
constraint). Certified: problems solved to optimality; BOLT optimal: certified problems in
which BOLT returned the optimum; max gap: largest difference between the best solution
found and BOLT; cert.\ gap: difference between the certified bound and BOLT (mean and
maximum); times: means. Gaps in terms of $\bar\tau_X$.}
\label{tab:cert}
\centering
\resizebox{1\textwidth}{!}{%
\begin{tabular}{@{}rlrrrrrrr@{}}
\toprule
\multicolumn{1}{c}{\multirow{2}{*}{$n$}} & \multicolumn{1}{c}{\multirow{2}{*}{constraint}} & \multicolumn{1}{c}{\multirow{2}{*}{certified}} & \multicolumn{1}{c}{\multirow{2}{*}{\begin{tabular}[c]{@{}c@{}}BOLT \\ optimal\end{tabular}}} & \multicolumn{1}{c}{\multirow{2}{*}{\begin{tabular}[c]{@{}c@{}}max \\ gap\end{tabular}}} & \multicolumn{1}{c}{\multirow{2}{*}{\begin{tabular}[c]{@{}c@{}}mean \\ cert. gap\end{tabular}}} & \multicolumn{1}{c}{\multirow{2}{*}{\begin{tabular}[c]{@{}c@{}}max \\ cert. gap\end{tabular}}} & \multicolumn{1}{c}{\multirow{2}{*}{\begin{tabular}[c]{@{}c@{}}BOLT \\ (s)\end{tabular}}} & \multicolumn{1}{c}{\multirow{2}{*}{\begin{tabular}[c]{@{}c@{}}BLP \\ (s)\end{tabular}}} \\
\multicolumn{1}{c}{} & \multicolumn{1}{c}{} & \multicolumn{1}{c}{} & \multicolumn{1}{c}{} & \multicolumn{1}{c}{} & \multicolumn{1}{c}{} & \multicolumn{1}{c}{} & \multicolumn{1}{c}{} & \multicolumn{1}{c}{} \\ \midrule
20 & complete pattern & 10 & 10 & 0 & 0 & 0 & 0.01 & 1 \\
20 & partial pattern & 10 & 8 & 4.2e-04 & 6.3e-05 & 4.2e-04 & 0.01 & 8 \\
20 & $b=5$ buckets & 10 & 10 & 0 & 0 & 0 & 0.01 & 10 \\
20 & top-5 list & 10 & 10 & 0 & 0 & 0 & 0.01 & 0 \\
50 & complete pattern & 9 & 8 & 1.8e-04 & 1.9e-05 & 1.8e-04 & 0.02 & 71 \\
50 & partial pattern & 3 & 2 & 2.3e-04 & 1.7e-03 & 4.8e-03 & 0.01 & 545 \\
50 & $b=5$ buckets & 7 & 7 & 0 & 2.1e-03 & 8.3e-03 & 0.02 & 442 \\
50 & top-5 list & 10 & 10 & 0 & 0 & 0 & 0.01 & 1 \\ \bottomrule
\end{tabular}%
}
\end{table}

\section{Applications}\label{sec:applications}

In this section we show how the constrained problems of this paper arise in real data, and
what the choice of the convention for the objects not ranked implies.

\subsection{A forced distribution: Q methodology}\label{sec:qmethod}

In Q methodology \citep{stephenson1935,brown1980,wattsstenner2012} each participant
sorts a set of statements into a grid with a prescribed number of statements per
column, from the most agree to the most disagree, so that every Q-sort is a bucket order
with the pattern of the grid. We analyzed the Lipset data \citep{lipset1963,brown1980},
available in the R package qmethod \citep{zabala2014}: $9$ participants sorted $33$
statements into the grid from $+4$ to $-4$ with pattern $(2,3,4,5,5,5,4,3,2)$. Since all the
input rankings have this pattern, a consensus Q-sort should have it as well, and it is the
median ranking with the pattern of the grid.

\begin{table}[h!]
\centering\tablesize
\caption{Lipset Q-sorts ($n=33$, $m=9$): median ranking with the pattern of the grid
(exact solution and BOLT) and MCBO with $b=9$ buckets (BOLT).}
\label{tab:qsortall}
\begin{tabular}{llrr}
\toprule
Solution & Bucket sizes & $\bar\tau_X$ & $\tau_X$ with the exact solution\\
\midrule
Grid pattern, exact & 2, 3, 4, 5, 5, 5, 4, 3, 2 & 0.2471 & 1.0000\\
Grid pattern, BOLT & 2, 3, 4, 5, 5, 5, 4, 3, 2 & 0.2458 & 0.8750\\
$b=9$, BOLT & 6, 5, 1, 2, 1, 11, 4, 2, 1 & 0.2534 & 0.7803\\
\bottomrule
\end{tabular}
\end{table}

Table~\ref{tab:qsortall} shows a case in which the exact procedure is worth running. With
a complete pattern, the positional formulation of Section~\ref{sec:blpconstraints}
certified the optimum in about five minutes: the consensus Q-sort of the $9$ participants
has an average $\tauX$ equal to $0.2471$, while the best solution of BOLT, found in $0.25$
seconds, has $0.2458$, i.e., it is $6$ points of distance farther from the judges. The
participants disagree considerably and, as in the simulations of
Section~\ref{sec:simexp}, BOLT can miss the optimum even in a small problem. In Q studies
the statements are usually a few dozen and the grid is a complete pattern, so that the
consensus Q-sort can be certified in practice, with the solution of BOLT as the starting
solution. Without the constraint of the grid, i.e., with $b=9$ buckets of free size, the
average $\tauX$ is only slightly larger ($0.2534$), but the bucket sizes are very different
from the grid, with a bucket of $11$ statements, and the solution cannot be read as a
Q-sort.

The standard Q procedure summarizes the Q-sorts that load on the same factor by a
\emph{factor array}, obtained by computing a weighted average of the standardized scores of
the statements and by reporting the result into the grid. This is an ad hoc procedure, of
the Borda type, that does not optimize any criterion. We extracted three factors with
qmethod (principal components, varimax rotation, automatic flagging), and for each factor
we computed the median ranking with the pattern of the grid of the Q-sorts flagged on that
factor (three for each factor), after reversing the Q-sorts with a negative loading; in the
three cases the exact procedure certified that BOLT returned the optimum. For each factor,
the median ranking has a larger average $\tauX$ with the flagged Q-sorts than the factor
array ($0.6187$ against $0.6136$, $0.5896$ against $0.5783$, $0.5391$ against $0.5240$), and
the two solutions are similar but not identical, with a $\tauX$ between them from $0.81$ to
$0.90$.

\subsection{Average $\tauX$ on the observed pairs}\label{sec:tauxobs}

In most of the data sets of the following sections, each judge ranks only some of the
objects, and we state each time which of the conventions of Section~\ref{sec:missing} is
used. With objects not ranked, the average $\tauX$ in Eq.~\ref{eq:avgtaux} attributes
$n(n-1)$ ordered pairs to each judge, also the pairs that are not counted, so that with
sparse data it is close to zero for any solution. Only in order to report a value that can
be read as a degree of agreement, we also compute the average $\tauX$ on the observed pairs.
Let $n_h$ be the number of objects ranked in $X^{(h)}$, and let
$T=\sum_{h}w_h\sum_{i\neq j}I^{(h)}_{ij}$ be the number of ordered pairs counted in
Eq.~\ref{eq:cijdelta}: $T=\sum_{h}w_h\,n_h(n_h-1)$ with the first convention and
$T=\sum_{h}w_h\,[n(n-1)-(n-n_h)(n-n_h-1)]$ with the third one. We define
\begin{equation}\label{eq:tauxobs}
\tauX^{obs}(\C,Y) = \frac{\sum_{i,j=1}^{n} c_{ij}\,y_{ij}}{T}.
\end{equation}
With the first convention, let $\tauX^{(h)}(Y)$ be the $\tauX$ between $X^{(h)}$ and the
restriction of $Y$ to the $n_h$ objects ranked in $X^{(h)}$. Since
$\sum_{i\neq j}I^{(h)}_{ij}x^{(h)}_{ij}y_{ij}=n_h(n_h-1)\,\tauX^{(h)}(Y)$, we obtain
\[
\tauX^{obs}(\C,Y)=\frac{\sum_h w_h\,n_h(n_h-1)\,\tauX^{(h)}(Y)}{\sum_h w_h\,n_h(n_h-1)},
\]
i.e., $\tauX^{obs}$ is the weighted average of the $\tauX$ of the judges computed on the
objects that each of them ranked, with weights equal to the numbers of observed pairs; it
is equal to $1$ only if the solution agrees with every judge on every observed pair.
$\tauX^{obs}$ is not a new objective function: $T$ does not depend on $Y$, so that the
average $\tauX$ and $\tauX^{obs}$ have the same solutions, with any constraint, while the
simple average of the $\tauX^{(h)}(Y)$ would be a different objective function. With
complete data and with the second convention, $T=m\,n(n-1)$ and $\tauX^{obs}$ coincides
with the average $\tauX$.

\subsection{Median top-$k$ lists}\label{sec:topkapp}

In this section we show three applications of the median top-$k$ list, in which the
number of objects ranges from $7$ to $613$.

\paragraph{Eurobarometer: trust in the sources of information} In the Eurobarometer 55.2
survey \citep{eurobarometer552}, $12{,}765$ respondents chose, in order of trust, the three
sources that they would trust most to explain the reasons for a major disaster, among
$n=7$ sources: scientists, journalists, consumer organisations, environmental protection
associations, government representatives, medical doctors and companies. Not choosing a
source is a judgement on the chosen ones, but it says nothing about the order of the four
sources that are not chosen, and we used the third convention of Section~\ref{sec:missing}.
The median top-$3$ list, certified, is scientists, environmental associations and doctors
($\tauX^{obs}=0.4372$), and it coincides with the usual summaries of this kind of question
(first mentions, total mentions, scores $3$, $2$, $1$). The convention matters when the
solution can tie the sources not chosen. With the third convention the MCBO with $b=3$
buckets is (scientists, environmental associations, doctors), (journalists, consumer
organisations, government), companies, and it singles out the companies as the least
trusted source; with the second convention it puts the four sources chosen less often in
a single last bucket, as predicted by Eq.~\ref{eq:conv23}. The question, the sample and the
complete results are given in the supplementary material.

\paragraph{Eurovision Song Contest} Since 2016, in the Eurovision Song Contest each
participating country has a jury of five members, and each juror ranks all the songs of a
show except the song of his or her own country, which cannot be voted. The rankings of the
five jurors are combined into a ranking of the national jury, and the first ten songs of
this ranking receive $12$, $10$, $8$, $7$, \dots, $1$ points; the jury ranking of the show is
given by the sum of the points over the countries. We used the rankings of the individual
jurors from the MIRoVision data set \citep{burgoyneetal2023}, for the $18$ shows (finals and
semi-finals) from 2016 to 2022 (the contest was not held in 2020), with $n$ from $16$ to $26$
songs and $m$ from $75$ to $215$ jurors. The song of the country of the juror is a missing
object, and we used the first convention of Section~\ref{sec:missing}: placing it in the
last position would mean that the juror considers it the worst song. In the final of 2019
and in the final and the second semi-final of 2022, some national juries were annulled and
replaced by aggregated results, so that their rankings are not available: for the
comparison we used the points of the juries whose rankings are in the data.

For each show we computed the median top-$10$ list of all the jurors, which is the natural
counterpart of the ten songs that receive points from each jury, and we compared it with
the top-$10$ list of the jury points. Table~\ref{tab:eurovision} shows that the median
top-$10$ list was certified in all the shows, with $15$ to $19$ candidates and in at most
$0.1$ seconds. The two lists have $8$ to $10$ songs in common and the same winner in $16$
shows out of $18$; in the first semi-final of 2016 and in the second semi-final of 2019 the
median top-$10$ list is led by a different song. By construction, the median top-$10$ list
has a larger agreement with the jurors than the list of the jury points, which is the
result of a two-stage procedure of the Borda type: $\tauX^{obs}$ ranges from $0.19$ to
$0.41$ over the shows, and it exceeds the one of the list of the jury points by $0.0006$ to
$0.018$. In the final of 2019 the winner of the median top-$10$ list (Sweden) is also the
winner of the points of the juries in the data, while the official jury ranking, which
includes the aggregated result that replaced an annulled jury, was won by North Macedonia.

\begin{table}[h!]
\centering
\tablesize
\caption{Eurovision Song Contest 2016--2022: median top-10 list of the jurors, all certified, compared with the top-10 list of the jury points (candidates of the exact procedure, songs in common to the two lists, winners)}
\label{tab:eurovision}
\begin{tabular}{lrrrrll}
\toprule
& & & & & \multicolumn{2}{c}{Winner} \\ \cmidrule(lr){6-7}
Show & $n$ & $m$ & Candidates & Common & Median & Jury \\
\midrule
2016 final & 26 & 210 & 17 & 8 & AU & AU \\
2016 semi-final 1 & 18 & 104 & 18 & 9 & RU & MT \\
2016 semi-final 2 & 18 & 105 & 18 & 10 & AU & AU \\
2017 final & 26 & 210 & 18 & 9 & PT & PT \\
2017 semi-final 1 & 18 & 105 & 18 & 8 & PT & PT \\
2017 semi-final 2 & 18 & 105 & 18 & 9 & BG & BG \\
2018 final & 26 & 215 & 19 & 8 & AT & AT \\
2018 semi-final 1 & 19 & 110 & 18 & 9 & IL & IL \\
2018 semi-final 2 & 18 & 105 & 18 & 10 & SE & SE \\
2019 final & 26 & 200 & 15 & 10 & SE & SE \\
2019 semi-final 1 & 17 & 100 & 17 & 9 & CZ & CZ \\
2019 semi-final 2 & 18 & 105 & 18 & 10 & SE & MK \\
2021 final & 26 & 195 & 18 & 8 & CH & CH \\
2021 semi-final 1 & 16 & 95 & 16 & 9 & MT & MT \\
2021 semi-final 2 & 17 & 100 & 17 & 9 & CH & CH \\
2022 final & 25 & 170 & 19 & 9 & GB & GB \\
2022 semi-final 1 & 17 & 95 & 17 & 9 & GR & GR \\
2022 semi-final 2 & 18 & 75 & 18 & 9 & SE & SE \\
\bottomrule
\end{tabular}
\end{table}

\paragraph{AAMAS bidding data} The AAMAS 2015 bidding data, available in the PrefLib
repository \citep{matteiwalsh2013}, contain the bids of $m=201$ reviewers on $n=613$ papers,
each placed by every reviewer into one of four categories: yes, maybe, no answer and no.
There are no missing objects. Each reviewer places most of the papers in the same category
(no answer), so that a tie is the most frequent relation between two papers, and every
entry of the combined input matrix is positive: $c_{ij}$ ranges from $111$ to $201$, with
$m=201$. Since $y_{ij}$ is either $1$ or $-1$, each term $c_{ij}y_{ij}$ of the average
$\tauX$ is maximized by $y_{ij}=1$, and all of them are maximized together by the solution
in which all the papers are tied, which is therefore the unique median ranking when no
constraint is imposed. A constraint on the solution is then the only way to obtain an
informative consensus, and the median top-$k$ lists single out the papers that are most
wanted by the reviewers. Table~\ref{tab:aamas} shows that the median top-$5$ and top-$10$
lists are certified and computed in less than $0.1$ seconds; with $k=20$ and $k=50$ the
candidates are $23$ and $55$, the solutions cannot be certified, and the solutions of BOLT
were computed in about $0.6$ and $6$ seconds. The table also reports the MCBO with $b=4$
buckets, as many as the categories: BOLT took $0.3$ seconds, against $114$ seconds of QUICK
\citep{dambrosioetal2019}, which found a slightly better solution (average $\tauX$ equal to
$0.862295$ against $0.862293$).

\begin{table}[h!]
\centering\tablesize
\caption{AAMAS 2015 bidding data ($n=613$ papers, $m=201$ reviewers): MCBOs with $b=4$
buckets and median top-$k$ lists.}
\label{tab:aamas}
\begin{tabular}{lrrlrr}
\toprule
Problem & Buckets & $\bar\tau_X$ & Certified & Candidates & Time (s)\\
\midrule
$b=4$ buckets (BOLT) & 4 & 0.862293 & -- & -- & 0.34\\
$b=4$ buckets (QUICK) & 4 & 0.862295 & -- & -- & 114.38\\
Top-5 & 6 & 0.857560 & yes & 5 & 0.06\\
Top-10 & 11 & 0.845570 & yes & 10 & 0.03\\
Top-20 & 21 & 0.821302 & no & 23 & 0.62\\
Top-50 & 51 & 0.748848 & no & 55 & 5.60\\
\bottomrule
\end{tabular}
\end{table}

\subsection{Large and sparse data sets}\label{sec:large}

In this section we consider two data sets with a large number of judges, in which the
judges express their preferences through ratings. A rating on a scale with $L$ levels gives
a bucket order with at most $L$ buckets of the objects rated by the judge; for the objects
that are not rated we report the results obtained with the first and the second
convention of Section~\ref{sec:missing}.

\paragraph{Sushi data with scores} \citet{dambrosioetal2019} analyzed a version of the
sushi data \citep{kamishima2003} in which each of $5000$ respondents gave a score from $0$
to $4$ to $10$ sushi out of $100$, and they computed the MCBO with $b=5$ buckets by DECoR,
in $674$ seconds, with average $\tauX$ equal to $0.0032$. Table~\ref{tab:sushicost} shows
that, with the sushi that are not rated treated as missing objects, as in
\citet{dambrosioetal2019}, BOLT finds a better solution ($\bar\tau_X=0.0033517$, i.e.,
$\tauX^{obs}=0.3687$ against $0.3522$) in $0.03$ seconds, while QUICK finds a solution with a
slightly smaller average $\tauX$ in $0.9$ seconds. With the sushi that are not rated tied
in the last position (second convention), the solution has sizes $(1,1,1,1,96)$: the four
sushi rated by the largest numbers of respondents, followed by all the others.

\begin{table}[h!]
\centering\tablesize
\caption{Sushi data with scores ($10$ sushi scored by each respondent out of $100$), with
the sushi not rated as missing objects: MCBO with $b=5$ buckets.}
\label{tab:sushicost}
\begin{tabular}{llrrr}
\toprule
Solution & Bucket sizes & $\bar\tau_X$ & $\tau_X^{obs}$ & Time (s)\\
\midrule
Published (DECoR) & 1, 44, 9, 40, 6 & 0.0032017 & 0.3522 & 674.00\\
BOLT & 44, 53, 1, 1, 1 & 0.0033517 & 0.3687 & 0.03\\
QUICK & 42, 1, 55, 1, 1 & 0.0033516 & 0.3687 & 0.91\\
\bottomrule
\end{tabular}
\end{table}

\paragraph{MovieLens data} The MovieLens 1M data set \citep{harperkonstan2015} contains
$1{,}000{,}209$ ratings, from $1$ to $5$ stars, of $3706$ movies by $6040$ users, each of whom
rated at least $20$ movies. We computed the median top-$10$ list and the MCBO with $b=5$
buckets (as many as the levels of the ratings) of the $N$ most rated movies, with
$N=500$, $1000$ and all the $3706$ movies, with the first and the second convention of
Section~\ref{sec:missing}. Each user gives a bucket order of the movies that he or she
rated. The data are very sparse. With $N=500$, $17.6\%$ of the ratings are observed, and
every pair of movies was rated by at least $100$ users in $99.5\%$ of the cases. With
$N=1000$ the observed ratings are $12.4\%$, and $74\%$ of the pairs were rated by at least
$100$ users. With all the movies, only $4.5\%$ of the ratings and $0.5\%$ of the ordered
pairs are observed, $18\%$ of the pairs of movies were never rated by the same user, and the
median number of users who rated both movies of a pair is $6$. The combined input matrix
was computed in $1.5$, $2.2$ and $10$ seconds, respectively.

Table~\ref{tab:movielens} shows the results. With the movies not rated as missing objects,
the median top-$10$ list was certified for all the values of $N$, with $10$ to $15$
candidates, in at most $1.5$ seconds. With all the movies, the MCBO with $b=5$ buckets took
$79$ seconds. The average $\tauX$ ranges from $0.0014$ to $0.0214$, while $\tauX^{obs}$ ranges
from $0.30$ to $0.42$. With the movies not rated tied in the last position, the solutions are
very different: the solution with $b=5$ buckets has four single movies followed by all the
others. The average $\tauX$ is large (from $0.72$ to $0.92$) because most pairs of movies are
tied for most users, since both movies were not rated.

\begin{table}[h!]
\centering\tablesize
\caption{MovieLens 1M, the $N$ most rated movies: median top-10 list and MCBO with $b=5$
buckets, with the movies not rated as missing objects and tied in the last position ($b$:
number of buckets of the solution).}
\label{tab:movielens}
\begin{tabular}{rlrrrrrrr}
\toprule
& & \multicolumn{4}{c}{Not rated: missing} & \multicolumn{3}{c}{Not rated: tied last} \\
\cmidrule(lr){3-6} \cmidrule(lr){7-9}
$N$ & Solution & $b$ & $\bar\tau_X$ & $\tau_X^{obs}$ & Time (s) & $b$ & $\bar\tau_X$ & Time (s)\\
\midrule
500 & Top-10 & 11 & 0.01877 & 0.3282 & 0.11 & 11 & 0.7168 & 0.02\\
500 & $b=5$ & 5 & 0.02141 & 0.3745 & 0.38 & 5 & 0.7228 & 0.26\\
1,000 & Top-10 & 11 & 0.00993 & 0.3168 & 0.07 & 11 & 0.7903 & 0.08\\
1,000 & $b=5$ & 5 & 0.01228 & 0.3919 & 2.47 & 5 & 0.7929 & 1.48\\
3,706 & Top-10 & 11 & 0.00142 & 0.3021 & 1.45 & 11 & 0.9162 & 1.18\\
3,706 & $b=5$ & 5 & 0.00195 & 0.4150 & 78.60 & 5 & 0.9167 & 23.79\\
\bottomrule
\end{tabular}
\end{table}

Table~\ref{tab:movielens-top10} compares the two median top-$10$ lists of all the movies.
With the movies not rated as missing objects, the list contains movies that are both
highly rated and rated by many users (The Godfather, The Shawshank Redemption, Schindler's
List, Star Wars, The Usual Suspects, \dots), and it is the same with $N=1000$ and with all
the movies. With the movies not rated tied in the last position, the list contains the ten
movies with the largest numbers of ratings (nine of them with $N=500$ and $N=1000$), and it
has $3$ to $5$ movies in common with the first one. Again, the first convention summarizes
the preferences expressed among the movies that each user watched, while the second one
mainly rewards the movies that were watched by many users. For comparison, the ten movies
with the highest mean rating are movies with very few ratings, and none of them is in the
two median top-$10$ lists of all the movies: with the first convention, each pair of movies
counts as many times as the number of users who rated both, so that a movie rated by a few
enthusiastic users cannot enter a median top-$10$ list.

\begin{table}[h!]
\centering
\tablesize
\caption{MovieLens 1M, all the 3{,}706 movies: median top-10 lists with the movies not rated as missing objects and tied in the last position}
\label{tab:movielens-top10}
\begin{tabular}{rll}
\toprule
 & Not rated as missing & Not rated tied last \\
\midrule
1 & The Godfather (1972) & American Beauty (1999) \\
2 & The Shawshank Redemption (1994) & Star Wars IV (1977) \\
3 & Schindler's List (1993) & Star Wars V (1980) \\
4 & Star Wars IV (1977) & Saving Private Ryan (1998) \\
5 & The Usual Suspects (1995) & The Silence of the Lambs (1991) \\
6 & Raiders of the Lost Ark (1981) & The Matrix (1999) \\
7 & The Silence of the Lambs (1991) & Star Wars VI (1983) \\
8 & American Beauty (1999) & Terminator 2: Judgment Day (1991) \\
9 & The Sixth Sense (1999) & Back to the Future (1985) \\
10 & Star Wars V (1980) & Jurassic Park (1993) \\
\bottomrule
\end{tabular}
\end{table}

\section{Concluding remarks}\label{sec:conclusions}

The median ranking and its constrained versions have been treated in this paper as one
problem: the search, among the rankings with an admissible pattern, of the ranking that
maximizes the average $\tauX$, written in terms of the combined input matrix. The same
heuristic, BOLT, and the same exact formulation apply to every set of admissible
patterns, and only the partition step and the local moves change with the constraint.

The experiments show that BOLT is accurate, with some limits that are worth stating. On
the $83$ real and simulated problems with certified optimum, BOLT returned the optimum in
$72$, and its largest gap from the optimum was $8.3\times10^{-5}$ in terms of the average
$\tauX$. On the benchmark of \citet{akbariescobedo2023a}, with up to $2{,}819$ objects, it
was better than the published ACP solutions in $75$ of the $85$ data sets and equal within
the rounding in the others. On the data of \citet{badalsingh2026} it was within a few
points of distance from SIgFUR, or better, in less than a second against minutes or hours
on the data sets with up to $240$ objects, and it was better than the best published
solution on the largest one. With nearly random judges, however, the objective function is
flat and the optimum is hard to find: on the simulated problems with ties BOLT found the
optimum in $433$ of the $733$ certified cases, the share falling below $10\%$ with $100$
objects or more and $\theta=0.02$, although the mean gap remained of the order of
$6\times10^{-5}$. In these problems the optimum has few ties, and the complete rankings
added to the starting solutions keep BOLT from stopping in local optima with too many ties;
the ablation also shows that both the local search and the optimal tie partitioning
matter, and that no single starting solution is enough. With constraints, the exact
formulation certified all the problems with $20$ objects, where BOLT returned the optimum
in $38$ of $40$ cases, the two exceptions having a partial pattern. With $50$ objects it
certified, within ten minutes, all the top-$5$ lists and nine of the ten complete
patterns, but only $7$ of the $10$ problems with $b=5$ buckets and $3$ of the $10$ with a
partial pattern; BOLT returned the optimum in $27$ of the $29$ certified problems, and in
the others the exact formulation improved it by less than $2.3\times10^{-4}$.

These results suggest a simple use of the two tools. BOLT is the method of choice for
problems of any size: once the combined input matrix has been computed, its computing time
does not depend on the number of judges, except through the rankings of the judges used as
starting solutions, whose number can be limited when they are many. The exact formulation
is the tool to certify a solution, or to bound its distance from the optimum, when the
problem is small enough: up to a few hundred objects without constraints, and up to about
fifty objects with a complete pattern, depending on how much the judges disagree. In this
range it is worth running also in practice: in the Q-sort of Section~\ref{sec:qmethod},
with $33$ statements and nine participants who disagree considerably, it certified in a
few minutes a consensus that BOLT had missed.

Several questions remain open. The exact formulation is weak when the number of buckets is
free, and stronger formulations or decomposition methods
\citep{yooescobedo2021,akbariescobedo2023a,kawashimasukegawa2026} could extend the
certification of the constrained problems. The treatment of the objects not ranked by the
judges is a choice on the combined input matrix, and the conventions of
Section~\ref{sec:missing}, which already lead to different solutions in the applications of
Section~\ref{sec:applications}, deserve a comparison of their own \citep{andrieuetal2023}.
Finally, when the judges are heterogeneous a single consensus may be a poor summary of
their preferences, and median rankings of groups of judges are a natural extension of this
work.

\section*{Supplementary material}
The supplementary material contains the derivations, the checks and the pseudocode of the
components of BOLT (Appendix~A), and the information needed to reproduce the experiments,
the details of the comparison with MorK and of the Eurobarometer data, and the complete
tables of the experiments (Appendix~B).

\section*{Software and data availability}
The code of BOLT, of the exact procedure for the median top-$k$ list and of the algorithms
for the median constrained bucket orders, and the scripts that reproduce all the
experiments, with the results and the description of the computing environment, are
available upon request to the authors.

\appendix
\renewcommand{\thetable}{\Alph{section}.\arabic{table}}
\setcounter{table}{0}
\makeatletter\@addtoreset{table}{section}\makeatother
\renewcommand{\thealgorithm}{S.\arabic{algorithm}}
\setcounter{algorithm}{0}

\section{Methods}\label{app:methods}

\subsection{Cardinality of the sub-universes}\label{app:card}
Table~\ref{tab:cardinality} shows the decomposition of the universe of rankings of
Section~\ref{sec:patterns} of the paper for $n=5$ and $n=6$ objects. For $n=5$, the last
column reproduces the corresponding row of Table~1 in \citet{dambrosioetal2019}.

\begin{table}[h!]
\centering
\tablesize
\caption{Cardinality of the sub-universes of rankings with $n=5$ and $n=6$ objects,
decomposed by number of buckets $b$, shape $\lambda$ and pattern. For each shape, the
table reports the number of patterns (npat), the number of rankings with each pattern
($|\Znnu|$) and the number of rankings with that shape ($|\Znlambda|$). The last column
reports $|\Znb|=b!\stirling{n}{b}$.}
\label{tab:cardinality}
\begin{tabular}{cclrrrr}
\toprule
$n$ & $b$ & shape $\lambda$ & npat & $|\Znnu|$ & $|\Znlambda|$ & $|\Znb|$\\
\midrule
5 & 1 & $(5)$ & 1 & 1 & 1 & 1\\
  & 2 & $(4,1)$ & 2 & 5 & 10 & 30\\
  &   & $(3,2)$ & 2 & 10 & 20 & \\
  & 3 & $(3,1,1)$ & 3 & 20 & 60 & 150\\
  &   & $(2,2,1)$ & 3 & 30 & 90 & \\
  & 4 & $(2,1,1,1)$ & 4 & 60 & 240 & 240\\
  & 5 & $(1,1,1,1,1)$ & 1 & 120 & 120 & 120\\
\cmidrule{2-7}
  & \multicolumn{5}{l}{Total: $|\mathcal{Z}^{5}|$} & 541\\
\midrule
6 & 1 & $(6)$ & 1 & 1 & 1 & 1\\
  & 2 & $(5,1)$ & 2 & 6 & 12 & 62\\
  &   & $(4,2)$ & 2 & 15 & 30 & \\
  &   & $(3,3)$ & 1 & 20 & 20 & \\
  & 3 & $(4,1,1)$ & 3 & 30 & 90 & 540\\
  &   & $(3,2,1)$ & 6 & 60 & 360 & \\
  &   & $(2,2,2)$ & 1 & 90 & 90 & \\
  & 4 & $(3,1,1,1)$ & 4 & 120 & 480 & 1,560\\
  &   & $(2,2,1,1)$ & 6 & 180 & 1,080 & \\
  & 5 & $(2,1,1,1,1)$ & 5 & 360 & 1,800 & 1,800\\
  & 6 & $(1,1,1,1,1,1)$ & 1 & 720 & 720 & 720\\
\cmidrule{2-7}
  & \multicolumn{5}{l}{Total: $|\mathcal{Z}^{6}|$} & 4,683\\
\bottomrule
\end{tabular}
\end{table}

\subsection{Two checks of the partition step}\label{app:checkdp}
The recursion of Eq.~\ref{eq:dp} was checked against the enumeration of all the
$2^{n-1}$ partitions of a fixed order, on $200$ random problems with $n=4$ to $8$ objects
and judges with ties. With all the patterns admissible, the recursion and the optimal tie
partitioning of our implementation returned the optimum in all the problems, and so did the
recursion with at most $b$ buckets, with $b$ drawn at random. We also checked the third
consequence of Section~\ref{sec:consequences}: with judges without ties, the enumeration
of $\Zn$ on $300$ random problems for each $n=3,4,5$ found median rankings with ties in
$353$ of the $900$ problems, but never only median rankings with ties, so that a median
ranking without ties was always available.

\subsection{A check of the special cases of BOLT}\label{app:special}
Table~\ref{tab:special} details the check of the special cases summarized in
Section~\ref{sec:complete} of the paper. We generated $60$ data sets with $n$ drawn from
$\{4,\dots,7\}$ and $m$ from $\{3,\dots,12\}$: each judge ranks the objects in a random
order and ties consecutive objects with probability $0.6$; in one data set out of three
each judge has one missing object, and in one out of two the judges have non-integer
weights drawn from $[0.5,3]$. The complete pattern was drawn at random among the patterns
with $2$ to $n-1$ buckets, and the partial pattern was obtained from it by leaving the
second size free, when the pattern has at least three buckets. For each data set and each
set of admissible rankings, the solution of BOLT was compared with the optimum obtained by
enumerating all the admissible rankings. BOLT always returned an admissible ranking, and
it missed the optimum in three problems, which come from two data sets with weights and
$n=6,7$: a top-$3$ list, and the complete and the partial pattern of the same data set.

\begin{table}[h!]
\centering\tablesize
\caption{BOLT against the enumeration of the admissible rankings ($60$ data sets, $n=4$
to $7$): number of problems, of problems in which the solution of BOLT is admissible and
of problems in which it is optimal.}\label{tab:special}
\begin{tabular}{lrrr}
\toprule
Admissible rankings & problems & admissible & optimal\\
\midrule
no constraint ($\Zn$) & 60 & 60 & 60\\
permutations ($\Sn$) & 60 & 60 & 60\\
$b$ buckets, $b=2,\dots,n-1$ & 220 & 220 & 220\\
complete pattern (at random) & 60 & 60 & 59\\
partial pattern (one free size) & 54 & 54 & 53\\
top-$k$ list, $k=1,\dots,n-2$ & 220 & 220 & 219\\
\bottomrule
\end{tabular}
\end{table}

\subsection{Derivation of the objective function of the top-$k$ lists}\label{app:topk}
Let $Y\in\Tnk$ be a top-$k$ list in which the set $K$ of the first $k$ objects is ranked
as $a_1\pre_Y a_2\pre_Y\cdots\pre_Y a_k$, and let $\gamma_a=\sum_{j\neq a}c_{ja}$. By
Eq.~\ref{eq:ordered} of the paper, Eq.~\ref{eq:topk} holds if the penalty of $Y$ is
$P(Y)=\sum_{a\in K}\gamma_a-\sum_{s<t}c_{a_sa_t}$. The ordered pairs of $Y$ are the pairs
$(a_s,a_t)$ with $s<t$, which contribute $c_{a_ta_s}$, and the pairs $(a,j)$ with $a\in K$
and $j$ in the last bucket, which contribute $c_{ja}$. For each $a\in K$,
$\sum_{j\notin K} c_{ja}=\gamma_a-\sum_{a'\in K,\,a'\neq a}c_{a'a}$, so that
\[
P(Y)=\sum_{s<t}c_{a_ta_s}+\sum_{a\in K}\gamma_a-\sum_{a\neq a'\in K}c_{a'a}
=\sum_{a\in K}\gamma_a-\sum_{s<t}c_{a_sa_t},
\]
because the sum over the ordered pairs of distinct objects of $K$ is
$\sum_{s<t}(c_{a_sa_t}+c_{a_ta_s})$.

\subsection{Pseudocode of the components of BOLT}\label{app:pseudo}

The algorithms below give the details of the components described in
Sections~\ref{sec:insertion}--\ref{sec:exacttopk} of the paper, as implemented in
our code. In all of them, a move or a choice is applied only if it improves the
current value by more than a tolerance equal to $10^{-10}\max\{1,\sum_{i,j}|c_{ij}|\}$, so
that rounding errors cannot produce cycles; the objects of each bucket are listed by their
index, and the objects of a ranking are visited in the order that they have at the
beginning of each pass, with ties broken by the index.

\begin{algorithm}[t]
\caption{Tie-aware insertion local search (Section~\ref{sec:insertion})}\label{alg:insertion}
\small
\begin{algorithmic}[1]
\Require ranking $Y$, combined input matrix $\C$, flag \textit{ties} (false for the local
search without ties)
\Repeat
  \State \textit{improved} $\gets$ false
  \For{each object $z$, in the order of $Y$ at the beginning of the pass}
    \State remove $z$ from $Y$; let $B'_1,\dots,B'_{b'}$ be the buckets left
    \State $U_{zt}\gets\sum_{j\in B'_t}c_{zj}$ and $V_{zt}\gets\sum_{j\in B'_t}c_{jz}$, $t=1,\dots,b'$,
    with their cumulative sums
    \State $P_0\gets$ contribution of $z$ to the penalty in its current position
    \State $P^\ast\gets\min_{t=0,\dots,b'}\{\sum_{s\leq t}U_{zs}+\sum_{s>t}V_{zs}\}$
    \Comment{$z$ alone in a gap}
    \If{\textit{ties}}
      \State $P^\ast_{\text{tie}}\gets\min_{t=1,\dots,b'}\{\sum_{s<t}U_{zs}+\sum_{s>t}V_{zs}\}$
      \Comment{$z$ tied with $B'_t$}
      \State if $P^\ast_{\text{tie}}<P^\ast$, the best position is the tie, and $P^\ast\gets P^\ast_{\text{tie}}$
    \EndIf
    \If{$P^\ast<P_0$}
      \State put $z$ in the best position; \textit{improved} $\gets$ true
    \Else
      \State put $z$ back in its position
    \EndIf
  \EndFor
\Until{not \textit{improved}}
\Ensure a local optimum $Y$
\end{algorithmic}
\end{algorithm}

With a fixed number of buckets (Section~\ref{sec:ls}), the same algorithm is used with
one change: an object that is alone in its bucket is placed only in the gaps, and any
other object only in the existing buckets, so that the number of buckets does not change.

\begin{algorithm}[t]
\caption{Best partition of a fixed order (Section~\ref{sec:dp})}\label{alg:partition}
\small
\begin{algorithmic}[1]
\Require order $\mathbf{o}=\langle o_1,\dots,o_n\rangle$, $\C$, sets $A_t$ of admissible
sizes and set $\mathcal{B}$ of admissible numbers of buckets
\State $D_0(0)\gets 0$; $D_t(e)\gets-\infty$ for all the other $t$ and $e$
\For{$e=1,\dots,n$}
  \State $\sigma\gets 0$
  \For{$s=e,e-1,\dots,1$} \Comment{all the values $g(s,e)$ in $O(n)$ operations}
    \State $g(s,e)\gets g(s,e-1)+\sigma$, with $g(e,e-1)=0$
    \State $\sigma\gets\sigma+c_{o_e o_{s-1}}$
  \EndFor
  \For{$t=1,\dots,\max\mathcal{B}$}
    \State $D_t(e)\gets\max\{D_{t-1}(s-1)+g(s,e):\ e-s+1\in A_t\}$, and $S_t(e)\gets$ the
    maximizing $s$ \Comment{$-\infty$ if no $s$ is admissible}
  \EndFor
\EndFor
\State $t^\ast\gets\arg\max_{t\in\mathcal{B}}D_t(n)$
\If{$D_{t^\ast}(n)=-\infty$}
  \State \Return no ranking compatible with $\mathbf{o}$ is admissible
\EndIf
\State $e\gets n$
\For{$t=t^\ast,t^\ast-1,\dots,1$} \Comment{backtracking}
  \State the $t$th bucket contains the objects in the positions $S_t(e),\dots,e$; $e\gets S_t(e)-1$
\EndFor
\Ensure the admissible ranking compatible with $\mathbf{o}$ with the smallest penalty
\end{algorithmic}
\end{algorithm}

For the median ranking every size is admissible, the index $t$ can be dropped, and the
algorithm becomes the optimal tie partitioning (OTP), with $D(e)=\max_s\{D(s-1)+g(s,e)\}$.

\begin{algorithm}[t]
\caption{Local search that preserves the pattern (Section~\ref{sec:ls})}\label{alg:cascade}
\scriptsize
\begin{algorithmic}[1]
\Require ranking $Y$ with buckets $B_1,\dots,B_b$, $\C$
\State $U_{zu}\gets\sum_{j\in B_u,\,j\neq z}c_{zj}$ and $V_{zu}\gets\sum_{j\in B_u,\,j\neq z}c_{jz}$
for every object $z$ and bucket $u$
\Repeat
  \State \Call{Cascade}{}; \Call{Exchange}{}
\Until{\Call{Exchange}{} has applied no move}
\Ensure a local optimum $Y$ with the same pattern
\Statex
\Procedure{Cascade}{}
  \Repeat
    \For{each object $z$, in the order of $Y$ at the beginning of the pass; let $z\in B_u$}
      \For{each bucket $t\neq u$}
        \State $M\gets\{z: u\to t\}$, plus: if $t>u$, the first object of $B_r$, $r\to r-1$, for
        $r=u+1,\dots,t$; if $t<u$, the last object of $B_r$, $r\to r+1$, for $r=t,\dots,u-1$
        \State $\Delta_t\gets$ change of the penalty: for each object of $M$, the change of its
        contribution with respect to the objects that do not move, from its sums $U$ and
        $V$; plus the change of the pairs of objects of $M$ \Comment{$O(b^2)$}
      \EndFor
      \If{$\min_t\Delta_t<0$}
        \State apply the move with the smallest $\Delta_t$; update $U$ and $V$ ($O(n)$ for each
        moved object)
      \EndIf
    \EndFor
  \Until{no cascade move has been applied in the pass}
\EndProcedure
\Statex
\Procedure{Exchange}{}
  \Repeat
    \For{each object $x$, in the order of $Y$ at the beginning of the pass}
      \For{each object $y$ in a bucket different from that of $x$}
        \State let $s<t$ be the buckets of the two objects, $p\in B_s$ and $q\in B_t$
        \State $\Delta_y\gets F_p(s,t)-F_q(s,t)+c_{pq}+c_{qp}$, with
        $F_z(s,t)=\sum_{s\leq w<t}U_{zw}-\sum_{s<w\leq t}V_{zw}$ \Comment{$O(1)$}
      \EndFor
      \If{$\min_y\Delta_y<0$}
        \State exchange $x$ with the object $y$ with the smallest $\Delta_y$; update $U$, $V$ and
        their cumulative sums
      \EndIf
    \EndFor
  \Until{no exchange has been applied in the pass}
\EndProcedure
\end{algorithmic}
\end{algorithm}

With a fixed number of buckets, \textsc{Cascade} is replaced by the insertion local search
with a fixed number of buckets (Algorithm~\ref{alg:insertion}), and it is alternated with
\textsc{Exchange} in the same way.

\begin{algorithm}[t]
\caption{Exact median top-$k$ list (Section~\ref{sec:exacttopk})}\label{alg:topk}
\small
\begin{algorithmic}[1]
\Require $\C$, $k$, largest number of candidates $n_{\max}$ (default $22$)
\State $\gamma_a\gets\sum_{j\neq a}c_{ja}$ for every object $a$
\State $S\gets$ \Call{Candidates}{$-\infty$}
\If{$|S|>n_{\max}$}
  \State $Y_0\gets$ solution of BOLT with the pattern $(1,\dots,1,n-k)$; $F_0\gets$ its value
  of Eq.~\ref{eq:topkobj}
  \State $S\gets$ \Call{Candidates}{$F_0$}
  \If{$|S|>n_{\max}$}
    \State \Return $Y_0$, not certified
  \EndIf
\EndIf
\State $\Phi(\emptyset)\gets 0$
\For{every subset $R\subseteq S$ with $1\leq|R|\leq k$, in increasing order of its binary code}
  \State $\Phi(R)\gets\max_{a\in R}\{\Phi(R\setminus\{a\})-2\gamma_a+2\sum_{j\in R\setminus\{a\}}c_{ja}\}$
\EndFor
\State $F^\ast\gets\max\{\Phi(R):\ |R|=k\}$
\State \Return the top-$k$ lists with value $F^\ast$, by backtracking (at most $1000$), certified
\Statex
\Procedure{Candidates}{$F_0$}
  \State $S\gets$ all the objects
  \Repeat
    \For{every object $x$}
      \State $\eta^-_x\gets$ sum of the $k-1$ smallest values of $\min(c_{xj},c_{jx})$, $j\in S\setminus\{x\}$
      \State $\eta^+_x\gets$ sum of the $k-1$ largest values of $\max(c_{xj},c_{jx})$, $j\in S\setminus\{x\}$
      \State $L_x\gets-2\gamma_x+2\eta^-_x$; $H_x\gets-2\gamma_x+2\eta^+_x$; $W_x\gets-2\gamma_x+\eta^+_x$
    \EndFor
    \For{every $a\in S$} \Comment{with the values computed above}
      \State remove $a$ if $H_a$ is smaller than the $k$th largest $L_j$, $j\neq a$
      \Comment{exchange test}
      \State remove $a$ if $W_a$ plus the $k-1$ largest $W_x$, $x\in S\setminus\{a\}$, is smaller than $F_0$
      \Comment{bound test}
    \EndFor
  \Until{$S$ does not change}
  \State \Return $S$
\EndProcedure
\end{algorithmic}
\end{algorithm}

\clearpage

\section{Experiments}\label{app:experiments}

\subsection{Reproducibility}\label{app:repro}
All the computations were run on a desktop computer with an Intel Core i5-6500 processor
(four cores, 3.2 GHz) and 16 GB of memory, under Windows 11, with R~4.6.1. The exact
formulations were solved with HiGHS~1.14.0 through the R package highs~1.14.0.2. SIgFUR and
FUR were run in MATLAB R2023a with the code made available by the authors
(\url{https://github.com/prakashvs613/RankAggSIgFUR-tied-matlab}).

Every experiment is a commented R script that fixes the random number generator and its
seed at the beginning, and saves its results together with the description of the R
session; the long runs save one file per problem. The summaries and the tables of the
paper and of this supplementary material are produced from the saved results, without
repeating the computations. BOLT is deterministic, so that all the results that do not
involve computing times are reproduced exactly. The permutations of the simulation
studies were drawn from the Mallows model with the Kendall distance through the
representation of \citet{flignerverducci1986}; the sampler was checked against the exact
probabilities of all the permutations with $n=4$ and $n=5$ objects and three values of
$\theta$, with $50{,}000$ draws each, and the $p$-values of the chi-squared tests ranged
from $0.33$ to $0.97$. The data sets used in the paper are publicly available, but the
Eurobarometer and MovieLens data cannot be redistributed: for them we provide the scripts
that read the original files.

\subsection{Ablation on the problems with certified optimum}\label{app:ablreal}
Table~\ref{tab:ablationreal} repeats the ablation of Table~\ref{tab:ablation} of the paper
on the $83$ real and simulated problems with certified optimum of
Section~\ref{sec:realexp}.

\begin{table}[h!]
\centering\tablesize
\caption{Ablation of BOLT on the $83$ problems with certified optimum: number of optima,
gap from the optimum (in terms of $\bar\tau_X$) and number of problems in which the
variant is worse than BOLT. Top: removal of a component; bottom: a single starting
solution (with its complete ranking from the local search without ties).}
\label{tab:ablationreal}
\centering
\begin{tabular}{@{}lrrrr@{}}
\toprule
\multicolumn{1}{c}{\multirow{2}{*}{Variant}} & \multicolumn{1}{c}{\multirow{2}{*}{\begin{tabular}[c]{@{}c@{}}optima \\ (of 83)\end{tabular}}} & \multicolumn{1}{c}{\multirow{2}{*}{\begin{tabular}[c]{@{}c@{}}mean \\ gap\end{tabular}}} & \multicolumn{1}{c}{\multirow{2}{*}{\begin{tabular}[c]{@{}c@{}}max \\ gap\end{tabular}}} & \multicolumn{1}{c}{\multirow{2}{*}{\begin{tabular}[c]{@{}c@{}}worse than BOLT \\ (of 83)\end{tabular}}} \\
\multicolumn{1}{c}{} & \multicolumn{1}{c}{} & \multicolumn{1}{c}{} & \multicolumn{1}{c}{} & \multicolumn{1}{c}{} \\ \midrule
All the components & 72 & 7.0e-06 & 8.3e-05 & 0 \\
Without the complete rankings & 72 & 8.2e-06 & 1.6e-04 & 2 \\
Only the complete rankings & 68 & 1.2e-05 & 1.6e-04 & 7 \\
Only the starting solutions from $\C$ & 65 & 2.1e-05 & 2.1e-04 & 11 \\
Only the rankings of the judges & 71 & 8.6e-06 & 1.7e-04 & 3 \\
Without OTP & 68 & 2.2e-05 & 5.3e-04 & 6 \\
Only OTP (no local search) & 63 & 8.6e-05 & 4.5e-03 & 15 \\ \midrule
Net flow alone & 50 & 1.0e-04 & 1.0e-03 & 32 \\
Copeland alone & 58 & 5.1e-05 & 4.9e-04 & 24 \\
Markov chain alone & 51 & 9.7e-05 & 1.1e-03 & 30 \\
Bucket pivot alone & 52 & 1.2e-04 & 1.6e-03 & 29 \\
Branch-and-bound start alone & 53 & 6.5e-05 & 6.4e-04 & 27 \\
Best judge alone & 71 & 8.6e-06 & 1.7e-04 & 3 \\
Median judge alone & 54 & 6.8e-05 & 4.6e-04 & 29 \\
Worst judge alone & 27 & 3.3e-04 & 2.6e-03 & 56 \\ \bottomrule
\end{tabular}
\end{table}

\subsection{Certification of the constrained problems}\label{app:certdet}
The problems of Table~\ref{tab:cert} of the paper were generated as follows. For each
$n\in\{20,50\}$ and each of the ten replications, the rankings of $m=100$ judges, without
ties, were drawn from a mixture of two Mallows models with $\theta=0.2$ and equal weights,
centred on the identity and on a random permutation, and the four constraints were
applied to the same data. The exact formulation of Section~\ref{sec:blpconstraints}, with
the strengthening inequalities and the cuts described there, was started from the
solution of BOLT, with four threads and a time limit of $600$ seconds, and the constraint
was checked on all the solutions. In $4$ of the $11$ problems not solved within the time
limit, all with a partial pattern, the exact formulation improved the solution of BOLT.
The results problem by problem are in Table~\ref{tab:instcert}.

\subsection{Comparison with MorK}\label{app:mork}
\citet{azzinietal2026} generated, for each number of objects $n=10,\dots,50$, problems with
$m=10$ judges whose rankings are random permutations, with random weights: $100$ problems
for $n\leq40$, $67$ for $41\leq n\leq45$ and $33$ for $46\leq n\leq50$, $3600$ in all.
Since their problems are not available, we generated $3600$ problems in the same way, with
the weights drawn from the uniform distribution on $(0,1)$ and normalized. All the problems
were solved exactly by the formulation of Section~\ref{sec:blp} with $y_{ij}+y_{ji}=1$, in
at most $25$ seconds; for $n\leq30$ the optimum was also confirmed by our implementation of
MorK-Exact. The gap of a solution is $1-\phi/\phi^\ast$, where $\phi$ is its score and
$\phi^\ast$ the score of the optimum, with the score $\phi^{(E)}$ of
\citet[Eq.~11]{azzinietal2026}; their Table~3 does not state which score is used, and we
assume that it is the same. Table~\ref{tab:mork} reports, for each $n$, the number of
problems in which BOLT returned the optimum and its mean gap, together with the mean gap of
MorK-Heuristic published by \citet{azzinietal2026}. Two cautions apply: the problems come
from the same distribution but are not the same, and when MorK-Exact did not solve all the
problems within one hour ($n=43$, $45$ and $46$ to $50$) the published means refer to the
solved problems only ($17$ of $33$ with $n=50$).

\begin{table}[h!]
\centering\tablesize
\caption{Comparison with MorK: for each $n$, number of problems, number of problems in
which BOLT returned the optimum, and mean gap ($\times10^{-5}$) of BOLT and of
MorK-Heuristic (published values).}
\label{tab:mork}
\resizebox{\textwidth}{!}{\begin{tabular}{rrrrr|rrrrr}
\toprule
$n$ & problems & BOLT opt. & BOLT & MorK-H & $n$ & problems & BOLT opt. & BOLT & MorK-H\\
\midrule
10 & 100 & 100 & 0.00 & 0.00 & 31 & 100 & 85 & 1.60 & 8.70\\
11 & 100 & 100 & 0.00 & 1.32 & 32 & 100 & 84 & 2.38 & 13.27\\
12 & 100 & 99 & 0.56 & 2.18 & 33 & 100 & 81 & 2.74 & 10.40\\
13 & 100 & 100 & 0.00 & 3.40 & 34 & 100 & 82 & 2.60 & 12.58\\
14 & 100 & 99 & 1.07 & 3.02 & 35 & 100 & 77 & 3.65 & 11.36\\
15 & 100 & 98 & 1.22 & 4.66 & 36 & 100 & 73 & 2.76 & 13.14\\
16 & 100 & 99 & 0.38 & 4.79 & 37 & 100 & 70 & 3.77 & 10.10\\
17 & 100 & 100 & 0.00 & 10.49 & 38 & 100 & 76 & 2.86 & 9.72\\
18 & 100 & 97 & 1.11 & 11.15 & 39 & 100 & 65 & 4.43 & 16.16\\
19 & 100 & 98 & 0.69 & 5.42 & 40 & 100 & 61 & 5.76 & 17.09\\
20 & 100 & 98 & 0.57 & 6.98 & 41 & 67 & 48 & 3.78 & 13.80\\
21 & 100 & 99 & 0.22 & 11.28 & 42 & 67 & 35 & 5.17 & 10.25\\
22 & 100 & 96 & 1.02 & 4.32 & 43 & 67 & 34 & 5.57 & 14.24\\
23 & 100 & 96 & 1.35 & 6.38 & 44 & 67 & 37 & 5.24 & 12.00\\
24 & 100 & 96 & 1.02 & 13.34 & 45 & 67 & 26 & 7.69 & 13.57\\
25 & 100 & 90 & 2.26 & 11.64 & 46 & 33 & 17 & 5.99 & 9.58\\
26 & 100 & 90 & 2.17 & 11.52 & 47 & 33 & 8 & 7.78 & 12.85\\
27 & 100 & 96 & 0.49 & 8.63 & 48 & 33 & 18 & 4.07 & 11.11\\
28 & 100 & 92 & 0.82 & 10.10 & 49 & 33 & 14 & 6.68 & 15.83\\
29 & 100 & 88 & 1.25 & 16.61 & 50 & 33 & 16 & 6.58 & 5.77\\
30 & 100 & 79 & 1.87 & 12.32 & & & & &\\
\midrule
\multicolumn{10}{l}{Mean over the $41$ values of $n$: BOLT 2.66, MorK-Heuristic 9.78; BOLT optimal in 3017 of 3600 problems.}\\
\bottomrule
\end{tabular}}
\end{table}

\subsection{Eurobarometer: question, sample and complete results}\label{app:eurobarometer}
The Eurobarometer 55.2 survey \citep{eurobarometer552} was conducted in May and June 2001 in
the $15$ Member States of the European Union, with $16{,}130$ respondents. Question Q21 was:
``Suppose there was a major disaster in your area. Who would you trust most to explain the
reasons for the disaster?'', followed by ``And the next most?'' and ``And the third
most?''. The respondents had to choose among $n=7$ sources: scientists (Sci), journalists
(Jou), consumer organisations (Con), environmental protection associations (Env),
government representatives (Gov), medical doctors (Doc) and companies (Com); the answers
``other'', ``none of these'' and ``don't know'' were spontaneous. We used the
$m=12{,}765$ respondents who chose three of the seven sources, so that each of them gives a
top-$3$ list. The data are distributed by GESIS (study ZA3509, doi:10.4232/1.10943) and
cannot be redistributed; the script that reads the original file is available from the
authors.

Table~\ref{tab:eboverall} reports, with the third convention, the median top-$3$ list,
certified with $3$ candidates, the top-$3$ lists of the usual summaries and two other
constrained median rankings, with their $\tauX^{obs}$. Table~\ref{tab:ebconventions}
compares the solutions obtained with the second and the third convention: each solution is
evaluated with the objective function of both conventions, the average $\tauX$ for the
second one and $\tauX^{obs}$ for the third one. The median top-$3$ list and the median
ranking with pattern $(3,4)$, i.e., the three most trusted sources without their order, do
not depend on the convention, while the MCBO with $b=3$ buckets does: with the second
convention the ties between the sources not chosen count as agreement and push the
solution towards a large last bucket (Eq.~\ref{eq:conv23}); with the third one the less
trusted sources are separated only by the comparisons between the sources chosen and the
sources not chosen, and the companies are singled out as the least trusted source. On these
data we also checked that, with the third convention, the score matrix of each respondent
is the average of the score matrices of all the complete rankings compatible with the
top-$3$ list, and that the combined input matrices of the two conventions differ by
$u_{ij}$.

\begin{table}[h!]
\centering\tablesize
\caption{Eurobarometer 55.2 ($m=12{,}765$ top-$3$ lists of $n=7$ sources), third
convention: median top-$3$ list, top-$3$ lists of the usual summaries and other
constrained median rankings, with their average $\tauX$ on the observed pairs.}
\label{tab:eboverall}
\begin{tabular}{llr}
\toprule
Solution & Ranking & $\tau_X^{obs}$ \\
\midrule
Median top-3 list & Sci $>$ Env $>$ Doc $>$ (Jou, Con, Gov, Com) & 0.4372 \\
Top-3 list of the first mentions & Sci $>$ Env $>$ Doc $>$ (Jou, Con, Gov, Com) & 0.4372 \\
Top-3 list of the total mentions & Sci $>$ Env $>$ Doc $>$ (Jou, Con, Gov, Com) & 0.4372 \\
Top-3 list of the scores 3-2-1 & Sci $>$ Env $>$ Doc $>$ (Jou, Con, Gov, Com) & 0.4372 \\
Median top-3 set, pattern $(3,4)$ & (Sci, Env, Doc) $>$ (Jou, Con, Gov, Com) & 0.4048 \\
MCBO, $b=3$ & (Sci, Env, Doc) $>$ (Jou, Con, Gov) $>$ Com & 0.4492 \\
\bottomrule
\end{tabular}
\end{table}

\begin{table}[h!]
\centering\tablesize
\caption{Eurobarometer 55.2: median top-$3$ list, median ranking with pattern $(3,4)$ (top-$3$
set) and MCBO with $b=3$ buckets with the second and the third convention (in parentheses,
the conventions that give the solution), each evaluated with the objective function of
both conventions.}
\label{tab:ebconventions}
\begin{tabular}{llrr}
\toprule
Solution & Ranking & $\bar\tau_X$ (second) & $\tau_X^{obs}$ (third) \\
\midrule
Top-3 list (both) & Sci $>$ Env $>$ Doc $>$ (Jou, Con, Gov, Com) & 0.4745 & 0.4372 \\
Top-3 set (both) & (Sci, Env, Doc) $>$ (Jou, Con, Gov, Com) & 0.4598 & 0.4048 \\
$b=3$ (second) & Sci $>$ (Env, Doc) $>$ (Jou, Con, Gov, Com) & 0.4742 & 0.4320 \\
$b=3$ (third) & (Sci, Env, Doc) $>$ (Jou, Con, Gov) $>$ Com & 0.3973 & 0.4492 \\
\bottomrule
\end{tabular}
\end{table}

\subsection{Simulation study for the constrained problems}\label{app:simcons}

\paragraph{Aims and design.} A heuristic has to be assessed on two grounds: how often it
finds the optimum, and at which computational cost. The two aspects cannot be assessed
under the same conditions, since the optimum is known only when the sub-universe of the
admissible rankings can be enumerated, while the computing time matters when it cannot.
By Eq.~\ref{eq:card-pattern}, the pattern $(1,3,4,3,1)$ with $n=12$ objects gives
$554{,}400$ bucket orders, which are enumerated in about $7$ seconds, but the pattern
$(2,4,8,4,2)$ with $n=20$ objects gives about $2.5\times10^{10}$ of them. For this reason
each experiment has a part with few objects, in which BOLT is compared with the optimum
obtained by enumeration, and, when it is relevant, a part with many objects. There is one
experiment for each kind of constraint. For the median top-$k$ list, the main question is
how often the exact procedure of Section~\ref{sec:exacttopk} certifies the solution; for
the median ranking with a prescribed pattern no other algorithm is available, and BOLT is
compared with the optimum; for the MCBO with $b$ buckets of free size, BOLT is also
compared with the QUICK and DECoR algorithms of \citet{dambrosioetal2019}.

\paragraph{Generation of the data.} We sampled permutations, or their first positions, from
two standard models for rankings, and we created the buckets and the ties afterwards. The
first one is the Mallows model with the Kendall distance \citep{mallows1957}, centred on
$(1,2,\dots,n)$, with $\theta=0.05$, $0.2$ and $1$ (the larger $\theta$, the more the judges
agree), sampled through the representation of \citet{flignerverducci1986}. The second one
is the Plackett--Luce model \citep{luce1959,plackett1975}, in which the objects have worths
$\exp(u_1),\dots,\exp(u_n)$, with $u_1>\dots>u_n$ drawn from a normal distribution with mean
zero and standard deviation $s$, and each judge chooses the objects one at a time, with
probabilities proportional to the worths of the objects not yet chosen; the larger $s$, the
more the judges agree. When all the judges come from a single Mallows model, for most pairs
of objects a clear majority agrees with the centre, and in our pilot runs all the
algorithms found the optimum in almost all the data sets. We therefore also sampled the
data from a mixture of two Mallows models with the same $\theta$ and equal weights, the
first one centred on $(1,2,\dots,n)$ and the second one on a random permutation. On average
the two centres order half of the pairs of objects in opposite ways, the elements $c_{ij}$
of these pairs are close to zero, and the penalty becomes flat, with many local optima: as
in Section~\ref{sec:simexp} of the paper, the mixture is a stress test for the algorithms.
Since the mixture has no single centre, the recovery of the true centre is studied only
with the single Mallows model.

\paragraph{Median top-$k$ lists.} Each of $m=500$ judges gives a top-$20$ list, i.e., the
first $20$ positions of a permutation sampled from one of the two models, with the other
objects tied in the last position (second convention of Section~\ref{sec:missing}). We
considered $n=100$, $500$, $1000$ and $5000$ objects, $\theta=0.05$, $0.2$, $1$ and $s=0.5$,
$1$, $2$, with $10$ data sets for each setting, and we computed the median top-$k$ lists with
$k=5$, $10$ and $20$, for a total of $720$ problems. The key quantity is the share of
certified solutions: the exact procedure certifies the solution when the candidates, i.e.,
the only objects that can belong to an optimal top-$k$ list, are at most $22$, without
enumerating the $n!/(n-k)!$ top-$k$ lists (about $10^{37}$ with $n=5000$ and $k=10$);
otherwise it returns the solution of BOLT.

Figure~\ref{fig:simtopk}(a) shows that the exact procedure certified the solution in
$94.2\%$ of the problems: always with $k=5$ and $k=10$, and in $82.5\%$ of the problems with
$k=20$ ($96.7\%$ with the Mallows model and $68.3\%$ with the Plackett--Luce model). The
problems not certified occur mainly with $n=100$ and with the smallest agreement of the
Plackett--Luce model, $s=0.5$, where several objects have column totals close to the one of
the $20$th object and cannot be excluded. When the solution was certified, BOLT found the
optimum in all the problems but two, with relative differences of at most $3.7\times10^{-8}$
in the average $\tauX$, so that it is a reliable fallback when the solution cannot be certified.
The exact procedure is also fast (Figure~\ref{fig:simtopk}(b)): the median computing time
of the certified problems was below $0.1$ seconds with up to $1000$ objects and $2.2$
seconds with $5000$ objects, while BOLT took $22.5$ seconds with $5000$ objects. As
expected, the recovery of the true top-$k$ list depends on the agreement among the judges:
with $\theta=1$ the median top-$k$ list was always the true one; with $\theta=0.2$ it
contained the true $k$ objects in $76.7\%$ of the problems, and in the true order in
$65.8\%$ of them; with $\theta=0.05$ and with the Plackett--Luce model it contained the right
objects less often, and in the right order only rarely. With $s=0.5$ the co-optimal
top-$k$ lists can be very many, because the objects that are never included in a list have
identical columns in the combined input matrix.

\begin{figure}[t]
\centering
\resizebox{\textwidth}{!}{%

}
\caption{Simulation, median top-$k$ lists ($m=500$ judges giving top-$20$ lists, $10$ data
sets per setting). (a) Percentage of solutions certified by the exact procedure of
Section~\ref{sec:exacttopk}. (b) Computing time of the exact procedure for the certified
solutions, all the models together; times below the resolution of the timer ($0.01$ s) are
drawn at $0.005$ s.}\label{fig:simtopk}
\end{figure}

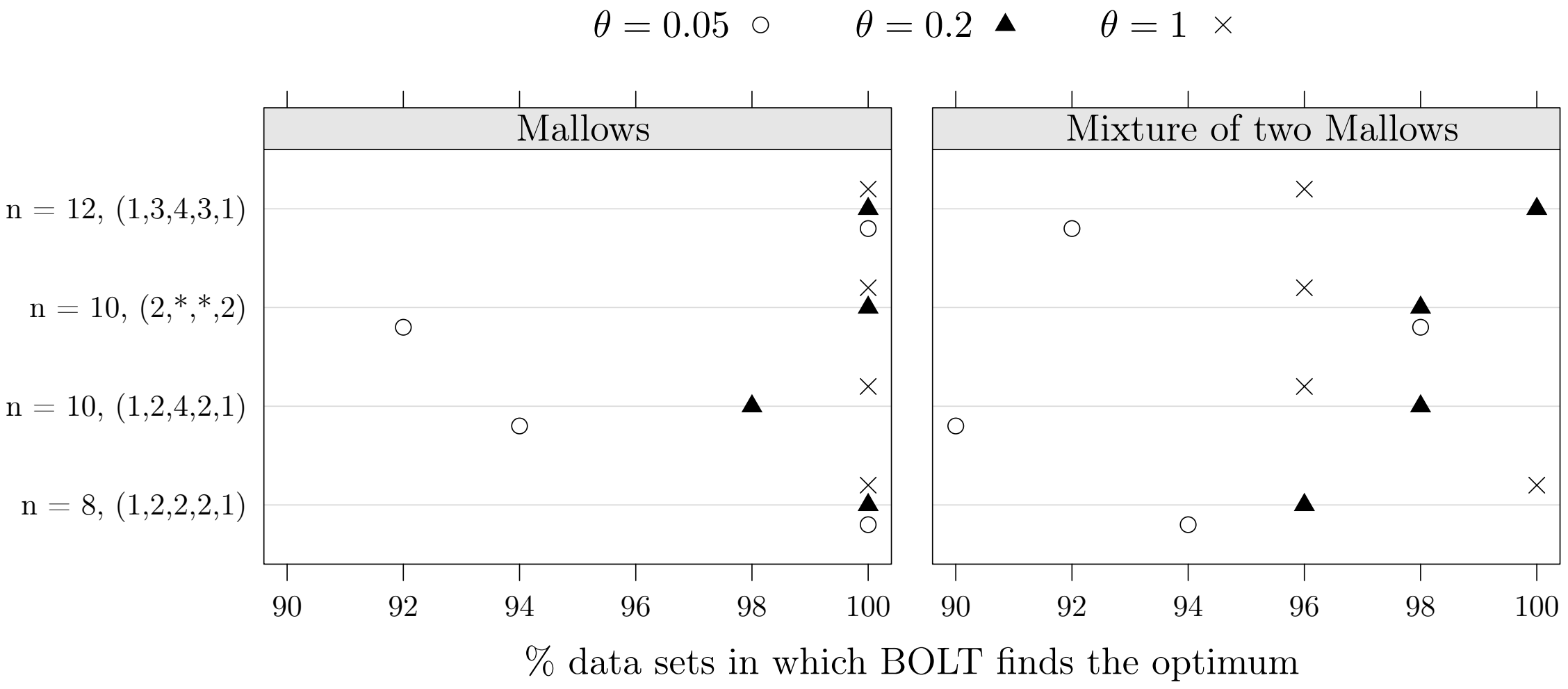
\begin{figure}[t]
\centering
\resizebox{\textwidth}{!}{%
\begin{tikzpicture}[x=1pt,y=1pt]
\definecolor{fillColor}{RGB}{255,255,255}
\path[use as bounding box,fill=fillColor,fill opacity=0.00] (0,0) rectangle (578.16,260.17);
\begin{scope}
\path[clip] (  0.00,  0.00) rectangle (578.16,260.17);

\path[] (  0.00,  0.00) rectangle (578.16,260.17);
\end{scope}
\begin{scope}
\path[clip] (  0.00,  0.00) rectangle (578.16,260.17);
\definecolor{drawColor}{RGB}{0,0,0}

\node[text=drawColor,anchor=base,inner sep=0pt, outer sep=0pt, scale=  1.20] at (330.48, 18.07) {\% data sets in which BOLT finds the optimum};
\end{scope}
\begin{scope}
\path[clip] (  0.00,  0.00) rectangle (578.16,260.17);
\definecolor{drawColor}{RGB}{0,0,0}

\path[draw=drawColor,line width= 0.4pt,line join=round,line cap=round] (114.27,213.88) -- (114.27,219.57);

\path[draw=drawColor,line width= 0.4pt,line join=round,line cap=round] (154.48,213.88) -- (154.48,219.57);

\path[draw=drawColor,line width= 0.4pt,line join=round,line cap=round] (194.69,213.88) -- (194.69,219.57);

\path[draw=drawColor,line width= 0.4pt,line join=round,line cap=round] (234.91,213.88) -- (234.91,219.57);

\path[draw=drawColor,line width= 0.4pt,line join=round,line cap=round] (275.12,213.88) -- (275.12,219.57);

\path[draw=drawColor,line width= 0.4pt,line join=round,line cap=round] (315.33,213.88) -- (315.33,219.57);
\end{scope}
\begin{scope}
\path[clip] (  0.00,  0.00) rectangle (578.16,260.17);
\definecolor{drawColor}{RGB}{0,0,0}

\node[text=drawColor,anchor=base east,inner sep=0pt, outer sep=0pt, scale=  0.96] at (100.54, 73.22) {n = 8, (1,2,2,2,1)};

\node[text=drawColor,anchor=base east,inner sep=0pt, outer sep=0pt, scale=  0.96] at (100.54,107.36) {n = 10, (1,2,4,2,1)};

\node[text=drawColor,anchor=base east,inner sep=0pt, outer sep=0pt, scale=  0.96] at (100.54,141.49) {n = 10, (2,*,*,2)};

\node[text=drawColor,anchor=base east,inner sep=0pt, outer sep=0pt, scale=  0.96] at (100.54,175.63) {n = 12, (1,3,4,3,1)};
\end{scope}
\begin{scope}
\path[clip] (  0.00,  0.00) rectangle (578.16,260.17);
\definecolor{drawColor}{RGB}{0,0,0}

\path[draw=drawColor,line width= 0.4pt,line join=round,line cap=round] (114.27, 56.04) -- (114.27, 50.35);

\path[draw=drawColor,line width= 0.4pt,line join=round,line cap=round] (154.48, 56.04) -- (154.48, 50.35);

\path[draw=drawColor,line width= 0.4pt,line join=round,line cap=round] (194.69, 56.04) -- (194.69, 50.35);

\path[draw=drawColor,line width= 0.4pt,line join=round,line cap=round] (234.91, 56.04) -- (234.91, 50.35);

\path[draw=drawColor,line width= 0.4pt,line join=round,line cap=round] (275.12, 56.04) -- (275.12, 50.35);

\path[draw=drawColor,line width= 0.4pt,line join=round,line cap=round] (315.33, 56.04) -- (315.33, 50.35);

\node[text=drawColor,anchor=base,inner sep=0pt, outer sep=0pt, scale=  0.96] at (114.27, 38.05) {90};

\node[text=drawColor,anchor=base,inner sep=0pt, outer sep=0pt, scale=  0.96] at (154.48, 38.05) {92};

\node[text=drawColor,anchor=base,inner sep=0pt, outer sep=0pt, scale=  0.96] at (194.69, 38.05) {94};

\node[text=drawColor,anchor=base,inner sep=0pt, outer sep=0pt, scale=  0.96] at (234.91, 38.05) {96};

\node[text=drawColor,anchor=base,inner sep=0pt, outer sep=0pt, scale=  0.96] at (275.12, 38.05) {98};

\node[text=drawColor,anchor=base,inner sep=0pt, outer sep=0pt, scale=  0.96] at (315.33, 38.05) {100};
\end{scope}
\begin{scope}
\path[clip] (106.23, 56.04) rectangle (323.37,199.42);
\definecolor{drawColor}{gray}{0.85}

\path[draw=drawColor,line width= 0.4pt,line join=round,line cap=round] (106.23, 76.52) -- (323.37, 76.52);

\path[draw=drawColor,line width= 0.4pt,line join=round,line cap=round] (106.23,110.66) -- (323.37,110.66);

\path[draw=drawColor,line width= 0.4pt,line join=round,line cap=round] (106.23,144.80) -- (323.37,144.80);

\path[draw=drawColor,line width= 0.4pt,line join=round,line cap=round] (106.23,178.94) -- (323.37,178.94);
\definecolor{drawColor}{RGB}{0,0,0}

\path[draw=drawColor,line width= 0.4pt,line join=round,line cap=round] (315.33, 69.69) circle (  2.71);

\path[draw=drawColor,line width= 0.4pt,line join=round,line cap=round] (194.69,103.83) circle (  2.71);

\path[draw=drawColor,line width= 0.4pt,line join=round,line cap=round] (154.48,137.97) circle (  2.71);

\path[draw=drawColor,line width= 0.4pt,line join=round,line cap=round] (315.33,172.11) circle (  2.71);
\definecolor{fillColor}{RGB}{0,0,0}

\path[fill=fillColor] (315.33, 80.74) --
	(318.98, 74.41) --
	(311.68, 74.41) --
	cycle;

\path[fill=fillColor] (275.12,114.88) --
	(278.77,108.55) --
	(271.47,108.55) --
	cycle;

\path[fill=fillColor] (315.33,149.01) --
	(318.98,142.69) --
	(311.68,142.69) --
	cycle;

\path[fill=fillColor] (315.33,183.15) --
	(318.98,176.83) --
	(311.68,176.83) --
	cycle;

\path[draw=drawColor,line width= 0.4pt,line join=round,line cap=round] (312.62, 80.64) -- (318.04, 86.06);

\path[draw=drawColor,line width= 0.4pt,line join=round,line cap=round] (312.62, 86.06) -- (318.04, 80.64);

\path[draw=drawColor,line width= 0.4pt,line join=round,line cap=round] (312.62,114.78) -- (318.04,120.20);

\path[draw=drawColor,line width= 0.4pt,line join=round,line cap=round] (312.62,120.20) -- (318.04,114.78);

\path[draw=drawColor,line width= 0.4pt,line join=round,line cap=round] (312.62,148.92) -- (318.04,154.34);

\path[draw=drawColor,line width= 0.4pt,line join=round,line cap=round] (312.62,154.34) -- (318.04,148.92);

\path[draw=drawColor,line width= 0.4pt,line join=round,line cap=round] (312.62,183.06) -- (318.04,188.48);

\path[draw=drawColor,line width= 0.4pt,line join=round,line cap=round] (312.62,188.48) -- (318.04,183.06);
\end{scope}
\begin{scope}
\path[clip] (  0.00,  0.00) rectangle (578.16,260.17);
\definecolor{drawColor}{RGB}{0,0,0}

\path[draw=drawColor,line width= 0.4pt,line join=round,line cap=round] (106.23, 56.04) rectangle (323.37,199.42);
\end{scope}
\begin{scope}
\path[clip] (106.23,199.42) rectangle (323.37,213.88);
\definecolor{drawColor}{gray}{0.90}
\definecolor{fillColor}{gray}{0.90}

\path[draw=drawColor,line width= 0.4pt,line join=round,line cap=round,fill=fillColor] (106.23,199.42) rectangle (323.37,213.88);
\definecolor{drawColor}{RGB}{0,0,0}

\node[text=drawColor,anchor=base west,inner sep=0pt, outer sep=0pt, scale=  1.20] at (193.44,202.52) {Mallows};
\end{scope}
\begin{scope}
\path[clip] (  0.00,  0.00) rectangle (578.16,260.17);
\definecolor{drawColor}{RGB}{0,0,0}

\path[draw=drawColor,line width= 0.4pt,line join=round,line cap=round] (106.23,199.42) rectangle (323.37,213.88);
\end{scope}
\begin{scope}
\path[clip] (  0.00,  0.00) rectangle (578.16,260.17);
\definecolor{drawColor}{RGB}{0,0,0}

\path[draw=drawColor,line width= 0.4pt,line join=round,line cap=round] (345.64,213.88) -- (345.64,219.57);

\path[draw=drawColor,line width= 0.4pt,line join=round,line cap=round] (385.85,213.88) -- (385.85,219.57);

\path[draw=drawColor,line width= 0.4pt,line join=round,line cap=round] (426.06,213.88) -- (426.06,219.57);

\path[draw=drawColor,line width= 0.4pt,line join=round,line cap=round] (466.27,213.88) -- (466.27,219.57);

\path[draw=drawColor,line width= 0.4pt,line join=round,line cap=round] (506.48,213.88) -- (506.48,219.57);

\path[draw=drawColor,line width= 0.4pt,line join=round,line cap=round] (546.69,213.88) -- (546.69,219.57);
\end{scope}
\begin{scope}
\path[clip] (  0.00,  0.00) rectangle (578.16,260.17);
\definecolor{drawColor}{RGB}{0,0,0}

\path[draw=drawColor,line width= 0.4pt,line join=round,line cap=round] (345.64, 56.04) -- (345.64, 50.35);

\path[draw=drawColor,line width= 0.4pt,line join=round,line cap=round] (385.85, 56.04) -- (385.85, 50.35);

\path[draw=drawColor,line width= 0.4pt,line join=round,line cap=round] (426.06, 56.04) -- (426.06, 50.35);

\path[draw=drawColor,line width= 0.4pt,line join=round,line cap=round] (466.27, 56.04) -- (466.27, 50.35);

\path[draw=drawColor,line width= 0.4pt,line join=round,line cap=round] (506.48, 56.04) -- (506.48, 50.35);

\path[draw=drawColor,line width= 0.4pt,line join=round,line cap=round] (546.69, 56.04) -- (546.69, 50.35);

\node[text=drawColor,anchor=base,inner sep=0pt, outer sep=0pt, scale=  0.96] at (345.64, 38.05) {90};

\node[text=drawColor,anchor=base,inner sep=0pt, outer sep=0pt, scale=  0.96] at (385.85, 38.05) {92};

\node[text=drawColor,anchor=base,inner sep=0pt, outer sep=0pt, scale=  0.96] at (426.06, 38.05) {94};

\node[text=drawColor,anchor=base,inner sep=0pt, outer sep=0pt, scale=  0.96] at (466.27, 38.05) {96};

\node[text=drawColor,anchor=base,inner sep=0pt, outer sep=0pt, scale=  0.96] at (506.48, 38.05) {98};

\node[text=drawColor,anchor=base,inner sep=0pt, outer sep=0pt, scale=  0.96] at (546.69, 38.05) {100};
\end{scope}
\begin{scope}
\path[clip] (337.60, 56.04) rectangle (554.73,199.42);
\definecolor{drawColor}{gray}{0.85}

\path[draw=drawColor,line width= 0.4pt,line join=round,line cap=round] (337.60, 76.52) -- (554.73, 76.52);

\path[draw=drawColor,line width= 0.4pt,line join=round,line cap=round] (337.60,110.66) -- (554.73,110.66);

\path[draw=drawColor,line width= 0.4pt,line join=round,line cap=round] (337.60,144.80) -- (554.73,144.80);

\path[draw=drawColor,line width= 0.4pt,line join=round,line cap=round] (337.60,178.94) -- (554.73,178.94);
\definecolor{drawColor}{RGB}{0,0,0}

\path[draw=drawColor,line width= 0.4pt,line join=round,line cap=round] (426.06, 69.69) circle (  2.71);

\path[draw=drawColor,line width= 0.4pt,line join=round,line cap=round] (345.64,103.83) circle (  2.71);

\path[draw=drawColor,line width= 0.4pt,line join=round,line cap=round] (506.48,137.97) circle (  2.71);

\path[draw=drawColor,line width= 0.4pt,line join=round,line cap=round] (385.85,172.11) circle (  2.71);
\definecolor{fillColor}{RGB}{0,0,0}

\path[fill=fillColor] (466.27, 80.74) --
	(469.92, 74.41) --
	(462.62, 74.41) --
	cycle;

\path[fill=fillColor] (506.48,114.88) --
	(510.13,108.55) --
	(502.83,108.55) --
	cycle;

\path[fill=fillColor] (506.48,149.01) --
	(510.13,142.69) --
	(502.83,142.69) --
	cycle;

\path[fill=fillColor] (546.69,183.15) --
	(550.34,176.83) --
	(543.04,176.83) --
	cycle;

\path[draw=drawColor,line width= 0.4pt,line join=round,line cap=round] (543.98, 80.64) -- (549.40, 86.06);

\path[draw=drawColor,line width= 0.4pt,line join=round,line cap=round] (543.98, 86.06) -- (549.40, 80.64);

\path[draw=drawColor,line width= 0.4pt,line join=round,line cap=round] (463.56,114.78) -- (468.98,120.20);

\path[draw=drawColor,line width= 0.4pt,line join=round,line cap=round] (463.56,120.20) -- (468.98,114.78);

\path[draw=drawColor,line width= 0.4pt,line join=round,line cap=round] (463.56,148.92) -- (468.98,154.34);

\path[draw=drawColor,line width= 0.4pt,line join=round,line cap=round] (463.56,154.34) -- (468.98,148.92);

\path[draw=drawColor,line width= 0.4pt,line join=round,line cap=round] (463.56,183.06) -- (468.98,188.48);

\path[draw=drawColor,line width= 0.4pt,line join=round,line cap=round] (463.56,188.48) -- (468.98,183.06);
\end{scope}
\begin{scope}
\path[clip] (  0.00,  0.00) rectangle (578.16,260.17);
\definecolor{drawColor}{RGB}{0,0,0}

\path[draw=drawColor,line width= 0.4pt,line join=round,line cap=round] (337.60, 56.04) rectangle (554.73,199.42);
\end{scope}
\begin{scope}
\path[clip] (337.60,199.42) rectangle (554.73,213.88);
\definecolor{drawColor}{gray}{0.90}
\definecolor{fillColor}{gray}{0.90}

\path[draw=drawColor,line width= 0.4pt,line join=round,line cap=round,fill=fillColor] (337.60,199.42) rectangle (554.73,213.88);
\definecolor{drawColor}{RGB}{0,0,0}

\node[text=drawColor,anchor=base west,inner sep=0pt, outer sep=0pt, scale=  1.20] at (383.63,202.52) {Mixture of two Mallows};
\end{scope}
\begin{scope}
\path[clip] (  0.00,  0.00) rectangle (578.16,260.17);
\definecolor{drawColor}{RGB}{0,0,0}

\path[draw=drawColor,line width= 0.4pt,line join=round,line cap=round] (337.60,199.42) rectangle (554.73,213.88);
\end{scope}
\begin{scope}
\path[clip] (  0.00,  0.00) rectangle (578.16,260.17);

\path[] (214.11,236.97) rectangle (446.85,248.13);
\end{scope}
\begin{scope}
\path[clip] (  0.00,  0.00) rectangle (578.16,260.17);
\definecolor{drawColor}{RGB}{0,0,0}

\node[text=drawColor,anchor=base west,inner sep=0pt, outer sep=0pt, scale=  1.20] at (220.11,238.42) {$\theta = 0.05$};
\end{scope}
\begin{scope}
\path[clip] (  0.00,  0.00) rectangle (578.16,260.17);
\definecolor{drawColor}{RGB}{0,0,0}

\node[text=drawColor,anchor=base west,inner sep=0pt, outer sep=0pt, scale=  1.20] at (310.80,238.42) {$\theta = 0.2$};
\end{scope}
\begin{scope}
\path[clip] (  0.00,  0.00) rectangle (578.16,260.17);
\definecolor{drawColor}{RGB}{0,0,0}

\node[text=drawColor,anchor=base west,inner sep=0pt, outer sep=0pt, scale=  1.20] at (395.49,238.42) {$\theta = 1$};
\end{scope}
\begin{scope}
\path[clip] (  0.00,  0.00) rectangle (578.16,260.17);
\definecolor{drawColor}{RGB}{0,0,0}

\path[draw=drawColor,line width= 0.4pt,line join=round,line cap=round] (278.11,242.55) circle (  2.71);
\end{scope}
\begin{scope}
\path[clip] (  0.00,  0.00) rectangle (578.16,260.17);
\definecolor{fillColor}{RGB}{0,0,0}

\path[fill=fillColor] (362.80,246.76) --
	(366.45,240.44) --
	(359.15,240.44) --
	cycle;
\end{scope}
\begin{scope}
\path[clip] (  0.00,  0.00) rectangle (578.16,260.17);
\definecolor{drawColor}{RGB}{0,0,0}

\path[draw=drawColor,line width= 0.4pt,line join=round,line cap=round] (435.44,239.84) -- (440.86,245.26);

\path[draw=drawColor,line width= 0.4pt,line join=round,line cap=round] (435.44,245.26) -- (440.86,239.84);
\end{scope}
\end{tikzpicture}
}
\caption{Simulation, prescribed patterns ($m=100$ judges, $50$ data sets per setting):
percentage of data sets in which BOLT finds the optimal solution, computed by enumerating
the sub-universe $\Znnu$.}\label{fig:simpattern}
\end{figure}

\paragraph{Prescribed patterns.} Each judge gives a bucket order obtained by cutting a
permutation, sampled from the Mallows model or from the mixture, into buckets of prescribed
sizes, as in a Q-sort; the true centre, cut in the same way, has the same pattern. In the
first part of the experiment the optimum is found by enumerating the sub-universe
$\Znnu$: $n=8$ with pattern $(1,2,2,2,1)$ ($5040$ bucket orders), $n=10$ with pattern
$(1,2,4,2,1)$ ($37{,}800$), $n=10$ with the partial pattern $(2,\free,\free,2)$ ($78{,}120$;
the data are cut by the sizes $(2,3,3,2)$) and $n=12$ with pattern $(1,3,4,3,1)$
($554{,}400$), with $50$ data sets of $m=100$ judges for each setting.
Figure~\ref{fig:simpattern} shows that BOLT found the optimum in $97.4\%$ of the $1200$
data sets, $98.7\%$ with the Mallows model and $96.2\%$ with the mixture; the share is
lowest with $\theta=0.05$ and with the mixture, i.e., when the penalty is flat. In the $31$
data sets in which the solution was not optimal, the relative difference with the optimal
average $\tauX$ was $1.1\%$ on average and $3.5\%$ at most, and BOLT took about $0.005$
seconds per problem. With the single Mallows model the true centre is a median ranking in
all the data sets with $\theta=1$, in $37.5\%$ of them with $\theta=0.2$ and in only $1\%$
with $\theta=0.05$: with a large dispersion, the median ranking is not the centre of the
model that generated the data, even when the pattern is right.

In the second part we considered $n=50$, $200$ and $1000$ objects, sorted into a grid with
seven columns whose sizes are proportional to $(1,2,3,4,3,2,1)$, and $m=500$ judges from
the Mallows model, with $10$ data sets for each setting. Here the optimum cannot be
computed, and we compared the median ranking with the pattern of the grid with the true
centre and with the MCBO with $b=7$ buckets of free size. The average $\tauX$ of the
solution with the pattern was never smaller than the one of the true centre, and the two
solutions are very close ($\tauX$ between them from $0.91$ to $1$). The price of the
pattern, i.e., the difference between the average $\tauX$ of the MCBO with $b=7$ buckets
and the one of the solution with the pattern, is small (at most $0.006$, with $n=50$ and
$\theta=0.05$), and it vanishes when the judges agree. With $1000$ objects BOLT took less
than one second.

\paragraph{Median constrained bucket orders.} In the last experiment the judges give
permutations sampled from the Mallows model or from the mixture, and the solution must
have exactly $b$ buckets of free size. For the optimum, the branch-and-bound algorithm is
too slow already with $n=10$ objects, and we enumerated $\Znb$ with $(n,b)=(8,4)$, $(9,5)$
and $(10,4)$, i.e., $40{,}824$, $834{,}120$ and $818{,}520$ bucket orders, with $50$ data sets
of $m=100$ judges for each setting. In the second part we considered $n=20$, $50$, $100$,
$200$ and $500$ objects, $b=5$ buckets and $m=100$ judges, with $10$ data sets for each
setting, and we compared each algorithm with the best solution found by any of the three.

\begin{figure}[h!]
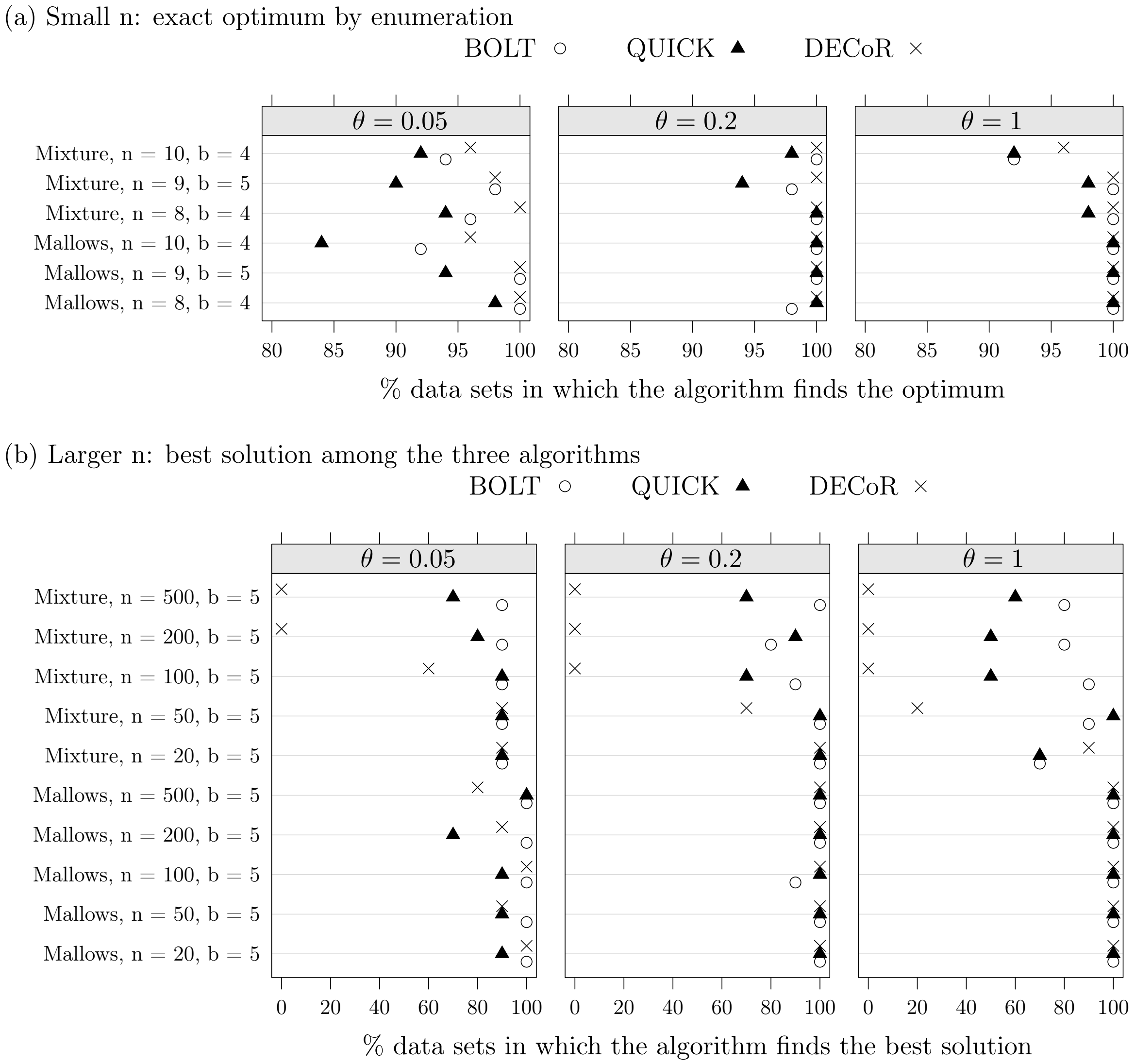

\centering
\resizebox{\textwidth}{!}{%

}
\caption{Simulation, MCBO with $b$ buckets of free size ($m=100$ judges). (a) Percentage
of data sets in which each algorithm finds the optimal solution, computed by enumerating
$\Znb$ ($50$ data sets per setting). (b) Percentage of data sets in which each algorithm
finds the best solution among the three ($10$ data sets per setting).}\label{fig:simmcbo}
\end{figure}

Figure~\ref{fig:simmcbo}(a) shows that, with few objects, the three algorithms found the
optimum in most of the data sets: BOLT in $98.2\%$ of them, QUICK in $96.2\%$ and DECoR in
$99.2\%$, with largest relative differences from the optimal average $\tauX$ of $1.7\%$,
$2.8\%$ and $1.4\%$. The hardest setting is the single Mallows model with $\theta=0.05$,
$n=10$ and $b=4$, where the shares were $92\%$, $84\%$ and $96\%$. BOLT and QUICK took about
$0.01$ seconds per problem and DECoR about $6$ seconds: with few objects DECoR is slightly
more accurate than BOLT, at a much larger cost. The picture changes when the number of
objects grows (Figure~\ref{fig:simmcbo}(b)). With the single Mallows model the three
algorithms behave similarly (BOLT found the best solution in $99.3\%$ of the data sets,
QUICK in $96\%$ and DECoR in $97.3\%$), but with the mixture the shares were $88.7\%$,
$78.7\%$ and $34.7\%$. DECoR deteriorates quickly with $n$, and with the mixture and
$n\geq200$ it never found the best solution, while BOLT found it in $83$ to $93\%$ of the
data sets for every $n$. With $500$ objects BOLT took on average $0.5$ seconds, QUICK $163$
seconds and DECoR $154$ seconds.

\subsection{Complete results}\label{app:complete}
Tables~\ref{tab:instreal}--\ref{tab:instcert} report the results problem by problem, or
cell by cell for the simulation study, from which the summaries of
Section~\ref{sec:experiments} of the paper are computed. The $64$ simulated problems of
Table~\ref{tab:instreal} have $m=20$ judges, whose permutations were drawn from the
Mallows model with the Kendall distance, with $\theta=0.05$ and $0.2$, and whose ties were
introduced by closing each gap between two consecutive objects with probability $0.3$;
there are five problems for each $n=20,30,40,50$ and three for each $n=60,70,80,90$.

{\scriptsize\setlength{\tabcolsep}{4pt}
\begin{longtable}{lrrrrrrrr}
\caption{The $83$ problems with certified optimum of Section~\ref{sec:realexp}: average $\tauX$ of the optimum and gap from the optimum of each method (PrefLib data sets by their code; randomized methods: best of $100$ runs; NSKS: NonStrictKwikSort).}\label{tab:instreal}\\
\toprule
Problem & $n$ & $m$ & $\bar\tau_X$ opt. & BOLT & BioConsert & NSKS & KwikSort & Fagin \\
\midrule
\endfirsthead
\multicolumn{9}{l}{\emph{(continued)}}\\
\toprule
Problem & $n$ & $m$ & $\bar\tau_X$ opt. & BOLT & BioConsert & NSKS & KwikSort & Fagin \\
\midrule
\endhead
\bottomrule
\endfoot
00028-00000001 &   5 & 18,723 & 0.23869 & 0 & 0 & 0 & 8.4e-02 & 0 \\
ML algorithms &  12 & 8 & 0.72159 & 0 & 0 & 0 & 3.8e-03 & 0 \\
00001-00000001 &  12 & 43,942 & 0.39751 & 0 & 0 & 0 & 1.6e-01 & 2.3e-03 \\
EMD &  15 & 112 & 0.16556 & 0 & 0 & 0 & 2.4e-03 & 1.2e-01 \\
00033-00000002 &  17 & 65 & 0.60928 & 0 & 0 & 0 & 4.5e-01 & 2.3e-02 \\
00006-00000019 &  28 & 9 & 0.86949 & 0 & 0 & 0 & 0 & 1.2e-02 \\
00006-00000013 &  29 & 9 & 0.85769 & 0 & 0 & 5.5e-04 & 5.5e-04 & 1.3e-02 \\
00006-00000001 &  30 & 9 & 0.88455 & 0 & 0 & 0 & 2.6e-04 & 7.4e-03 \\
00006-00000039 &  30 & 9 & 0.86820 & 0 & 0 & 0 & 0 & 1.7e-02 \\
00006-00000041 &  30 & 9 & 0.84751 & 0 & 0 & 0 & 0 & 1.4e-02 \\
00038-00000003 & 102 & 32 & 0.90390 & 0 & 0 & 0 & 8.4e-01 & 0 \\
00038-00000005 & 103 & 31 & 0.90482 & 0 & 0 & 0 & 8.5e-01 & 0 \\
00038-00000006 & 133 & 38 & 0.92595 & 0 & 0 & 0 & 8.8e-01 & 0 \\
00038-00000008 & 147 & 51 & 0.92028 & 0 & 0 & 0 & 8.7e-01 & 0 \\
00038-00000007 & 155 & 51 & 0.93632 & 0 & 0 & 0 & 8.9e-01 & 0 \\
Universities & 229 & 10 & 0.72026 & 7.7e-06 & 6.9e-05 & 4.2e-03 & 3.6e-03 & 5.9e-03 \\
Universities, research & 229 & 4.5 & 0.73866 & 6.8e-05 & 1.3e-04 & 7.7e-03 & 8.6e-03 & 7.4e-03 \\
Universities, teaching & 229 & 3 & 0.65611 & 0 & 0 & 3.9e-03 & 6.5e-02 & 1.3e-02 \\
Cities & 240 & 4 & 0.74793 & 7.0e-05 & 2.1e-04 & 3.5e-03 & 4.3e-03 & 1.1e-02 \\
Simulated, $\theta=0.05$, rep.~1 &  20 & 20 & 0.19921 & 0 & 0 & 1.6e-03 & 4.7e-03 & 1.0e-01 \\
Simulated, $\theta=0.05$, rep.~2 &  20 & 20 & 0.17816 & 0 & 0 & 3.2e-03 & 5.3e-03 & 1.4e-01 \\
Simulated, $\theta=0.05$, rep.~3 &  20 & 20 & 0.19947 & 0 & 0 & 5.8e-03 & 3.7e-03 & 1.0e-01 \\
Simulated, $\theta=0.05$, rep.~4 &  20 & 20 & 0.19632 & 0 & 0 & 2.6e-03 & 2.1e-03 & 1.5e-01 \\
Simulated, $\theta=0.05$, rep.~5 &  20 & 20 & 0.21632 & 0 & 0 & 5.3e-03 & 3.7e-03 & 1.3e-01 \\
Simulated, $\theta=0.20$, rep.~1 &  20 & 20 & 0.45895 & 0 & 0 & 0 & 5.3e-04 & 3.6e-02 \\
Simulated, $\theta=0.20$, rep.~2 &  20 & 20 & 0.46474 & 0 & 0 & 5.3e-04 & 1.1e-03 & 1.7e-02 \\
Simulated, $\theta=0.20$, rep.~3 &  20 & 20 & 0.40158 & 0 & 0 & 3.2e-03 & 2.1e-03 & 2.5e-02 \\
Simulated, $\theta=0.20$, rep.~4 &  20 & 20 & 0.46684 & 0 & 0 & 0 & 5.3e-04 & 2.5e-02 \\
Simulated, $\theta=0.20$, rep.~5 &  20 & 20 & 0.49079 & 0 & 0 & 0 & 1.1e-03 & 3.3e-02 \\
Simulated, $\theta=0.05$, rep.~1 &  30 & 20 & 0.20115 & 0 & 0 & 8.0e-03 & 6.7e-03 & 1.6e-01 \\
Simulated, $\theta=0.05$, rep.~2 &  30 & 20 & 0.26483 & 0 & 0 & 6.2e-03 & 1.4e-03 & 1.0e-01 \\
Simulated, $\theta=0.05$, rep.~3 &  30 & 20 & 0.27414 & 0 & 0 & 3.4e-03 & 5.1e-03 & 3.8e-02 \\
Simulated, $\theta=0.05$, rep.~4 &  30 & 20 & 0.25851 & 0 & 0 & 3.2e-03 & 3.2e-03 & 1.4e-01 \\
Simulated, $\theta=0.05$, rep.~5 &  30 & 20 & 0.26402 & 0 & 0 & 2.1e-03 & 1.8e-03 & 1.7e-01 \\
Simulated, $\theta=0.20$, rep.~1 &  30 & 20 & 0.55471 & 0 & 0 & 1.4e-03 & 9.2e-04 & 2.5e-02 \\
Simulated, $\theta=0.20$, rep.~2 &  30 & 20 & 0.53529 & 0 & 0 & 6.9e-04 & 6.9e-04 & 3.1e-02 \\
Simulated, $\theta=0.20$, rep.~3 &  30 & 20 & 0.54862 & 0 & 0 & 4.6e-04 & 1.6e-03 & 1.9e-02 \\
Simulated, $\theta=0.20$, rep.~4 &  30 & 20 & 0.58448 & 0 & 0 & 6.9e-04 & 0 & 1.6e-02 \\
Simulated, $\theta=0.20$, rep.~5 &  30 & 20 & 0.55161 & 0 & 0 & 9.2e-04 & 1.6e-03 & 2.6e-02 \\
Simulated, $\theta=0.05$, rep.~1 &  40 & 20 & 0.25532 & 0 & 1.3e-04 & 3.6e-03 & 4.4e-03 & 1.4e-01 \\
Simulated, $\theta=0.05$, rep.~2 &  40 & 20 & 0.26526 & 0 & 0 & 6.7e-03 & 4.9e-03 & 7.6e-02 \\
Simulated, $\theta=0.05$, rep.~3 &  40 & 20 & 0.27949 & 0 & 0 & 3.7e-03 & 3.2e-03 & 9.2e-02 \\
Simulated, $\theta=0.05$, rep.~4 &  40 & 20 & 0.31276 & 0 & 0 & 5.8e-03 & 3.6e-03 & 4.8e-02 \\
Simulated, $\theta=0.05$, rep.~5 &  40 & 20 & 0.26577 & 0 & 1.3e-04 & 5.0e-03 & 4.0e-03 & 8.4e-02 \\
Simulated, $\theta=0.20$, rep.~1 &  40 & 20 & 0.61904 & 0 & 0 & 6.4e-04 & 5.1e-04 & 1.1e-02 \\
Simulated, $\theta=0.20$, rep.~2 &  40 & 20 & 0.64712 & 0 & 0 & 9.0e-04 & 6.4e-04 & 1.3e-02 \\
Simulated, $\theta=0.20$, rep.~3 &  40 & 20 & 0.65692 & 0 & 0 & 2.6e-04 & 7.7e-04 & 9.1e-03 \\
Simulated, $\theta=0.20$, rep.~4 &  40 & 20 & 0.62295 & 0 & 0 & 5.1e-04 & 6.4e-04 & 1.7e-02 \\
Simulated, $\theta=0.20$, rep.~5 &  40 & 20 & 0.64295 & 0 & 0 & 3.8e-04 & 3.8e-04 & 1.1e-02 \\
Simulated, $\theta=0.05$, rep.~1 &  50 & 20 & 0.31792 & 0 & 1.6e-04 & 5.6e-03 & 2.9e-03 & 3.8e-02 \\
Simulated, $\theta=0.05$, rep.~2 &  50 & 20 & 0.30163 & 0 & 0 & 6.4e-03 & 5.6e-03 & 3.5e-02 \\
Simulated, $\theta=0.05$, rep.~3 &  50 & 20 & 0.31314 & 0 & 0 & 8.2e-03 & 4.8e-03 & 4.5e-02 \\
Simulated, $\theta=0.05$, rep.~4 &  50 & 20 & 0.29571 & 8.2e-05 & 8.2e-05 & 7.6e-03 & 5.6e-03 & 8.8e-02 \\
Simulated, $\theta=0.05$, rep.~5 &  50 & 20 & 0.30722 & 0 & 0 & 6.9e-03 & 4.2e-03 & 4.3e-02 \\
Simulated, $\theta=0.20$, rep.~1 &  50 & 20 & 0.70237 & 0 & 0 & 9.0e-04 & 2.4e-04 & 9.1e-03 \\
Simulated, $\theta=0.20$, rep.~2 &  50 & 20 & 0.69073 & 0 & 0 & 1.2e-03 & 7.3e-04 & 1.1e-02 \\
Simulated, $\theta=0.20$, rep.~3 &  50 & 20 & 0.69233 & 0 & 0 & 9.8e-04 & 5.7e-04 & 1.2e-02 \\
Simulated, $\theta=0.20$, rep.~4 &  50 & 20 & 0.72249 & 0 & 0 & 5.7e-04 & 4.9e-04 & 1.1e-02 \\
Simulated, $\theta=0.20$, rep.~5 &  50 & 20 & 0.69673 & 0 & 0 & 2.4e-04 & 4.1e-04 & 1.0e-02 \\
Simulated, $\theta=0.05$, rep.~1 &  60 & 20 & 0.35181 & 0 & 5.6e-05 & 7.5e-03 & 5.4e-03 & 3.4e-02 \\
Simulated, $\theta=0.05$, rep.~2 &  60 & 20 & 0.36014 & 0 & 0 & 6.4e-03 & 4.6e-03 & 3.9e-02 \\
Simulated, $\theta=0.05$, rep.~3 &  60 & 20 & 0.35056 & 5.6e-05 & 1.7e-04 & 6.7e-03 & 4.7e-03 & 4.0e-02 \\
Simulated, $\theta=0.20$, rep.~1 &  60 & 20 & 0.74712 & 0 & 0 & 9.0e-04 & 8.5e-04 & 8.6e-03 \\
Simulated, $\theta=0.20$, rep.~2 &  60 & 20 & 0.74347 & 0 & 0 & 1.1e-03 & 7.3e-04 & 6.2e-03 \\
Simulated, $\theta=0.20$, rep.~3 &  60 & 20 & 0.75749 & 0 & 0 & 7.9e-04 & 4.0e-04 & 8.2e-03 \\
Simulated, $\theta=0.05$, rep.~1 &  70 & 20 & 0.40118 & 4.1e-05 & 8.3e-05 & 4.3e-03 & 4.8e-03 & 3.0e-02 \\
Simulated, $\theta=0.05$, rep.~2 &  70 & 20 & 0.38665 & 8.3e-05 & 0 & 5.3e-03 & 4.2e-03 & 3.0e-02 \\
Simulated, $\theta=0.05$, rep.~3 &  70 & 20 & 0.36072 & 0 & 8.3e-05 & 7.0e-03 & 6.4e-03 & 3.6e-02 \\
Simulated, $\theta=0.20$, rep.~1 &  70 & 20 & 0.77739 & 0 & 0 & 4.6e-04 & 4.6e-04 & 4.9e-03 \\
Simulated, $\theta=0.20$, rep.~2 &  70 & 20 & 0.77660 & 0 & 0 & 5.0e-04 & 7.0e-04 & 5.3e-03 \\
Simulated, $\theta=0.20$, rep.~3 &  70 & 20 & 0.76867 & 0 & 0 & 7.0e-04 & 1.7e-04 & 4.9e-03 \\
Simulated, $\theta=0.05$, rep.~1 &  80 & 20 & 0.44502 & 6.3e-05 & 3.2e-05 & 5.2e-03 & 3.8e-03 & 1.8e-02 \\
Simulated, $\theta=0.05$, rep.~2 &  80 & 20 & 0.43285 & 3.2e-05 & 9.5e-05 & 4.8e-03 & 3.7e-03 & 3.1e-02 \\
Simulated, $\theta=0.05$, rep.~3 &  80 & 20 & 0.41983 & 0 & 1.3e-04 & 4.8e-03 & 4.9e-03 & 3.2e-02 \\
Simulated, $\theta=0.20$, rep.~1 &  80 & 20 & 0.79437 & 0 & 0 & 1.2e-03 & 5.4e-04 & 4.1e-03 \\
Simulated, $\theta=0.20$, rep.~2 &  80 & 20 & 0.79965 & 0 & 0 & 5.1e-04 & 4.7e-04 & 5.3e-03 \\
Simulated, $\theta=0.20$, rep.~3 &  80 & 20 & 0.79389 & 0 & 0 & 4.7e-04 & 6.0e-04 & 3.7e-03 \\
Simulated, $\theta=0.05$, rep.~1 &  90 & 20 & 0.47099 & 2.5e-05 & 1.2e-04 & 4.8e-03 & 2.9e-03 & 2.0e-02 \\
Simulated, $\theta=0.05$, rep.~2 &  90 & 20 & 0.45019 & 0 & 1.2e-04 & 4.8e-03 & 5.3e-03 & 2.1e-02 \\
Simulated, $\theta=0.05$, rep.~3 &  90 & 20 & 0.45838 & 5.0e-05 & 2.5e-05 & 5.2e-03 & 4.2e-03 & 2.2e-02 \\
Simulated, $\theta=0.20$, rep.~1 &  90 & 20 & 0.81094 & 0 & 0 & 8.5e-04 & 5.7e-04 & 4.7e-03 \\
Simulated, $\theta=0.20$, rep.~2 &  90 & 20 & 0.82037 & 0 & 0 & 4.5e-04 & 3.2e-04 & 3.5e-03 \\
Simulated, $\theta=0.20$, rep.~3 &  90 & 20 & 0.82235 & 0 & 0 & 7.0e-04 & 4.7e-04 & 3.9e-03 \\
\end{longtable}}
{\scriptsize\setlength{\tabcolsep}{4pt}
\begin{longtable}{lrrrrrr}
\caption{The $85$ PrefLib data sets of \citet{akbariescobedo2023a}: lower bound $P_{LB}$ (as a sum of distances), gap from the lower bound of BOLT and of the best published ACP solution (in \% of the bound, published with two decimals), and elapsed time of BOLT.}\label{tab:instacp}\\
\toprule
Data set & $n$ & $m$ & Lower bound & BOLT (\%) & ACP (\%) & Time (s) \\
\midrule
\endfirsthead
\multicolumn{7}{l}{\emph{(continued)}}\\
\toprule
Data set & $n$ & $m$ & Lower bound & BOLT (\%) & ACP (\%) & Time (s) \\
\midrule
\endhead
\bottomrule
\endfoot
ED-14-02 &  100 &  5000 & 4,725,000 & 0.00 & 0.00 & 8.58 \\
ED-14-03 &  100 &  5000 & 4,651,850 & 0.00 & 0.00 & 8.62 \\
MD-03-03 &  102 &    32 & 15,840 & 0.00 & 0.00 & 0.14 \\
MD-03-05 &  103 &    31 & 15,500 & 0.00 & 0.00 & 0.20 \\
MD-03-06 &  133 &    38 & 24,700 & 0.00 & 0.00 & 0.26 \\
MD-03-08 &  147 &    51 & 43,629 & 0.00 & 0.00 & 0.31 \\
MD-03-07 &  155 &    51 & 38,760 & 0.00 & 0.00 & 0.31 \\
ED-10-50 &  170 &     4 & 24,378 & 0.79 & 1.44 & 0.19 \\
ED-10-49 &  351 &     4 & 100,411 & 0.47 & 0.65 & 0.72 \\
ED-18-01 &  379 & 36655 & 30,454,824 & 0.00 & 0.00 & 13.07 \\
ED-18-03 &  477 & 32086 & 23,258,021 & 0.00 & 0.00 & 18.47 \\
ED-11-12 & 1210 &     4 & 1,590,332 & 1.10 & 2.77 & 19.26 \\
ED-11-31 & 1223 &     4 & 1,517,556 & 1.21 & 1.69 & 16.91 \\
ED-11-09 & 1272 &     4 & 1,561,690 & 1.44 & 3.39 & 18.08 \\
ED-11-23 & 1342 &     4 & 1,843,920 & 1.27 & 1.62 & 20.11 \\
ED-11-21 & 1347 &     4 & 1,834,227 & 1.27 & 2.66 & 27.31 \\
ED-11-37 & 1351 &     4 & 1,786,116 & 1.27 & 3.04 & 29.45 \\
ED-11-25 & 1356 &     4 & 1,850,744 & 1.31 & 2.27 & 20.33 \\
ED-11-13 & 1363 &     4 & 1,964,622 & 1.02 & 1.60 & 24.02 \\
ED-11-29 & 1368 &     4 & 1,750,605 & 0.85 & 3.31 & 25.50 \\
ED-11-14 & 1375 &     4 & 1,863,664 & 1.09 & 3.20 & 29.00 \\
ED-11-30 & 1386 &     4 & 1,828,275 & 0.86 & 2.38 & 21.42 \\
ED-11-06 & 1449 &     4 & 2,095,369 & 1.07 & 2.71 & 32.80 \\
ED-11-04 & 1467 &     4 & 2,226,074 & 1.18 & 2.59 & 35.19 \\
ED-11-07 & 1474 &     4 & 2,305,641 & 1.25 & 1.81 & 33.83 \\
ED-11-34 & 1509 &     4 & 2,372,642 & 0.93 & 1.50 & 29.83 \\
ED-11-22 & 1514 &     4 & 2,244,794 & 0.84 & 1.83 & 29.78 \\
ED-11-11 & 1545 &     4 & 2,272,137 & 1.05 & 2.01 & 39.01 \\
ED-11-15 & 1563 &     4 & 2,420,476 & 0.88 & 2.01 & 33.98 \\
ED-11-08 & 1572 &     4 & 2,704,150 & 1.07 & 1.26 & 32.88 \\
ED-11-28 & 1616 &     4 & 2,716,107 & 0.60 & 0.78 & 25.64 \\
ED-11-40 & 1623 &     4 & 2,482,556 & 0.26 & 0.32 & 31.72 \\
ED-11-36 & 1634 &     4 & 2,732,229 & 0.92 & 1.51 & 32.45 \\
ED-11-33 & 1646 &     4 & 2,752,549 & 0.95 & 1.85 & 45.97 \\
ED-11-05 & 1673 &     4 & 2,590,774 & 0.49 & 0.63 & 38.49 \\
ED-11-18 & 1681 &     4 & 2,932,501 & 1.04 & 1.94 & 39.17 \\
ED-11-16 & 1708 &     4 & 2,911,360 & 0.72 & 1.17 & 44.44 \\
ED-11-32 & 1751 &     4 & 3,267,824 & 0.91 & 1.68 & 45.44 \\
ED-11-38 & 1754 &     4 & 3,308,787 & 0.94 & 1.97 & 44.91 \\
ED-11-39 & 1788 &     4 & 2,940,244 & 0.29 & 0.35 & 34.58 \\
ED-11-68 & 1826 &     4 & 3,495,142 & 0.52 & 0.62 & 30.36 \\
ED-11-49 & 1845 &     4 & 3,911,132 & 0.69 & 0.90 & 67.14 \\
ED-11-20 & 1870 &     4 & 3,449,797 & 0.71 & 1.47 & 48.05 \\
ED-11-26 & 1931 &     4 & 3,710,296 & 0.75 & 1.33 & 47.18 \\
ED-11-35 & 1936 &     4 & 3,805,316 & 0.80 & 1.53 & 46.17 \\
ED-11-74 & 1976 &     4 & 4,176,464 & 0.59 & 1.13 & 45.29 \\
ED-11-60 & 1977 &     4 & 4,147,277 & 0.71 & 0.87 & 45.18 \\
ED-11-58 & 2011 &     4 & 4,366,852 & 0.75 & 1.09 & 59.47 \\
ED-11-62 & 2014 &     4 & 4,230,329 & 0.52 & 0.86 & 47.67 \\
ED-11-17 & 2015 &     4 & 4,056,848 & 0.75 & 1.26 & 64.61 \\
ED-11-66 & 2024 &     4 & 4,283,401 & 0.81 & 1.98 & 59.91 \\
ED-11-24 & 2049 &     4 & 4,290,818 & 0.71 & 1.41 & 69.87 \\
ED-11-67 & 2066 &     4 & 4,473,066 & 0.68 & 1.30 & 75.83 \\
ED-11-27 & 2092 &     4 & 4,647,453 & 0.70 & 1.11 & 51.46 \\
ED-11-10 & 2096 &     4 & 4,355,605 & 0.40 & 0.62 & 49.42 \\
ED-11-19 & 2104 &     4 & 4,369,932 & 0.66 & 1.24 & 53.30 \\
ED-11-50 & 2111 &     4 & 4,871,442 & 0.54 & 0.61 & 62.65 \\
ED-11-51 & 2112 &     4 & 4,755,457 & 0.74 & 1.51 & 66.12 \\
ED-11-65 & 2119 &     4 & 4,741,402 & 0.24 & 0.27 & 46.81 \\
ED-11-41 & 2123 &     4 & 5,017,714 & 0.91 & 1.04 & 93.83 \\
ED-11-71 & 2127 &     4 & 4,901,052 & 0.37 & 0.51 & 61.90 \\
ED-11-46 & 2133 &     4 & 4,835,028 & 0.94 & 1.36 & 76.72 \\
ED-11-43 & 2153 &     4 & 5,008,377 & 0.80 & 1.27 & 77.39 \\
ED-11-48 & 2194 &     4 & 5,060,729 & 0.68 & 1.12 & 60.53 \\
ED-11-52 & 2242 &     4 & 5,282,143 & 0.35 & 0.62 & 55.25 \\
ED-11-73 & 2258 &     4 & 5,409,851 & 0.20 & 0.29 & 59.15 \\
ED-11-45 & 2265 &     4 & 5,705,419 & 0.07 & 0.07 & 65.15 \\
ED-11-70 & 2276 &     4 & 5,624,119 & 0.28 & 0.49 & 65.88 \\
ED-11-59 & 2281 &     4 & 5,590,364 & 0.41 & 0.68 & 74.54 \\
ED-11-77 & 2317 &     4 & 4,946,315 & 0.07 & 0.08 & 78.53 \\
ED-11-53 & 2321 &     4 & 5,601,106 & 0.15 & 0.28 & 58.63 \\
ED-11-69 & 2338 &     4 & 5,955,107 & 0.37 & 0.52 & 79.25 \\
ED-11-55 & 2353 &     4 & 6,116,004 & 0.56 & 0.59 & 76.41 \\
ED-11-75 & 2391 &     4 & 6,379,179 & 0.61 & 0.76 & 93.75 \\
ED-11-44 & 2434 &     4 & 6,435,079 & 0.49 & 0.55 & 74.02 \\
ED-11-64 & 2446 &     4 & 6,416,412 & 0.45 & 0.59 & 141.55 \\
ED-11-72 & 2447 &     4 & 6,408,613 & 0.58 & 0.82 & 93.95 \\
ED-11-63 & 2510 &     4 & 6,550,637 & 0.45 & 0.65 & 81.47 \\
ED-11-54 & 2512 &     4 & 6,485,654 & 0.41 & 0.66 & 87.93 \\
ED-11-57 & 2559 &     4 & 6,822,253 & 0.44 & 0.76 & 125.44 \\
ED-11-76 & 2581 &     4 & 6,197,768 & 0.09 & 0.10 & 79.36 \\
ED-11-42 & 2598 &     4 & 6,411,396 & 0.13 & 0.15 & 86.55 \\
ED-11-56 & 2632 &     4 & 7,193,229 & 0.33 & 0.55 & 85.25 \\
ED-11-61 & 2726 &     4 & 7,751,722 & 0.69 & 0.96 & 116.95 \\
ED-11-47 & 2819 &     4 & 8,049,813 & 0.18 & 0.23 & 93.53 \\
\end{longtable}}
{\scriptsize\setlength{\tabcolsep}{4pt}
\begin{longtable}{rrrrlrrrrr}
\caption{Simulated problems with ties (Section~\ref{sec:simexp}), ten problems per cell: number of problems with certified optimum, number of them in which BOLT returned the optimum, mean and maximum gap from the optimum on the certified problems, and mean elapsed time of BOLT on all the problems of the cell (model: Mallows model or mixture of two Mallows models).}\label{tab:instsim}\\
\toprule
$n$ & $m$ & $\theta$ & $p_{tie}$ & Model & Certified & BOLT opt. & Mean gap & Max gap & Time (s) \\
\midrule
\endfirsthead
\multicolumn{10}{l}{\emph{(continued)}}\\
\toprule
$n$ & $m$ & $\theta$ & $p_{tie}$ & Model & Certified & BOLT opt. & Mean gap & Max gap & Time (s) \\
\midrule
\endhead
\bottomrule
\endfoot
 50 & 10 & 0.02 & 0.1 & mixture & 10 &  3 & 2.6e-04 & 4.9e-04 & 0.04 \\
 50 & 10 & 0.02 & 0.1 & single & 10 &  2 & 2.9e-04 & 8.2e-04 & 0.04 \\
 50 & 10 & 0.02 & 0.3 & mixture & 10 &  3 & 2.1e-04 & 4.9e-04 & 0.05 \\
 50 & 10 & 0.02 & 0.3 & single & 10 &  7 & 6.5e-05 & 3.3e-04 & 0.05 \\
 50 & 10 & 0.02 & 0.6 & mixture & 10 &  4 & 1.1e-04 & 3.3e-04 & 0.05 \\
 50 & 10 & 0.02 & 0.6 & single & 10 &  8 & 4.9e-05 & 3.3e-04 & 0.06 \\
 50 & 10 & 0.05 & 0.1 & mixture & 10 &  6 & 1.5e-04 & 9.8e-04 & 0.04 \\
 50 & 10 & 0.05 & 0.1 & single & 10 &  6 & 1.3e-04 & 4.9e-04 & 0.04 \\
 50 & 10 & 0.05 & 0.3 & mixture & 10 &  6 & 6.5e-05 & 1.6e-04 & 0.05 \\
 50 & 10 & 0.05 & 0.3 & single & 10 &  6 & 6.5e-05 & 1.6e-04 & 0.05 \\
 50 & 10 & 0.05 & 0.6 & mixture & 10 &  7 & 8.2e-05 & 4.9e-04 & 0.04 \\
 50 & 10 & 0.05 & 0.6 & single & 10 &  7 & 6.5e-05 & 3.3e-04 & 0.05 \\
 50 & 10 & 0.20 & 0.1 & mixture & 10 &  9 & 3.3e-05 & 3.3e-04 & 0.04 \\
 50 & 10 & 0.20 & 0.1 & single & 10 & 10 & 0 & 0 & 0.05 \\
 50 & 10 & 0.20 & 0.3 & mixture & 10 & 10 & 0 & 0 & 0.05 \\
 50 & 10 & 0.20 & 0.3 & single & 10 & 10 & 0 & 0 & 0.05 \\
 50 & 10 & 0.20 & 0.6 & mixture & 10 & 10 & 0 & 0 & 0.05 \\
 50 & 10 & 0.20 & 0.6 & single & 10 & 10 & 0 & 0 & 0.04 \\
 50 & 50 & 0.02 & 0.1 & mixture & 10 &  8 & 6.5e-06 & 3.3e-05 & 0.11 \\
 50 & 50 & 0.02 & 0.1 & single & 10 &  8 & 1.3e-05 & 9.8e-05 & 0.11 \\
 50 & 50 & 0.02 & 0.3 & mixture & 10 &  6 & 1.3e-05 & 3.3e-05 & 0.11 \\
 50 & 50 & 0.02 & 0.3 & single & 10 &  7 & 9.8e-06 & 3.3e-05 & 0.12 \\
 50 & 50 & 0.02 & 0.6 & mixture & 10 &  9 & 3.3e-06 & 3.3e-05 & 0.15 \\
 50 & 50 & 0.02 & 0.6 & single & 10 & 10 & 0 & 0 & 0.13 \\
 50 & 50 & 0.05 & 0.1 & mixture & 10 & 10 & 0 & 0 & 0.11 \\
 50 & 50 & 0.05 & 0.1 & single & 10 & 10 & 0 & 0 & 0.11 \\
 50 & 50 & 0.05 & 0.3 & mixture & 10 &  9 & 3.3e-06 & 3.3e-05 & 0.12 \\
 50 & 50 & 0.05 & 0.3 & single & 10 & 10 & 0 & 0 & 0.11 \\
 50 & 50 & 0.05 & 0.6 & mixture & 10 &  9 & 3.3e-06 & 3.3e-05 & 0.13 \\
 50 & 50 & 0.05 & 0.6 & single & 10 &  8 & 6.5e-06 & 3.3e-05 & 0.14 \\
 50 & 50 & 0.20 & 0.1 & mixture & 10 & 10 & 0 & 0 & 0.10 \\
 50 & 50 & 0.20 & 0.1 & single & 10 & 10 & 0 & 0 & 0.12 \\
 50 & 50 & 0.20 & 0.3 & mixture & 10 & 10 & 0 & 0 & 0.12 \\
 50 & 50 & 0.20 & 0.3 & single & 10 & 10 & 0 & 0 & 0.11 \\
 50 & 50 & 0.20 & 0.6 & mixture & 10 & 10 & 0 & 0 & 0.12 \\
 50 & 50 & 0.20 & 0.6 & single & 10 & 10 & 0 & 0 & 0.11 \\
100 & 10 & 0.02 & 0.1 & mixture &  8 &  0 & 4.5e-04 & 5.3e-04 & 0.13 \\
100 & 10 & 0.02 & 0.1 & single & 10 &  0 & 3.5e-04 & 4.8e-04 & 0.11 \\
100 & 10 & 0.02 & 0.3 & mixture &  7 &  0 & 3.7e-04 & 6.5e-04 & 0.12 \\
100 & 10 & 0.02 & 0.3 & single &  9 &  0 & 2.9e-04 & 4.8e-04 & 0.16 \\
100 & 10 & 0.02 & 0.6 & mixture &  4 &  0 & 1.8e-04 & 2.0e-04 & 0.15 \\
100 & 10 & 0.02 & 0.6 & single &  9 &  0 & 1.9e-04 & 4.0e-04 & 0.16 \\
100 & 10 & 0.05 & 0.1 & mixture & 10 &  0 & 2.8e-04 & 6.1e-04 & 0.11 \\
100 & 10 & 0.05 & 0.1 & single & 10 &  1 & 1.5e-04 & 3.2e-04 & 0.10 \\
100 & 10 & 0.05 & 0.3 & mixture & 10 &  0 & 1.6e-04 & 3.2e-04 & 0.12 \\
100 & 10 & 0.05 & 0.3 & single & 10 &  1 & 1.1e-04 & 2.4e-04 & 0.11 \\
100 & 10 & 0.05 & 0.6 & mixture & 10 &  0 & 1.1e-04 & 2.4e-04 & 0.12 \\
100 & 10 & 0.05 & 0.6 & single & 10 &  3 & 6.5e-05 & 2.0e-04 & 0.13 \\
100 & 10 & 0.20 & 0.1 & mixture & 10 &  4 & 5.7e-05 & 2.0e-04 & 0.11 \\
100 & 10 & 0.20 & 0.1 & single & 10 &  9 & 4.0e-06 & 4.0e-05 & 0.10 \\
100 & 10 & 0.20 & 0.3 & mixture & 10 &  6 & 2.0e-05 & 8.1e-05 & 0.14 \\
100 & 10 & 0.20 & 0.3 & single & 10 & 10 & 0 & 0 & 0.09 \\
100 & 10 & 0.20 & 0.6 & mixture & 10 &  2 & 4.8e-05 & 1.2e-04 & 0.13 \\
100 & 10 & 0.20 & 0.6 & single & 10 & 10 & 0 & 0 & 0.10 \\
100 & 50 & 0.02 & 0.1 & mixture & 10 &  2 & 3.6e-05 & 8.1e-05 & 0.39 \\
100 & 50 & 0.02 & 0.1 & single & 10 &  1 & 1.5e-05 & 3.2e-05 & 0.30 \\
100 & 50 & 0.02 & 0.3 & mixture & 10 &  0 & 4.1e-05 & 8.1e-05 & 0.32 \\
100 & 50 & 0.02 & 0.3 & single & 10 &  1 & 2.3e-05 & 4.0e-05 & 0.40 \\
100 & 50 & 0.02 & 0.6 & mixture &  9 &  2 & 2.1e-05 & 4.0e-05 & 0.38 \\
100 & 50 & 0.02 & 0.6 & single & 10 &  1 & 1.5e-05 & 3.2e-05 & 0.39 \\
100 & 50 & 0.05 & 0.1 & mixture & 10 &  1 & 1.6e-05 & 4.0e-05 & 0.32 \\
100 & 50 & 0.05 & 0.1 & single & 10 &  8 & 2.4e-06 & 1.6e-05 & 0.29 \\
100 & 50 & 0.05 & 0.3 & mixture & 10 &  3 & 1.3e-05 & 3.2e-05 & 0.31 \\
100 & 50 & 0.05 & 0.3 & single & 10 & 10 & 0 & 0 & 0.31 \\
100 & 50 & 0.05 & 0.6 & mixture & 10 &  6 & 4.0e-06 & 1.6e-05 & 0.44 \\
100 & 50 & 0.05 & 0.6 & single & 10 &  6 & 9.7e-06 & 4.8e-05 & 0.40 \\
100 & 50 & 0.20 & 0.1 & mixture & 10 &  8 & 1.6e-06 & 8.1e-06 & 0.31 \\
100 & 50 & 0.20 & 0.1 & single & 10 & 10 & 0 & 0 & 0.28 \\
100 & 50 & 0.20 & 0.3 & mixture & 10 &  8 & 1.6e-06 & 8.1e-06 & 0.35 \\
100 & 50 & 0.20 & 0.3 & single & 10 & 10 & 0 & 0 & 0.31 \\
100 & 50 & 0.20 & 0.6 & mixture & 10 &  7 & 2.4e-06 & 8.1e-06 & 0.42 \\
100 & 50 & 0.20 & 0.6 & single & 10 &  8 & 1.6e-06 & 8.1e-06 & 0.31 \\
200 & 10 & 0.02 & 0.1 & mixture &  0 &  0 & -- & -- & 0.32 \\
200 & 10 & 0.02 & 0.1 & single &  1 &  0 & 2.8e-04 & 2.8e-04 & 0.40 \\
200 & 10 & 0.02 & 0.3 & mixture &  0 &  0 & -- & -- & 0.38 \\
200 & 10 & 0.02 & 0.3 & single &  0 &  0 & -- & -- & 0.40 \\
200 & 10 & 0.02 & 0.6 & mixture &  0 &  0 & -- & -- & 0.40 \\
200 & 10 & 0.02 & 0.6 & single &  0 &  0 & -- & -- & 0.52 \\
200 & 10 & 0.05 & 0.1 & mixture &  1 &  0 & 2.3e-04 & 2.3e-04 & 0.32 \\
200 & 10 & 0.05 & 0.1 & single &  1 &  0 & 1.3e-04 & 1.3e-04 & 0.37 \\
200 & 10 & 0.05 & 0.3 & mixture &  1 &  0 & 2.0e-04 & 2.0e-04 & 0.33 \\
200 & 10 & 0.05 & 0.3 & single &  1 &  0 & 1.5e-04 & 1.5e-04 & 0.42 \\
200 & 10 & 0.05 & 0.6 & mixture &  0 &  0 & -- & -- & 0.37 \\
200 & 10 & 0.05 & 0.6 & single &  1 &  0 & 1.2e-04 & 1.2e-04 & 0.51 \\
200 & 10 & 0.20 & 0.1 & mixture &  1 &  0 & 4.0e-05 & 4.0e-05 & 0.30 \\
200 & 10 & 0.20 & 0.1 & single &  1 &  1 & 0 & 0 & 0.39 \\
200 & 10 & 0.20 & 0.3 & mixture &  1 &  0 & 4.0e-05 & 4.0e-05 & 0.31 \\
200 & 10 & 0.20 & 0.3 & single &  1 &  1 & 0 & 0 & 0.39 \\
200 & 10 & 0.20 & 0.6 & mixture &  1 &  0 & 3.0e-05 & 3.0e-05 & 0.39 \\
200 & 10 & 0.20 & 0.6 & single &  1 &  1 & 0 & 0 & 0.42 \\
200 & 50 & 0.02 & 0.1 & mixture &  0 &  0 & -- & -- & 0.89 \\
200 & 50 & 0.02 & 0.1 & single &  1 &  0 & 3.8e-05 & 3.8e-05 & 1.06 \\
200 & 50 & 0.02 & 0.3 & mixture &  0 &  0 & -- & -- & 0.95 \\
200 & 50 & 0.02 & 0.3 & single &  1 &  0 & 2.8e-05 & 2.8e-05 & 1.15 \\
200 & 50 & 0.02 & 0.6 & mixture &  0 &  0 & -- & -- & 1.09 \\
200 & 50 & 0.02 & 0.6 & single &  1 &  0 & 1.2e-05 & 1.2e-05 & 1.37 \\
200 & 50 & 0.05 & 0.1 & mixture &  1 &  0 & 1.0e-05 & 1.0e-05 & 0.88 \\
200 & 50 & 0.05 & 0.1 & single &  1 &  0 & 8.0e-06 & 8.0e-06 & 0.96 \\
200 & 50 & 0.05 & 0.3 & mixture &  1 &  0 & 1.4e-05 & 1.4e-05 & 0.94 \\
200 & 50 & 0.05 & 0.3 & single &  1 &  0 & 4.0e-06 & 4.0e-06 & 1.07 \\
200 & 50 & 0.05 & 0.6 & mixture &  1 &  0 & 1.6e-05 & 1.6e-05 & 1.10 \\
200 & 50 & 0.05 & 0.6 & single &  1 &  0 & 4.0e-06 & 4.0e-06 & 1.33 \\
200 & 50 & 0.20 & 0.1 & mixture &  1 &  0 & 3.0e-05 & 3.0e-05 & 0.89 \\
200 & 50 & 0.20 & 0.1 & single &  1 &  1 & 0 & 0 & 0.86 \\
200 & 50 & 0.20 & 0.3 & mixture &  1 &  1 & 0 & 0 & 0.94 \\
200 & 50 & 0.20 & 0.3 & single &  1 &  1 & 0 & 0 & 0.91 \\
200 & 50 & 0.20 & 0.6 & mixture &  1 &  0 & 1.4e-05 & 1.4e-05 & 1.04 \\
200 & 50 & 0.20 & 0.6 & single &  1 &  1 & 0 & 0 & 0.92 \\
\end{longtable}}
{\scriptsize\setlength{\tabcolsep}{4pt}
\begin{longtable}{rlrlrrrrr}
\caption{Certification of BOLT for constrained problems (Section~\ref{sec:conexp}), problem by problem: status of the exact formulation within $600$ seconds, whether BOLT returned the optimum (certified problems only), difference between the best solution found by the exact formulation and BOLT, difference between the certified bound and BOLT (both in terms of $\bar\tau_X$), and elapsed times.}\label{tab:instcert}\\
\toprule
$n$ & Constraint & Rep. & Exact & BOLT opt. & Gap & Cert. gap & BOLT (s) & Exact (s) \\
\midrule
\endfirsthead
\multicolumn{9}{l}{\emph{(continued)}}\\
\toprule
$n$ & Constraint & Rep. & Exact & BOLT opt. & Gap & Cert. gap & BOLT (s) & Exact (s) \\
\midrule
\endhead
\bottomrule
\endfoot
20 & complete pattern &  1 & optimal & yes & 0 & 0 & 0.02 & 2.9 \\
20 & complete pattern &  2 & optimal & yes & 0 & 0 & 0.00 & 0.5 \\
20 & complete pattern &  3 & optimal & yes & 0 & 0 & 0.03 & 0.8 \\
20 & complete pattern &  4 & optimal & yes & 0 & 0 & 0.00 & 0.7 \\
20 & complete pattern &  5 & optimal & yes & 0 & 0 & 0.01 & 0.5 \\
20 & complete pattern &  6 & optimal & yes & 0 & 0 & 0.02 & 1.6 \\
20 & complete pattern &  7 & optimal & yes & 0 & 0 & 0.00 & 0.2 \\
20 & complete pattern &  8 & optimal & yes & 0 & 0 & 0.00 & 1.0 \\
20 & complete pattern &  9 & optimal & yes & 0 & 0 & 0.00 & 0.3 \\
20 & complete pattern & 10 & optimal & yes & 0 & 0 & 0.00 & 0.3 \\
20 & partial pattern &  1 & optimal & no & 4.2e-04 & 4.2e-04 & 0.01 & 15.2 \\
20 & partial pattern &  2 & optimal & yes & 0 & 0 & 0.02 & 5.4 \\
20 & partial pattern &  3 & optimal & yes & 0 & 0 & 0.01 & 4.2 \\
20 & partial pattern &  4 & optimal & yes & 0 & 0 & 0.00 & 9.5 \\
20 & partial pattern &  5 & optimal & yes & 0 & 0 & 0.02 & 6.0 \\
20 & partial pattern &  6 & optimal & yes & 0 & 0 & 0.00 & 3.0 \\
20 & partial pattern &  7 & optimal & yes & 0 & 0 & 0.00 & 7.2 \\
20 & partial pattern &  8 & optimal & yes & 0 & 0 & 0.00 & 12.8 \\
20 & partial pattern &  9 & optimal & yes & 0 & 0 & 0.00 & 5.5 \\
20 & partial pattern & 10 & optimal & no & 2.1e-04 & 2.1e-04 & 0.00 & 14.6 \\
20 & $b=5$ buckets &  1 & optimal & yes & 0 & 0 & 0.00 & 9.9 \\
20 & $b=5$ buckets &  2 & optimal & yes & 0 & 0 & 0.00 & 7.7 \\
20 & $b=5$ buckets &  3 & optimal & yes & 0 & 0 & 0.00 & 7.6 \\
20 & $b=5$ buckets &  4 & optimal & yes & 0 & 0 & 0.02 & 5.1 \\
20 & $b=5$ buckets &  5 & optimal & yes & 0 & 0 & 0.00 & 14.6 \\
20 & $b=5$ buckets &  6 & optimal & yes & 0 & 0 & 0.00 & 13.0 \\
20 & $b=5$ buckets &  7 & optimal & yes & 0 & 0 & 0.01 & 12.0 \\
20 & $b=5$ buckets &  8 & optimal & yes & 0 & 0 & 0.02 & 7.5 \\
20 & $b=5$ buckets &  9 & optimal & yes & 0 & 0 & 0.02 & 6.8 \\
20 & $b=5$ buckets & 10 & optimal & yes & 0 & 0 & 0.00 & 11.0 \\
20 & top-5 list &  1 & optimal & yes & 0 & 0 & 0.00 & 0.1 \\
20 & top-5 list &  2 & optimal & yes & 0 & 0 & 0.00 & 0.2 \\
20 & top-5 list &  3 & optimal & yes & 0 & 0 & 0.00 & 0.6 \\
20 & top-5 list &  4 & optimal & yes & 0 & 0 & 0.00 & 0.5 \\
20 & top-5 list &  5 & optimal & yes & 0 & 0 & 0.01 & 0.2 \\
20 & top-5 list &  6 & optimal & yes & 0 & 0 & 0.00 & 0.1 \\
20 & top-5 list &  7 & optimal & yes & 0 & 0 & 0.02 & 0.1 \\
20 & top-5 list &  8 & optimal & yes & 0 & 0 & 0.00 & 0.1 \\
20 & top-5 list &  9 & optimal & yes & 0 & 0 & 0.00 & 0.1 \\
20 & top-5 list & 10 & optimal & yes & 0 & 0 & 0.02 & 0.1 \\
50 & complete pattern &  1 & time limit & -- & 0 & 1.5e-05 & 0.02 & 600.0 \\
50 & complete pattern &  2 & optimal & yes & 0 & 0 & 0.02 & 5.9 \\
50 & complete pattern &  3 & optimal & yes & 0 & 0 & 0.02 & 23.8 \\
50 & complete pattern &  4 & optimal & yes & 0 & 0 & 0.02 & 2.8 \\
50 & complete pattern &  5 & optimal & no & 1.8e-04 & 1.8e-04 & 0.02 & 10.6 \\
50 & complete pattern &  6 & optimal & yes & 0 & 0 & 0.02 & 15.8 \\
50 & complete pattern &  7 & optimal & yes & 0 & 0 & 0.02 & 23.9 \\
50 & complete pattern &  8 & optimal & yes & 0 & 0 & 0.02 & 3.6 \\
50 & complete pattern &  9 & optimal & yes & 0 & 0 & 0.00 & 12.7 \\
50 & complete pattern & 10 & optimal & yes & 0 & 0 & 0.01 & 6.1 \\
50 & partial pattern &  1 & optimal & yes & 0 & 0 & 0.02 & 378.3 \\
50 & partial pattern &  2 & time limit & -- & 0 & 1.8e-03 & 0.02 & 600.1 \\
50 & partial pattern &  3 & optimal & yes & 0 & 0 & 0.00 & 564.1 \\
50 & partial pattern &  4 & time limit & -- & 8.2e-05 & 1.3e-03 & 0.01 & 600.3 \\
50 & partial pattern &  5 & time limit & -- & 3.3e-05 & 1.0e-03 & 0.01 & 600.1 \\
50 & partial pattern &  6 & time limit & -- & 0 & 4.8e-03 & 0.01 & 600.1 \\
50 & partial pattern &  7 & time limit & -- & 0 & 2.5e-03 & 0.01 & 600.2 \\
50 & partial pattern &  8 & time limit & -- & 3.3e-05 & 3.7e-03 & 0.02 & 600.1 \\
50 & partial pattern &  9 & optimal & no & 2.3e-04 & 2.3e-04 & 0.00 & 303.2 \\
50 & partial pattern & 10 & time limit & -- & 1.6e-04 & 1.3e-03 & 0.01 & 600.1 \\
50 & $b=5$ buckets &  1 & optimal & yes & 0 & 0 & 0.00 & 448.3 \\
50 & $b=5$ buckets &  2 & optimal & yes & 0 & 0 & 0.01 & 186.6 \\
50 & $b=5$ buckets &  3 & optimal & yes & 0 & 0 & 0.01 & 326.9 \\
50 & $b=5$ buckets &  4 & optimal & yes & 0 & 0 & 0.02 & 473.5 \\
50 & $b=5$ buckets &  5 & optimal & yes & 0 & 0 & 0.01 & 526.7 \\
50 & $b=5$ buckets &  6 & time limit & -- & 0 & 8.3e-03 & 0.02 & 600.2 \\
50 & $b=5$ buckets &  7 & time limit & -- & 0 & 6.5e-03 & 0.02 & 600.1 \\
50 & $b=5$ buckets &  8 & optimal & yes & 0 & 0 & 0.02 & 311.9 \\
50 & $b=5$ buckets &  9 & optimal & yes & 0 & 0 & 0.02 & 349.4 \\
50 & $b=5$ buckets & 10 & time limit & -- & 0 & 5.8e-03 & 0.02 & 600.0 \\
50 & top-5 list &  1 & optimal & yes & 0 & 0 & 0.02 & 0.7 \\
50 & top-5 list &  2 & optimal & yes & 0 & 0 & 0.00 & 0.9 \\
50 & top-5 list &  3 & optimal & yes & 0 & 0 & 0.00 & 1.1 \\
50 & top-5 list &  4 & optimal & yes & 0 & 0 & 0.00 & 0.8 \\
50 & top-5 list &  5 & optimal & yes & 0 & 0 & 0.02 & 0.8 \\
50 & top-5 list &  6 & optimal & yes & 0 & 0 & 0.02 & 0.8 \\
50 & top-5 list &  7 & optimal & yes & 0 & 0 & 0.00 & 0.7 \\
50 & top-5 list &  8 & optimal & yes & 0 & 0 & 0.01 & 1.1 \\
50 & top-5 list &  9 & optimal & yes & 0 & 0 & 0.01 & 0.7 \\
50 & top-5 list & 10 & optimal & yes & 0 & 0 & 0.01 & 1.2 \\
\end{longtable}}


\begin{thebibliography}{73}
\expandafter\ifx\csname natexlab\endcsname\relax\def\natexlab#1{#1}\fi
\providecommand{\url}[1]{\texttt{#1}}
\providecommand{\href}[2]{#2}
\providecommand{\path}[1]{#1}
\providecommand{\DOIprefix}{doi:}
\providecommand{\ArXivprefix}{arXiv:}
\providecommand{\URLprefix}{URL: }
\providecommand{\Pubmedprefix}{pmid:}
\providecommand{\doi}[1]{\href{http://dx.doi.org/#1}{\path{#1}}}
\providecommand{\Pubmed}[1]{\href{pmid:#1}{\path{#1}}}
\providecommand{\bibinfo}[2]{#2}
\ifx\xfnm\relax \def\xfnm[#1]{\unskip,\space#1}\fi
\bibitem[{Ailon(2010)}]{ailon2010}
\bibinfo{author}{Ailon, N.}, \bibinfo{year}{2010}.
\newblock \bibinfo{title}{Aggregation of partial rankings, p-ratings and top-m
  lists}.
\newblock \bibinfo{journal}{Algorithmica} \bibinfo{volume}{57},
  \bibinfo{pages}{284--300}.
\bibitem[{Ailon et~al.(2008)Ailon, Charikar and
  Newman}]{ailoncharikarnewman2008}
\bibinfo{author}{Ailon, N.}, \bibinfo{author}{Charikar, M.},
  \bibinfo{author}{Newman, A.}, \bibinfo{year}{2008}.
\newblock \bibinfo{title}{Aggregating inconsistent information: ranking and
  clustering}.
\newblock \bibinfo{journal}{Journal of the ACM} \bibinfo{volume}{55},
  \bibinfo{pages}{23:1--23:27}.
\bibitem[{Akbari and Escobedo(2021)}]{akbariescobedo2021}
\bibinfo{author}{Akbari, S.}, \bibinfo{author}{Escobedo, A.R.},
  \bibinfo{year}{2021}.
\newblock \bibinfo{title}{Lower bounds on {K}emeny rank aggregation with
  non-strict rankings}, in: \bibinfo{booktitle}{2021 IEEE Symposium Series on
  Computational Intelligence (SSCI)}, \bibinfo{publisher}{IEEE}. pp.
  \bibinfo{pages}{1--8}.
\bibitem[{Akbari and Escobedo(2023a)}]{akbariescobedo2023a}
\bibinfo{author}{Akbari, S.}, \bibinfo{author}{Escobedo, A.R.},
  \bibinfo{year}{2023}a.
\newblock \bibinfo{title}{Approximate {C}ondorcet partitioning: solving
  large-scale rank aggregation problems}.
\newblock \bibinfo{journal}{Computers \& Operations Research}
  \bibinfo{volume}{153}, \bibinfo{pages}{106164}.
\bibitem[{Akbari and Escobedo(2023b)}]{akbariescobedo2023b}
\bibinfo{author}{Akbari, S.}, \bibinfo{author}{Escobedo, A.R.},
  \bibinfo{year}{2023}b.
\newblock \bibinfo{title}{Beyond {K}emeny rank aggregation: a
  parameterizable-penalty framework for robust ranking aggregation with ties}.
\newblock \bibinfo{journal}{Omega} \bibinfo{volume}{119},
  \bibinfo{pages}{102893}.
\bibitem[{Aledo et~al.(2025a)Aledo, Dom{\'i}nguez, Jaime-Alc{\'a}ntara and
  Landete}]{aledoetal2025}
\bibinfo{author}{Aledo, J.A.}, \bibinfo{author}{Dom{\'i}nguez, C.},
  \bibinfo{author}{Jaime-Alc{\'a}ntara, J.d.D.}, \bibinfo{author}{Landete, M.},
  \bibinfo{year}{2025}a.
\newblock \bibinfo{title}{An optimization-based approach to ranking aggregation
  with weak order outputs}.
\newblock \bibinfo{howpublished}{arXiv preprint arXiv:2511.19345 (version 2,
  July 2026)}.
\bibitem[{Aledo et~al.(2017)Aledo, G{\'a}mez and Rosete}]{aledoetal2017}
\bibinfo{author}{Aledo, J.A.}, \bibinfo{author}{G{\'a}mez, J.A.},
  \bibinfo{author}{Rosete, A.}, \bibinfo{year}{2017}.
\newblock \bibinfo{title}{Utopia in the solution of the bucket order problem}.
\newblock \bibinfo{journal}{Decision Support Systems} \bibinfo{volume}{97},
  \bibinfo{pages}{69--80}.
\bibitem[{Aledo et~al.(2018)Aledo, G{\'a}mez and Rosete}]{aledoetal2018}
\bibinfo{author}{Aledo, J.A.}, \bibinfo{author}{G{\'a}mez, J.A.},
  \bibinfo{author}{Rosete, A.}, \bibinfo{year}{2018}.
\newblock \bibinfo{title}{Approaching rank aggregation problems by using
  evolution strategies: the case of the optimal bucket order problem}.
\newblock \bibinfo{journal}{European Journal of Operational Research}
  \bibinfo{volume}{270}, \bibinfo{pages}{982--998}.
\bibitem[{Aledo et~al.(2021)Aledo, G{\'a}mez and Rosete}]{aledoetal2021}
\bibinfo{author}{Aledo, J.A.}, \bibinfo{author}{G{\'a}mez, J.A.},
  \bibinfo{author}{Rosete, A.}, \bibinfo{year}{2021}.
\newblock \bibinfo{title}{A highly scalable algorithm for weak rankings
  aggregation}.
\newblock \bibinfo{journal}{Information Sciences} \bibinfo{volume}{570},
  \bibinfo{pages}{144--171}.
\bibitem[{Aledo et~al.(2025b)Aledo, G{\'a}mez and
  Rosete}]{aledogamezrosete2025}
\bibinfo{author}{Aledo, J.A.}, \bibinfo{author}{G{\'a}mez, J.A.},
  \bibinfo{author}{Rosete, A.}, \bibinfo{year}{2025}b.
\newblock \bibinfo{title}{A consensus set for the aggregation of partial
  rankings: the case of the optimal set of bucket orders problem}.
\newblock \bibinfo{journal}{Information Fusion} \bibinfo{volume}{124},
  \bibinfo{pages}{103353}.
\newblock \DOIprefix\doi{10.1016/j.inffus.2025.103353}.
\bibitem[{Ali and Meil{\u{a}}(2012)}]{alimeila2012}
\bibinfo{author}{Ali, A.}, \bibinfo{author}{Meil{\u{a}}, M.},
  \bibinfo{year}{2012}.
\newblock \bibinfo{title}{Experiments with {K}emeny ranking: what works when?}
\newblock \bibinfo{journal}{Mathematical Social Sciences} \bibinfo{volume}{64},
  \bibinfo{pages}{28--40}.
\bibitem[{Alvo and Yu(2014)}]{alvoyu2014}
\bibinfo{author}{Alvo, M.}, \bibinfo{author}{Yu, P.L.H.}, \bibinfo{year}{2014}.
\newblock \bibinfo{title}{Statistical Methods for Ranking Data}.
\newblock \bibinfo{publisher}{Springer}, \bibinfo{address}{New York}.
\bibitem[{Amodio et~al.(2016)Amodio, D'Ambrosio and Siciliano}]{amodioetal2016}
\bibinfo{author}{Amodio, S.}, \bibinfo{author}{D'Ambrosio, A.},
  \bibinfo{author}{Siciliano, R.}, \bibinfo{year}{2016}.
\newblock \bibinfo{title}{Accurate algorithms for identifying the median
  ranking when dealing with weak and partial rankings under the {K}emeny
  axiomatic approach}.
\newblock \bibinfo{journal}{European Journal of Operational Research}
  \bibinfo{volume}{249}, \bibinfo{pages}{667--676}.
\bibitem[{Andrews(1976)}]{andrews1976}
\bibinfo{author}{Andrews, G.E.}, \bibinfo{year}{1976}.
\newblock \bibinfo{title}{The Theory of Partitions}. volume~\bibinfo{volume}{2}
  of \textit{\bibinfo{series}{Encyclopedia of Mathematics and its
  Applications}}.
\newblock \bibinfo{publisher}{Addison-Wesley}, \bibinfo{address}{Reading, MA}.
\bibitem[{Andrieu et~al.(2023)Andrieu, Cohen-Boulakia, Couceiro, Denise and
  Pierrot}]{andrieuetal2023}
\bibinfo{author}{Andrieu, P.}, \bibinfo{author}{Cohen-Boulakia, S.},
  \bibinfo{author}{Couceiro, M.}, \bibinfo{author}{Denise, A.},
  \bibinfo{author}{Pierrot, A.}, \bibinfo{year}{2023}.
\newblock \bibinfo{title}{A unifying rank aggregation framework to suitably and
  efficiently aggregate any kind of rankings}.
\newblock \bibinfo{journal}{International Journal of Approximate Reasoning}
  \bibinfo{volume}{162}, \bibinfo{pages}{109035}.
\newblock \DOIprefix\doi{10.1016/j.ijar.2023.109035}.
\bibitem[{Azzini et~al.(2027)Azzini, Morandi and Munda}]{azzinietal2026}
\bibinfo{author}{Azzini, I.}, \bibinfo{author}{Morandi, N.},
  \bibinfo{author}{Munda, G.}, \bibinfo{year}{2027}.
\newblock \bibinfo{title}{Further improvements of mork algorithms for
  identifying the kemeny median rankings}.
\newblock \bibinfo{journal}{European Journal of Operational Research}
  \bibinfo{volume}{336}, \bibinfo{pages}{605--616}.
\bibitem[{Azzini and Munda(2020)}]{azzinimunda2020}
\bibinfo{author}{Azzini, I.}, \bibinfo{author}{Munda, G.},
  \bibinfo{year}{2020}.
\newblock \bibinfo{title}{A new approach for identifying the {K}emeny median
  ranking}.
\newblock \bibinfo{journal}{European Journal of Operational Research}
  \bibinfo{volume}{281}, \bibinfo{pages}{388--401}.
\bibitem[{Badal and Das(2018)}]{badaldas2018}
\bibinfo{author}{Badal, P.S.}, \bibinfo{author}{Das, A.}, \bibinfo{year}{2018}.
\newblock \bibinfo{title}{Efficient algorithms using subiterative convergence
  for {K}emeny ranking problem}.
\newblock \bibinfo{journal}{Computers \& Operations Research}
  \bibinfo{volume}{98}, \bibinfo{pages}{198--210}.
\bibitem[{Badal and Singh(2026)}]{badalsingh2026}
\bibinfo{author}{Badal, P.S.}, \bibinfo{author}{Singh, R.},
  \bibinfo{year}{2026}.
\newblock \bibinfo{title}{Efficient heuristics for handling ties in {K}emeny
  rank aggregation}.
\newblock \bibinfo{journal}{Operations Research Perspectives}
  \bibinfo{volume}{17}, \bibinfo{pages}{100413}.
\newblock \DOIprefix\doi{10.1016/j.orp.2026.100413}.
\bibitem[{Barth{\'e}lemy et~al.(1989)Barth{\'e}lemy, Gu{\'e}noche and
  Hudry}]{barthelemyetal1989}
\bibinfo{author}{Barth{\'e}lemy, J.P.}, \bibinfo{author}{Gu{\'e}noche, A.},
  \bibinfo{author}{Hudry, O.}, \bibinfo{year}{1989}.
\newblock \bibinfo{title}{Median linear orders: heuristics and a branch and
  bound algorithm}.
\newblock \bibinfo{journal}{European Journal of Operational Research}
  \bibinfo{volume}{42}, \bibinfo{pages}{313--325}.
\newblock \DOIprefix\doi{10.1016/0377-2217(89)90442-6}.
\bibitem[{Bartholdi et~al.(1989)Bartholdi, Tovey and Trick}]{bartholdietal1989}
\bibinfo{author}{Bartholdi, J.}, \bibinfo{author}{Tovey, C.A.},
  \bibinfo{author}{Trick, M.A.}, \bibinfo{year}{1989}.
\newblock \bibinfo{title}{Voting schemes for which it can be difficult to tell
  who won the election}.
\newblock \bibinfo{journal}{Social Choice and Welfare} \bibinfo{volume}{6},
  \bibinfo{pages}{157--165}.
\bibitem[{Brancotte et~al.(2015)Brancotte, Yang, Blin, Cohen-Boulakia, Denise
  and Hamel}]{brancotteetal2015}
\bibinfo{author}{Brancotte, B.}, \bibinfo{author}{Yang, B.},
  \bibinfo{author}{Blin, G.}, \bibinfo{author}{Cohen-Boulakia, S.},
  \bibinfo{author}{Denise, A.}, \bibinfo{author}{Hamel, S.},
  \bibinfo{year}{2015}.
\newblock \bibinfo{title}{Rank aggregation with ties: experiments and
  analysis}.
\newblock \bibinfo{journal}{Proceedings of the VLDB Endowment}
  \bibinfo{volume}{8}, \bibinfo{pages}{1202--1213}.
\bibitem[{Brown(1980)}]{brown1980}
\bibinfo{author}{Brown, S.R.}, \bibinfo{year}{1980}.
\newblock \bibinfo{title}{Political Subjectivity: Applications of {Q}
  Methodology in Political Science}.
\newblock \bibinfo{publisher}{Yale University Press}, \bibinfo{address}{New
  Haven}.
\bibitem[{Burgoyne et~al.(2023)Burgoyne, Spijkervet and
  Baker}]{burgoyneetal2023}
\bibinfo{author}{Burgoyne, J.A.}, \bibinfo{author}{Spijkervet, J.},
  \bibinfo{author}{Baker, D.J.}, \bibinfo{year}{2023}.
\newblock \bibinfo{title}{Measuring the {E}urovision {S}ong {C}ontest: A living
  dataset for real-world {MIR}}, in: \bibinfo{booktitle}{Proceedings of the
  24th International Society for Music Information Retrieval Conference},
  \bibinfo{address}{Milan}. pp. \bibinfo{pages}{817--823}.
\bibitem[{Ceberio et~al.(2015)Ceberio, Mendiburu and Lozano}]{ceberioetal2015}
\bibinfo{author}{Ceberio, J.}, \bibinfo{author}{Mendiburu, A.},
  \bibinfo{author}{Lozano, J.A.}, \bibinfo{year}{2015}.
\newblock \bibinfo{title}{The linear ordering problem revisited}.
\newblock \bibinfo{journal}{European Journal of Operational Research}
  \bibinfo{volume}{241}, \bibinfo{pages}{686--696}.
\bibitem[{Cohen-Boulakia et~al.(2011)Cohen-Boulakia, Denise and
  Hamel}]{cohenboulakiaetal2011}
\bibinfo{author}{Cohen-Boulakia, S.}, \bibinfo{author}{Denise, A.},
  \bibinfo{author}{Hamel, S.}, \bibinfo{year}{2011}.
\newblock \bibinfo{title}{Using medians to generate consensus rankings for
  biological data}, in: \bibinfo{booktitle}{Scientific and Statistical Database
  Management (SSDBM 2011)}, \bibinfo{publisher}{Springer},
  \bibinfo{address}{Berlin, Heidelberg}. pp. \bibinfo{pages}{73--90}.
\newblock \DOIprefix\doi{10.1007/978-3-642-22351-8_5}.
\bibitem[{Cook(2006)}]{cook2006}
\bibinfo{author}{Cook, W.D.}, \bibinfo{year}{2006}.
\newblock \bibinfo{title}{Distance-based and ad hoc consensus models in ordinal
  preference ranking}.
\newblock \bibinfo{journal}{European Journal of Operational Research}
  \bibinfo{volume}{172}, \bibinfo{pages}{369--385}.
\bibitem[{Cook et~al.(2007)Cook, Golany, Penn and Raviv}]{Cook2007}
\bibinfo{author}{Cook, W.D.}, \bibinfo{author}{Golany, B.},
  \bibinfo{author}{Penn, M.}, \bibinfo{author}{Raviv, T.},
  \bibinfo{year}{2007}.
\newblock \bibinfo{title}{Creating a consensus ranking of proposals from
  reviewers' partial ordinal rankings}.
\newblock \bibinfo{journal}{Computers \& Operations Research}
  \bibinfo{volume}{34}, \bibinfo{pages}{954--965}.
\bibitem[{Cook and Saipe(1976)}]{cooksaipe1976}
\bibinfo{author}{Cook, W.D.}, \bibinfo{author}{Saipe, A.L.},
  \bibinfo{year}{1976}.
\newblock \bibinfo{title}{Committee approach to priority planning: the median
  ranking method}.
\newblock \bibinfo{journal}{Cahiers du Centre d'{\'E}tudes de Recherche
  Op{\'e}rationnelle} \bibinfo{volume}{18}, \bibinfo{pages}{337--351}.
\bibitem[{Cook and Seiford(1978)}]{Cook1978}
\bibinfo{author}{Cook, W.D.}, \bibinfo{author}{Seiford, L.M.},
  \bibinfo{year}{1978}.
\newblock \bibinfo{title}{Priority ranking and consensus formation}.
\newblock \bibinfo{journal}{Management Science} \bibinfo{volume}{24},
  \bibinfo{pages}{1721--1732}.
\bibitem[{Critchlow(1985)}]{critchlow1985}
\bibinfo{author}{Critchlow, D.E.}, \bibinfo{year}{1985}.
\newblock \bibinfo{title}{Metric Methods for Analyzing Partially Ranked Data}.
  volume~\bibinfo{volume}{34} of \textit{\bibinfo{series}{Lecture Notes in
  Statistics}}.
\newblock \bibinfo{publisher}{Springer}, \bibinfo{address}{New York}.
\bibitem[{D'Ambrosio et~al.(2026)D'Ambrosio, Amodio, Mazzeo, Albano and
  Plaia}]{consrank}
\bibinfo{author}{D'Ambrosio, A.}, \bibinfo{author}{Amodio, S.},
  \bibinfo{author}{Mazzeo, G.}, \bibinfo{author}{Albano, A.},
  \bibinfo{author}{Plaia, A.}, \bibinfo{year}{2026}.
\newblock \bibinfo{title}{{ConsRank}: Compute the Median Ranking(s) According
  to the {K}emeny's Axiomatic Approach}.
\newblock \bibinfo{note}{R package version 3.0}.
\newblock \URLprefix \url{https://CRAN.R-project.org/package=ConsRank}.
\bibitem[{D'Ambrosio et~al.(2019)D'Ambrosio, Iorio, Staiano and
  Siciliano}]{dambrosioetal2019}
\bibinfo{author}{D'Ambrosio, A.}, \bibinfo{author}{Iorio, C.},
  \bibinfo{author}{Staiano, M.}, \bibinfo{author}{Siciliano, R.},
  \bibinfo{year}{2019}.
\newblock \bibinfo{title}{Median constrained bucket order rank aggregation}.
\newblock \bibinfo{journal}{Computational Statistics} \bibinfo{volume}{34},
  \bibinfo{pages}{787--802}.
\newblock \DOIprefix\doi{10.1007/s00180-018-0858-z}.
\bibitem[{D'Ambrosio et~al.(2017)D'Ambrosio, Mazzeo, Iorio and
  Siciliano}]{dambrosioetal2017}
\bibinfo{author}{D'Ambrosio, A.}, \bibinfo{author}{Mazzeo, G.},
  \bibinfo{author}{Iorio, C.}, \bibinfo{author}{Siciliano, R.},
  \bibinfo{year}{2017}.
\newblock \bibinfo{title}{A differential evolution algorithm for finding the
  median ranking under the {K}emeny axiomatic approach}.
\newblock \bibinfo{journal}{Computers \& Operations Research}
  \bibinfo{volume}{82}, \bibinfo{pages}{126--138}.
\bibitem[{Dwork et~al.(2001)Dwork, Kumar, Naor and Sivakumar}]{dworketal2001}
\bibinfo{author}{Dwork, C.}, \bibinfo{author}{Kumar, R.},
  \bibinfo{author}{Naor, M.}, \bibinfo{author}{Sivakumar, D.},
  \bibinfo{year}{2001}.
\newblock \bibinfo{title}{Rank aggregation methods for the web}, in:
  \bibinfo{booktitle}{Proceedings of the 10th International Conference on World
  Wide Web}, \bibinfo{publisher}{ACM}, \bibinfo{address}{New York}. pp.
  \bibinfo{pages}{613--622}.
\bibitem[{Emond and Mason(2000)}]{emondmason2000}
\bibinfo{author}{Emond, E.J.}, \bibinfo{author}{Mason, D.W.},
  \bibinfo{year}{2000}.
\newblock \bibinfo{title}{A new technique for high level decision support}.
\newblock \bibinfo{type}{Technical Report}. Department of National Defence,
  Operational Research Division. \bibinfo{address}{Ottawa, Canada}.
\bibitem[{Emond and Mason(2002)}]{emondmason2002}
\bibinfo{author}{Emond, E.J.}, \bibinfo{author}{Mason, D.W.},
  \bibinfo{year}{2002}.
\newblock \bibinfo{title}{A new rank correlation coefficient with application
  to the consensus ranking problem}.
\newblock \bibinfo{journal}{Journal of Multi-Criteria Decision Analysis}
  \bibinfo{volume}{11}, \bibinfo{pages}{17--28}.
\bibitem[{{European Commission}(2012)}]{eurobarometer552}
\bibinfo{author}{{European Commission}}, \bibinfo{year}{2012}.
\newblock \bibinfo{title}{Eurobarometer 55.2 (may--june 2001)}.
\newblock \bibinfo{howpublished}{GESIS Data Archive, Cologne. ZA3509 Data file
  Version 1.0.1, doi:10.4232/1.10943}.
\bibitem[{Fagin et~al.(2004)Fagin, Kumar, Mahdian, Sivakumar and
  Vee}]{faginetal2004}
\bibinfo{author}{Fagin, R.}, \bibinfo{author}{Kumar, R.},
  \bibinfo{author}{Mahdian, M.}, \bibinfo{author}{Sivakumar, D.},
  \bibinfo{author}{Vee, E.}, \bibinfo{year}{2004}.
\newblock \bibinfo{title}{Comparing and aggregating rankings with ties}, in:
  \bibinfo{booktitle}{Proceedings of the 23rd ACM SIGMOD-SIGACT-SIGART
  Symposium on Principles of Database Systems}, pp. \bibinfo{pages}{47--58}.
\newblock \DOIprefix\doi{10.1145/1055558.1055568}.
\bibitem[{Fagin et~al.(2006)Fagin, Kumar, Mahdian, Sivakumar and
  Vee}]{faginetal2006}
\bibinfo{author}{Fagin, R.}, \bibinfo{author}{Kumar, R.},
  \bibinfo{author}{Mahdian, M.}, \bibinfo{author}{Sivakumar, D.},
  \bibinfo{author}{Vee, E.}, \bibinfo{year}{2006}.
\newblock \bibinfo{title}{Comparing partial rankings}.
\newblock \bibinfo{journal}{SIAM Journal on Discrete Mathematics}
  \bibinfo{volume}{20}, \bibinfo{pages}{628--648}.
\bibitem[{Fagin et~al.(2003)Fagin, Kumar and Sivakumar}]{faginetal2003}
\bibinfo{author}{Fagin, R.}, \bibinfo{author}{Kumar, R.},
  \bibinfo{author}{Sivakumar, D.}, \bibinfo{year}{2003}.
\newblock \bibinfo{title}{Comparing top $k$ lists}.
\newblock \bibinfo{journal}{SIAM Journal on Discrete Mathematics}
  \bibinfo{volume}{17}, \bibinfo{pages}{134--160}.
\bibitem[{Fisher(1958)}]{fisher1958}
\bibinfo{author}{Fisher, W.D.}, \bibinfo{year}{1958}.
\newblock \bibinfo{title}{On grouping for maximum homogeneity}.
\newblock \bibinfo{journal}{Journal of the American Statistical Association}
  \bibinfo{volume}{53}, \bibinfo{pages}{789--798}.
\bibitem[{Fligner and Verducci(1986)}]{flignerverducci1986}
\bibinfo{author}{Fligner, M.A.}, \bibinfo{author}{Verducci, J.S.},
  \bibinfo{year}{1986}.
\newblock \bibinfo{title}{Distance based ranking models}.
\newblock \bibinfo{journal}{Journal of the Royal Statistical Society: Series B}
  \bibinfo{volume}{48}, \bibinfo{pages}{359--369}.
\bibitem[{Gionis et~al.(2006)Gionis, Mannila, Puolam{\"a}ki and
  Ukkonen}]{gionisetal2006}
\bibinfo{author}{Gionis, A.}, \bibinfo{author}{Mannila, H.},
  \bibinfo{author}{Puolam{\"a}ki, K.}, \bibinfo{author}{Ukkonen, A.},
  \bibinfo{year}{2006}.
\newblock \bibinfo{title}{Algorithms for discovering bucket orders from data},
  in: \bibinfo{booktitle}{Proceedings of the 12th ACM SIGKDD International
  Conference on Knowledge Discovery and Data Mining}, \bibinfo{publisher}{ACM},
  \bibinfo{address}{New York}. pp. \bibinfo{pages}{561--566}.
\bibitem[{Good(1975)}]{good1975}
\bibinfo{author}{Good, I.J.}, \bibinfo{year}{1975}.
\newblock \bibinfo{title}{The number of orderings of $n$ candidates when ties
  are permitted}.
\newblock \bibinfo{journal}{Fibonacci Quarterly} \bibinfo{volume}{13},
  \bibinfo{pages}{11--18}.
\bibitem[{Gross(1962)}]{gross1962}
\bibinfo{author}{Gross, O.A.}, \bibinfo{year}{1962}.
\newblock \bibinfo{title}{Preferential arrangements}.
\newblock \bibinfo{journal}{The American Mathematical Monthly}
  \bibinfo{volume}{69}, \bibinfo{pages}{4--8}.
\bibitem[{Harper and Konstan(2015)}]{harperkonstan2015}
\bibinfo{author}{Harper, F.M.}, \bibinfo{author}{Konstan, J.A.},
  \bibinfo{year}{2015}.
\newblock \bibinfo{title}{The {MovieLens} datasets: History and context}.
\newblock \bibinfo{journal}{ACM Transactions on Interactive Intelligent
  Systems} \bibinfo{volume}{5}, \bibinfo{pages}{19:1--19:19}.
\bibitem[{Huangfu and Hall(2018)}]{huangfuhall2018}
\bibinfo{author}{Huangfu, Q.}, \bibinfo{author}{Hall, J.A.J.},
  \bibinfo{year}{2018}.
\newblock \bibinfo{title}{Parallelizing the dual revised simplex method}.
\newblock \bibinfo{journal}{Mathematical Programming Computation}
  \bibinfo{volume}{10}, \bibinfo{pages}{119--142}.
\bibitem[{Kamishima(2003)}]{kamishima2003}
\bibinfo{author}{Kamishima, T.}, \bibinfo{year}{2003}.
\newblock \bibinfo{title}{Nantonac collaborative filtering: Recommendation
  based on order responses}, in: \bibinfo{booktitle}{Proceedings of the Ninth
  ACM SIGKDD International Conference on Knowledge Discovery and Data Mining},
  pp. \bibinfo{pages}{583--588}.
\bibitem[{Kawashima and Sukegawa(2026)}]{kawashimasukegawa2026}
\bibinfo{author}{Kawashima, D.}, \bibinfo{author}{Sukegawa, N.},
  \bibinfo{year}{2026}.
\newblock \bibinfo{title}{Condorcet-type properties of the linear ordering
  problem with ties}.
\newblock \bibinfo{note}{ArXiv:2609.19593}.
\bibitem[{Kemeny(1959)}]{kemeny1959}
\bibinfo{author}{Kemeny, J.G.}, \bibinfo{year}{1959}.
\newblock \bibinfo{title}{Mathematics without numbers}.
\newblock \bibinfo{journal}{Daedalus} \bibinfo{volume}{88},
  \bibinfo{pages}{577--591}.
\bibitem[{Kemeny and Snell(1962)}]{kemenysnell1962}
\bibinfo{author}{Kemeny, J.G.}, \bibinfo{author}{Snell, J.L.},
  \bibinfo{year}{1962}.
\newblock \bibinfo{title}{Preference rankings: an axiomatic approach}, in:
  \bibinfo{booktitle}{Mathematical Models in the Social Sciences}.
  \bibinfo{publisher}{Blaisdell}, \bibinfo{address}{New York}, pp.
  \bibinfo{pages}{9--23}.
\bibitem[{Lipset(1963)}]{lipset1963}
\bibinfo{author}{Lipset, S.M.}, \bibinfo{year}{1963}.
\newblock \bibinfo{title}{The value patterns of democracy: A case study in
  comparative analysis}.
\newblock \bibinfo{journal}{American Sociological Review} \bibinfo{volume}{28},
  \bibinfo{pages}{515--531}.
\bibitem[{Lorena et~al.(2022)Lorena, Chaves, Mauri and Lorena}]{lorenaetal2022}
\bibinfo{author}{Lorena, L.H.N.}, \bibinfo{author}{Chaves, A.A.},
  \bibinfo{author}{Mauri, G.R.}, \bibinfo{author}{Lorena, L.A.N.},
  \bibinfo{year}{2022}.
\newblock \bibinfo{title}{An adaptive biased random-key genetic algorithm for
  rank aggregation with ties and incomplete rankings}, in:
  \bibinfo{booktitle}{2022 IEEE Congress on Evolutionary Computation (CEC)}.
\newblock \DOIprefix\doi{10.1109/CEC55065.2022.9870203}.
\bibitem[{Luce(1959)}]{luce1959}
\bibinfo{author}{Luce, R.D.}, \bibinfo{year}{1959}.
\newblock \bibinfo{title}{Individual Choice Behavior: A Theoretical Analysis}.
\newblock \bibinfo{publisher}{Wiley}, \bibinfo{address}{New York}.
\bibitem[{Mallows(1957)}]{mallows1957}
\bibinfo{author}{Mallows, C.L.}, \bibinfo{year}{1957}.
\newblock \bibinfo{title}{Non-null ranking models. {I}}.
\newblock \bibinfo{journal}{Biometrika} \bibinfo{volume}{44},
  \bibinfo{pages}{114--130}.
\bibitem[{Marden(1996)}]{marden1996}
\bibinfo{author}{Marden, J.I.}, \bibinfo{year}{1996}.
\newblock \bibinfo{title}{Analyzing and Modeling Rank Data}.
\newblock \bibinfo{publisher}{Chapman \& Hall}, \bibinfo{address}{London}.
\bibitem[{Mathieu and Mauras(2020)}]{mathieumauras2020}
\bibinfo{author}{Mathieu, C.}, \bibinfo{author}{Mauras, S.},
  \bibinfo{year}{2020}.
\newblock \bibinfo{title}{How to aggregate top-lists: approximation algorithms
  via scores and average ranks}, in: \bibinfo{booktitle}{Proceedings of the
  31st Annual ACM-SIAM Symposium on Discrete Algorithms},
  \bibinfo{publisher}{SIAM}.
\bibitem[{Mattei and Walsh(2013)}]{matteiwalsh2013}
\bibinfo{author}{Mattei, N.}, \bibinfo{author}{Walsh, T.},
  \bibinfo{year}{2013}.
\newblock \bibinfo{title}{{PrefLib}: A library for preferences}, in:
  \bibinfo{booktitle}{Algorithmic Decision Theory. ADT 2013},
  \bibinfo{publisher}{Springer}, \bibinfo{address}{Berlin, Heidelberg}. pp.
  \bibinfo{pages}{259--270}.
\bibitem[{Phung and Hamel(2025)}]{phunghamel2025}
\bibinfo{author}{Phung, X.K.}, \bibinfo{author}{Hamel, S.},
  \bibinfo{year}{2025}.
\newblock \bibinfo{title}{Efficient space reduction techniques by optimized
  majority rules for the {K}emeny aggregation problem}.
\newblock \bibinfo{note}{ArXiv:2506.15097}.
\bibitem[{Plackett(1975)}]{plackett1975}
\bibinfo{author}{Plackett, R.L.}, \bibinfo{year}{1975}.
\newblock \bibinfo{title}{The analysis of permutations}.
\newblock \bibinfo{journal}{Journal of the Royal Statistical Society. Series C
  (Applied Statistics)} \bibinfo{volume}{24}, \bibinfo{pages}{193--202}.
\bibitem[{Rico et~al.(2023)Rico, Vela and D{\'i}az}]{ricoetal2023}
\bibinfo{author}{Rico, N.}, \bibinfo{author}{Vela, C.R.},
  \bibinfo{author}{D{\'i}az, I.}, \bibinfo{year}{2023}.
\newblock \bibinfo{title}{Reducing the time required to find the {K}emeny
  ranking by exploiting a necessary condition for being a winner}.
\newblock \bibinfo{journal}{European Journal of Operational Research}
  \bibinfo{volume}{305}, \bibinfo{pages}{1323--1336}.
\newblock \DOIprefix\doi{10.1016/j.ejor.2022.07.031}.
\bibitem[{Romano et~al.(2025)Romano, Conversano, Siciliano and
  D'Ambrosio}]{romanoetal2025}
\bibinfo{author}{Romano, M.}, \bibinfo{author}{Conversano, C.},
  \bibinfo{author}{Siciliano, R.}, \bibinfo{author}{D'Ambrosio, A.},
  \bibinfo{year}{2025}.
\newblock \bibinfo{title}{Particle swarm optimization for preference rankings}.
\newblock \bibinfo{journal}{Advances in Data Analysis and Classification}
  \DOIprefix\doi{10.1007/s11634-025-00626-9}.
\bibitem[{Schiavinotto and St{\"u}tzle(2004)}]{schiavinottostutzle2004}
\bibinfo{author}{Schiavinotto, T.}, \bibinfo{author}{St{\"u}tzle, T.},
  \bibinfo{year}{2004}.
\newblock \bibinfo{title}{The linear ordering problem: instances, search space
  analysis and algorithms}.
\newblock \bibinfo{journal}{Journal of Mathematical Modelling and Algorithms}
  \bibinfo{volume}{3}, \bibinfo{pages}{367--402}.
\bibitem[{Stanley(2012)}]{stanley2012}
\bibinfo{author}{Stanley, R.P.}, \bibinfo{year}{2012}.
\newblock \bibinfo{title}{Enumerative Combinatorics}.
  volume~\bibinfo{volume}{1}.
\newblock \bibinfo{edition}{2} ed., \bibinfo{publisher}{Cambridge University
  Press}, \bibinfo{address}{Cambridge}.
\bibitem[{Stephenson(1935)}]{stephenson1935}
\bibinfo{author}{Stephenson, W.}, \bibinfo{year}{1935}.
\newblock \bibinfo{title}{Technique of factor analysis}.
\newblock \bibinfo{journal}{Nature} \bibinfo{volume}{136},
  \bibinfo{pages}{297}.
\bibitem[{Ukkonen et~al.(2009)Ukkonen, Puolam{\"a}ki, Gionis and
  Mannila}]{ukkonenetal2009}
\bibinfo{author}{Ukkonen, A.}, \bibinfo{author}{Puolam{\"a}ki, K.},
  \bibinfo{author}{Gionis, A.}, \bibinfo{author}{Mannila, H.},
  \bibinfo{year}{2009}.
\newblock \bibinfo{title}{A randomized approximation algorithm for computing
  bucket orders}.
\newblock \bibinfo{journal}{Information Processing Letters}
  \bibinfo{volume}{109}, \bibinfo{pages}{356--359}.
\bibitem[{Watts and Stenner(2012)}]{wattsstenner2012}
\bibinfo{author}{Watts, S.}, \bibinfo{author}{Stenner, P.},
  \bibinfo{year}{2012}.
\newblock \bibinfo{title}{Doing {Q} Methodological Research: Theory, Method and
  Interpretation}.
\newblock \bibinfo{publisher}{Sage}, \bibinfo{address}{London}.
\bibitem[{Yoo and Escobedo(2021)}]{yooescobedo2021}
\bibinfo{author}{Yoo, Y.}, \bibinfo{author}{Escobedo, A.R.},
  \bibinfo{year}{2021}.
\newblock \bibinfo{title}{A new binary programming formulation and social
  choice property for {K}emeny rank aggregation}.
\newblock \bibinfo{journal}{Decision Analysis} \bibinfo{volume}{18},
  \bibinfo{pages}{296--320}.
\newblock \DOIprefix\doi{10.1287/deca.2021.0433}.
\bibitem[{Yoo et~al.(2020)Yoo, Escobedo and Skolfield}]{yooetal2020}
\bibinfo{author}{Yoo, Y.}, \bibinfo{author}{Escobedo, A.R.},
  \bibinfo{author}{Skolfield, J.K.}, \bibinfo{year}{2020}.
\newblock \bibinfo{title}{A new correlation coefficient for comparing and
  aggregating non-strict and incomplete rankings}.
\newblock \bibinfo{journal}{European Journal of Operational Research}
  \bibinfo{volume}{285}, \bibinfo{pages}{1025--1041}.
\newblock \DOIprefix\doi{10.1016/j.ejor.2020.02.027}.
\bibitem[{Young and Levenglick(1978)}]{younglevenglick1978}
\bibinfo{author}{Young, H.P.}, \bibinfo{author}{Levenglick, A.},
  \bibinfo{year}{1978}.
\newblock \bibinfo{title}{A consistent extension of {C}ondorcet's election
  principle}.
\newblock \bibinfo{journal}{SIAM Journal on Applied Mathematics}
  \bibinfo{volume}{35}, \bibinfo{pages}{285--300}.
\newblock \DOIprefix\doi{10.1137/0135023}.
\bibitem[{Zabala(2014)}]{zabala2014}
\bibinfo{author}{Zabala, A.}, \bibinfo{year}{2014}.
\newblock \bibinfo{title}{qmethod: a package to explore human perspectives
  using {Q} methodology}.
\newblock \bibinfo{journal}{The R Journal} \bibinfo{volume}{6},
  \bibinfo{pages}{163--173}.
\bibitem[{van Zuylen and Williamson(2009)}]{vanzuylenwilliamson2009}
\bibinfo{author}{van Zuylen, A.}, \bibinfo{author}{Williamson, D.P.},
  \bibinfo{year}{2009}.
\newblock \bibinfo{title}{Deterministic pivoting algorithms for constrained
  ranking and clustering problems}.
\newblock \bibinfo{journal}{Mathematics of Operations Research}
  \bibinfo{volume}{34}, \bibinfo{pages}{594--620}.

\end{thebibliography}
\end{document}